\documentclass[epj]{svjour}
\usepackage{amssymb}
\usepackage{amsmath,bm}
\usepackage{graphicx}
\usepackage{epstopdf}
\usepackage{slashed}
\usepackage{xcolor}
\usepackage{url}
\usepackage{xurl}
\usepackage{hyperref}

\def\beq{\begin{equation}}
\def\eeq{\end{equation}}

\newcommand{\gt}{>}

\renewcommand{\v}[1]{\textbf{#1}}
\renewcommand{\rm}[1]{\textrm{#1}}

\def\esym{$E_{sym}(\rho)$~}

\def\es0{$E_{sym}(\rho_0)$~}

\def\us0{$U_{sym}(\rho_0,k_F)$~}

\def\l0{$L(\rho_0)$~}

\def\l0{$L(\rho_0)$~}

\def\D {$\Delta(1232)$~}
\def\rc {$\rho^{\textrm{crit}}_{\Delta^-}$~}

\usepackage[figuresright]{rotating}

\begin{document}

\title{Beyond $\rho^{2/3}$ Scaling: Microscopic Origins and Multimessengers of High-Density Nuclear Symmetry Energy\thanks{\bf Dedicated to the memory of Prof. Philip J. Siemens (1943-2023).
}}

\author{Bao-An Li (Bao-An.Li@etamu.edu) }                     
%
%
\institute{Department of Physics and Astronomy, East Texas A$\&$M University, Commerce, TX 75429, USA}
\date{Received: date / Revised version: date}
%
\abstract{The nature and Equation of State (EOS) of dense neutron-rich matter are still very poorly known, while they have broad impacts on many interesting issues in both astrophysics and nuclear physics. In particular, the nuclear symmetry energy $E_{\mathrm{sym}}(\rho)$ encoding the cost to make nuclear matter more neutron-rich has been the most uncertain component of the EOS of dense neutron-rich nucleonic matter. It significantly affects the radii, tidal deformations, cooling rates, and frequencies of various oscillation modes of isolated neutron stars as well as the strain amplitude and frequencies of gravitational waves from their mergers, besides its many effects on the structures of nuclei as well as the dynamics and observables of their collisions. Siemens (1970s) observed that $E_{\mathrm{sym}}(\rho)$ scales as $(\rho/\rho_0)^{2/3}$ near the saturation density $\rho_0$ of nuclear matter, since both the kinetic part and the potential contribution (quadratic in momentum) exhibit this dependence. The scaling holds if: (1) the nucleon isoscalar potential is quadratic in momentum, and (2) the isovector interaction is weakly density dependent. After examining many empirical evidences and understanding theoretical findings in the literature we conclude that: (1) Siemens' $\rho^{2/3}$ scaling is robust and serves as a valuable benchmark for both nuclear theories and experiments up to $2\rho_0$ but breaks down at higher densities, (2) Experimental and theoretical findings about $E_{\mathrm{sym}}(\rho)$ up to $2\rho_0$ are broadly consistent, but uncertainties remain large for its curvature $K_{\mathrm{sym}}$ and higher-order parameters, (3) Above $2\rho_0$, uncertainties grow due to poorly constrained spin-isospin dependent tensor and three-body forces as well as the resulting nucleon short-range correlations. Looking forward, combining signals from both observations of neutron stars and terrestrial heavy-ion reaction experiments is the most promising path to finally constraining the high-density $E_{\mathrm{sym}}(\rho)$ and the EOS of supradense neutron-rich matter. Multiple examples of community efforts to further constrain the high-density $E_{\mathrm{sym}}(\rho)$ using both real and mocked data of present and future high-precision observations of neutron stars, as well as heavy-ion collisions involving high-energy rare isotopes, are briefly reviewed.
}
\authorrunning{Li}
\titlerunning{Microscopic Origins and Probes of High-Density Symmetry Energy}
\maketitle
\setcounter{tocdepth}{3} 
\tableofcontents
\section{Introduction}
The Equation of State (EOS) of neutron-rich matter governs the dynamics of nuclear reactions, neutron star mergers, and the structure of isolated neutron stars. Owing to the limited understanding of nuclear interactions and correlations under extreme conditions of density, temperature, pressure, and isospin asymmetry, as well as the long-standing challenge of solving accurately many-body problems, the EOS of dense neutron-rich matter remains poorly constrained. In particular, the nuclear symmetry energy $E_{\text{sym}}(\rho)$, which quantifies the energy cost of increasing neutron excess in nuclear systems, represents the most uncertain component of the nucleonic EOS at supra-saturation densities. Given its profound implications for both nuclear physics and astrophysics, constraining the high-density behavior of $E_{\text{sym}}(\rho)$ has long been a central and shared objective of many terrestrial nuclear experiments and astrophysical observations. Over the past two decades, sustained efforts by many researchers in both communities have been devoted to narrowing these uncertainties and advancing our understanding of the symmetry energy at high densities, see e.g., Refs. \cite{LiBA98,Li-Udo,Lat04,Ste05,Bar05,LCK08,Tra12,Tsa12,Hor14,LiBA14,Bal16,Oer17,LiBA17} for earlier reviews.

In this brief review of recent developments in studying high-density nuclear symmetry energy, we highlight new advances and outline several major open questions, particularly in light of forthcoming opportunities with high-energy radioactive beam experiments, high-precision X-ray observations of neutron stars, and next-generation gravitational wave detectors capable of probing high-frequency signals from neutron star mergers. 

We begin by recalling the key physical ingredients that govern the density dependence of nuclear symmetry energy. We then summarize the physical conditions and empirical evidence supporting Siemens’ $\rho^{2/3}$ scaling of $E_{\text{sym}}(\rho)$ around the saturation density $\rho_0$ of symmetric nuclear matter (SNM) \cite{Siemens1971}. Moving beyond this $\rho^{2/3}$ benchmark, we discuss the microscopic origins of the very uncertain high-density $E_{\rm{sym}}(\rho)$ and illustrate how multimessengers from heavy-ion collisions and neutron star observations are advancing our understanding of the high-density behavior of $E_{\rm{sym}}(\rho)$. We end this brief review with a list of major challenges for the community to overcome to finally pin down the high-density $E_{\rm{sym}}(\rho)$.

\section{Symmetry energy in terms of the Lane potential according to the Hugenholtz-Van Hove (HVH) theorem}\label{s-HVH}
For asymmetric nucleonic matter (ANM) of isospin asymmetry $\delta=(\rho_n-\rho_p)/\rho$ where $\rho_n$, $\rho_p$, and $\rho=\rho_n+\rho_p$ are the densities of neutrons, protons, and nucleons, respectively, its EOS can be written as
\begin{equation}\label{eos1}
E(\rho ,\delta )=E_0(\rho)+E_{\rm{sym},2}(\rho )\delta ^{2} +E_{\rm{sym},4}(\rho ) \delta ^{4} +\mathcal{O}(\delta^6)
\end{equation}
in terms of the energy per nucleon $E_0(\rho)\equiv E(\rho ,\delta=0)$ in SNM, the isospin-quadratic symmetry energy $E_{\rm{sym},2}(\rho )$ and the 
isospin-quartic (fourth-order) symmetry energy $E_{\mathrm{sym,4}}(\rho)$. It is necessary to note that Eq. (\ref{eos1}) is valid under the assumption that the masses of protons and neutrons are identical, such that there is a proton-neutron exchange symmetry in ANM.  Otherwise, there are also terms with odd powers of $\delta$.
In the literature, the $E_{\rm{sym},2}(\rho )$ is often referred to as the nuclear symmetry energy, denoted often by $E_{\rm{sym}}(\rho )$ or $S$. In the following, we use the notations $E_{\rm{sym}}(\rho )$, $S$ or $E_{\rm{sym},2}(\rho )$ interchangeably for nuclear symmetry energy. The notation $E_{\rm{sym},2}(\rho )$ is mostly used when the symmetry energy appears in the same equation with the isospin-quartic symmetry energy $E_{\mathrm{sym,4}}(\rho)$. Specifically, they are defined as 
\begin{equation}
S=E_{\rm{sym}}(\rho )\equiv E_{\rm{sym},2}(\rho )\equiv\left.\frac{1}{2}\frac{\partial ^{2}E(\rho,\delta )}{\partial \delta ^{2}}\right|_{\delta=0}
\end{equation}
and 
\begin{equation}
E_{\mathrm{sym,4}}(\rho )\equiv \left.\frac{1}{24}\frac{\partial ^{4}E(\rho,\delta )}{\partial \delta ^{4}}\right|_{\delta=0} \label{Esyme4}.
\end{equation}
If the $E_{\mathrm{sym,4}}(\rho)$ is negligibly small, the Eq. (\ref{eos1}) is reduced to the so-called empirical parabolic approximation (PA) of nuclear EOS \cite{Bom91}. Then, the symmetry energy can be approximated by the difference between the energy per nucleon in pure neutron matter (PNM) and SNM, i.e., 
$
E_{\rm sym}(\rho )\approx E(\rho,1)-E(\rho,0).$

It is also well known that the single-particle potential $U_{n/p}(k,\rho,\delta)$ for a nucleon $\tau=n/p$ with the third component of its isospin quantum number $\tau_3=\pm$ and a momentum $k$ in ANM can be written as
\begin{eqnarray}\label{sp}
&&U_{\tau}(k,\rho,\delta)=U_0(k,\rho)+\tau_3 U_{\rm sym,1}(k,\rho)\cdot\delta\nonumber\\
&&+U_{\rm sym,2}(k,\rho)\cdot\delta^2+\tau_3 U_{\rm sym,3}(k,\rho)\cdot\delta^3+\mathcal{O}(\delta^4)
\end{eqnarray}
in terms of the isoscalar $U_0(k,\rho)$ and $U_{\rm sym,2}(k,\rho)$ as well as the isovector $U_{\rm sym,1}(k,\rho)$ and $U_{\rm sym,3}(k,\rho)$ potentials, respectively. Keeping only the zeroth and first order terms of this expansion, Eq. (\ref{sp}) reduces naturally to the classical Lane potential 
\begin{equation}
 U_{\tau}(k,\rho,\delta)=U_0(k,\rho)+\tau_3 U_{\rm sym,1}(k,\rho)\cdot\delta   
\end{equation}
from earlier optical model analyses of nucleon-nucleus scattering data \cite{Lan62}. The isospin-dependent $\tau_3$ term originates from the 
$\vec{\tau_1}\cdot\vec{\tau_2}$ form of nucleon-nucleon effective interactions in nuclear matter. In particular, using such interactions, the resulting direct term of the mean-field potential for the nucleon with $\tau_3$ is proportional to the product of $\tau_3$ and the isospin asymmetry $\delta$ of the medium. For detailed discussions, see, e.g., Section 7.4 ``Isospin Potential and Symmetry Energy" in the textbook by P.J. Siemens and A.S. Jensen \cite{SiemensBook}.

The Hugenholtz-Van Hove (HVH) theorem \cite{Hug58,Sat99}
\begin{equation}\label{HVH}
E_{\rm{F}}=\frac{d (\rho E)}{d \rho} =  E+ \rho\frac{d E}{d \rho} = E+ P/\rho
\end{equation}
governs the relation between the Fermi energy $E_{\rm{F}}$ and the average energy per nucleon $E$ in all Fermionic systems at an arbitrary density $\rho$ with pressure $P$ at zero temperature. Applying it to the special case of SNM at saturation density $\rho_0$ where $P=0$, one obtains the well-known result of $E_{\rm{F}}=E.$ Generally, this fundamental theorem provides a direct link between the EOS of Eq. (\ref{eos1}) and the single-nucleon potentials of Eq. (\ref{sp}) at the Fermi momenta of neutrons and protons. Applying it to ANM with energy density $\varepsilon (\rho ,\delta )=\rho E(\rho ,\delta )$,
one has \cite{XuC10,XuC11,Rchen,CXu14}
\begin{eqnarray}
t(k_{F_{n}})+U_{n}(\rho ,\delta ,k_{F_{n}}) &=&\frac{\partial \varepsilon
(\rho ,\delta )}{\partial \rho _{n}},  \label{cUn} \\
t(k_{F_{p}})+U_{p}(\rho ,\delta ,k_{F_{p}}) &=&\frac{\partial \varepsilon
(\rho ,\delta )}{\partial \rho _{p}},  \label{cUp}
\end{eqnarray}%
where $t(k_{F_{\tau }})=k_{F_{\tau }}^{2}/2m$ is the nucleon kinetic energy
at Fermi momentum $k_{F_{\tau }}=k_{F}(1+\tau \delta )^{1/3}$ with $%
k_{F}=(3\pi ^{2}\rho /2)^{1/3}$ being the Fermi momentum in symmetric
nuclear matter at density $\rho$. All terms on both sides of the above two equations can be expanded in terms of $\delta$. 
In particular, $t(k_{F_{\tau }})$ and $U_{\tau }(\rho ,\delta
,k_{F_{\tau }})$ can be expanded as a power series of $\delta $,
respectively, as \cite{XuC11,Rchen}
\begin{eqnarray}
&&t(k_{F_{\tau }})=t(k_{F})  \notag \\
&+&\frac{\partial t(k)}{\partial k}|_{k_{F}}\cdot \frac{1}{3}k_{F}(\tau
\delta )  \notag \\
&+&\frac{1}{2}\Big[\frac{k_{F}^{2}}{9}{\frac{\partial ^{2}t(k)}{\partial
k^{2}}}|_{k_{F}}-\frac{2k_{F}}{9}{\frac{\partial t(k)}{\partial k}}|_{k_{F}}%
\Big]\delta ^{2}  \notag \\
&+&\mathcal{O}(\delta ^{3}),  \label{taylort}
\end{eqnarray}%
and%
\begin{eqnarray}
&&U_{\tau }(\rho ,\delta ,k_{F_{\tau }})=U_{0}(\rho ,k_{F})  \notag \\
&+&\Big[\frac{k_{F}}{3}{\frac{\partial U_{0}(\rho ,k)}{\partial k}}%
|_{k_{F}}+U_{sym,1}(\rho ,k_{F})\Big](\tau \delta )  \notag \\
&+&\Big[\frac{k_{F}}{3}{\frac{\partial U_{sym,1}(\rho ,k)}{\partial k}}%
|_{k_{F}}+U_{sym,2}(\rho ,k_{F})\Big]\delta ^{2}  \notag \\
&+&\frac{1}{2}\Big[\frac{k_{F}^{2}}{9}{\frac{\partial ^{2}U_{0}(\rho ,k)}{%
\partial k^{2}}}|_{k_{F}}-\frac{2k_{F}}{9}{\frac{\partial U_{0}(\rho ,k)}{%
\partial k}}|_{k_{F}}\Big]\delta ^{2}  \notag \\
&+&\mathcal{O}(\delta ^{3}).  \label{taylorU}
\end{eqnarray}
Then, by comparing coefficients of the first-order and second-order $\delta $ terms on both left- and right-hand
sides of Eq. (\ref{cUn}) and Eq. (\ref{cUp}), respectively, the quadratic symmetry energy can be obtained as
\cite{XuC10,XuC11,Rchen,CXu14,FKW05}
\begin{equation}
E_{\rm{sym},2}(\rho) =\frac{1}{3} \frac{k_{\rm{F}}^2}{2 m} +\frac{1}{2} U_{\rm{sym},1}(\rho,k_{\rm{F}})+\frac{k_{\rm{F}}}{6}\left(\frac{\partial U_0}{\partial k}\right)_{k_{\rm{F}}}-\frac{1}{6}\frac{k^4_{\rm{F}}}{2m^3}.
\label{FKW}
\end{equation}
The last term is a very small relativistic correction \cite{FKW02}, and the remaining terms were also obtained within the Brueckner theory \cite{bru64,bru68,Dab73}.
Similarly, within the HVH theorem the quartic symmetry energy $E_{\rm{sym},4}(\rho)$ can be written as\cite{XuC11,Rchen},
\begin{eqnarray}\label{Esym4}
&&E_{\rm{sym},4}(\rho) = \frac{\hbar^2}{162m}
\left(\frac{3\pi^2}{2}\right)^{2/3} \rho^{2/3}\nonumber\\
&&+\Bigg[\frac{5}{324}\frac{\partial U_0(\rho,k)}{\partial k}
k - \frac{1}{108} \frac{\partial^2 U_0(\rho,k)}{\partial
k^2} k^2 +\frac{1}{648} \frac{\partial^3
U_0(\rho,k)}{\partial k^3}k^3\nonumber\\
&&-
\frac{1}{36} \frac{ \partial U_{\rm{sym},1}(\rho,k)}{\partial k}
k + \frac{1}{72} \frac{ \partial^2 U_{\rm{sym},1}(\rho,k)}{\partial
k^2} k^2\nonumber\\
&& + \frac{1}{12} \frac{\partial
U_{\rm{sym},2}(\rho,k)}{\partial k}k+ \frac{1}{4}
U_{\rm{sym},3}(\rho,k)\Bigg]_{k_{\rm{F}}}.
\end{eqnarray}
The slope $L(\rho) \equiv \left[3 \rho (\partial E_{\rm sym}/\partial \rho\right)]_{\rho}$
at an arbitrary density $\rho$ can be written as \cite{XuC11,Rchen},
\begin{eqnarray}
&&L(\rho) = \frac{2}{3} \frac{\hbar^2 k_F^2}{2 m_0^*} + \frac{3}{2} U_{\rm sym,1}(\rho,k_F)
- \frac{1}{6}\Big(\frac{\hbar^2 k^3}{{m_0^*}^2}\frac{\partial m_0^*}{\partial k} \Big)|_{k_F} \nonumber\\
&&+\frac{\partial U_{\rm sym,1}}{\partial k}|_{k_F} k_F
+ 3U_{\rm sym,2}(\rho,k_F), \label{Lexp2}
\end{eqnarray}
where 
\begin{equation}
m^*_0/m=(1+\frac{m}{\hbar^2k_{\rm F}}\partial U_0/\partial k)^{-1}|_{k_F} 
\end{equation}
is the nucleon isoscalar effective mass. In terms of the latter and neglecting the relativistic correction, the symmetry energy of Eq. (\ref{FKW}) can be reduced to
\begin{equation}
E_{\rm{sym}}(\rho) =\frac{1}{3} \frac{k_{\rm{F}}^2}{2 m^*_0(\rho,k_{\rm{F}})} +\frac{1}{2} U_{\rm{sym},1}(\rho,k_{\rm{F}}). \label{Esymexp2}
\end{equation}
The above decompositions of $S=E_{\rm{sym}}(\rho)=E_{\rm{sym},2}(\rho)$, $E_{\rm{sym},4}(\rho)$ and $L(\rho)$ in terms of the density and momentum dependence of single-nucleon potentials within the HVH theorem, it clearly reveals directly the microscopic physics underlying them at the non-relativistic mean-field level.

In this review and most of the work cited, a non-relativistic theoretical approach is considered in decomposing the single-particle potential $U_{n/p}$, nuclear symmetry energy, nucleon effective masses, and then relating their components consistently. It is interesting to note that in relativistic theories, there are different types of potentials/self-energies. In particular, the Schr\"{o}dinger equivalent potential $U_{\mathrm{SEP},J}$ determined by the nucleon scalar self-energy $\Sigma _{J}^{\rm{S}}$ and the time-like component of the vector self-energy $\Sigma _{J}^{0}$ \,\cite{Jaminon:1989wj} is most closely related to the 
$U_{n/p}$ discussed above \cite{Chen05}. The so-called Lorentz mass can then be defined by using $U_{\mathrm{SEP},J}$ for a nucleon $J$, similar to its non-relativistic effective mass defined above. Moreover, a correspondingly relativistic version of the decomposition of the $E_{\rm{sym}}(\rho)$ and $L(\rho)$ in terms of Lorentz covariant nucleon self-energies was given in Ref.\,\cite{Cai-EL}. We refer interested readers to the latter for more details or Section 6.5 on ``Decompositions of nuclear symmetry energy and neutron-proton effective mass splittings in relativistic approaches" of Ref. \cite{PPNP-Li} for a review.  

\section{Generalized isospin asymmetries and symmetry energies in clustered matter, strange matter, and quark matter}\label{G-Esym}
The EOS of uniform ANM described by Eq. (\ref{eos1}) and the associated definition of its symmetry energy term have their respective validity limits. On one hand, at low densities below the so-called Mott points, various clusters start forming in otherwise uniform ANM. It is then unnecessary and conceptually questionable to introduce a symmetry energy of clustered matter for describing its EOS in the same sense as for uniform ANM. On the other hand, at high densities, the presence of strangeness and other particles besides nucleons in hadronic matter may complicate the definition and extraction of the nucleonic symmetry energy. Moreover, quark matter (QM) has its own up-down quark asymmetry and the associated QM symmetry energy. Furthermore, in the hadron-quark mixed phase, the physics becomes even more complicated. The underlying physics, consistency, and smooth transition of using different asymmetries and corresponding symmetry energies (if they exist at all) in describing nucleonic matter, strange hadronic matter, hadron-quark mixed phase, and QM deserve further investigation. We make a few comments and provide some references below as possible starting points for future studies on the relevant questions.

In constructing the EOS of clustered matter, one has to go beyond the mean-field model by considering correlations/fluctuations and in-medium properties of mostly light nuclei, especially for stellar matter for astrophysical applications \cite{Lat-Eos,Shen,Hor06,Joe10,Hag12,Rop13,Typ14,Hag14}. 
It is known that correlations are strongly isospin-dependent. For example, nucleon short-range correlations are the strongest in SNM but are almost negligible in PNM. If one simply calculates a ``symmetry energy" for clustered matter by taking the difference in EOSs of PNM and SNM, the result would be strongly affected by the stronger correlations in SNM. As pointed out earlier in Ref. \cite{LiBA17}, for the clustered matter, because of the different binding energies of mirror nuclei, Coulomb interactions, and different locations of proton and neutron drip lines in the atomic chart, the clustered system no longer possesses a proton-neutron exchange symmetry. It is therefore conceptually ambiguous to define a symmetry energy for clustered matter in the same sense as for the uniform ANM. Moreover, different clusters in the medium have their own local internal isospin asymmetries and densities. In terms of the average density $\rho_{av}$ and the average isospin asymmetry $\delta_{av}$ of the whole system, the EOS of clustered matter has been found to have odd terms in $\delta_{av}$ that are clearly noticeable compared to the $\delta^2_{av}$ term \cite{Agr14,Fan14}. Nonetheless, in the literature, both the second-order derivative of energy per nucleon $e_{\rm{cluster}}(\rho_{av},\delta_{av})$ in clustered matter with respect to $\delta_{av}$, i.e., 
\begin{equation}\label{cluster1}
E^{\rm cluster}_{\rm sym}(\rho_{av})\equiv\frac{1}{2}[\partial ^{2}e_{\rm cluster}/\partial \delta_{av} ^{2}]_{\delta_{av} =0},
\end{equation}
and 
\begin{eqnarray}\label{cluster2}
E^{\rm cluster}_{\rm sym}(\rho_{av})&\equiv& 1/2[e_{\rm{cluster}}(\delta_{av}=1)+ e_{\rm{cluster}}(\delta_{av}=-1)\nonumber\\
&-&2e_{\rm{cluster}}(\delta_{av}=0)]
\end{eqnarray}
have been used in extracting the ``symmetry energy" of clustered matter. These two expressions give the same result regardless of whether there are odd terms in $\delta_{av}$ in the EOS of clustered matter. Interestingly, the $E^{\rm cluster}_{\rm sym}(\rho_{av})$ extracted in this way was found to stay finite at the limit of zero average density \cite{Joe10}. Nevertheless, it is important to stress that this $E^{\rm cluster}_{\rm sym}(\rho_{av})$ may have physical meanings very different from those of the \esym for uniform ANM.

As baryon density and/or temperature increase, various new particles besides nucleons start appearing in strange hadronic matter. There are many interesting questions associated with various symmetries that may appear or disappear as new particles emerge. In particular, how the existence of hyperons may affect the nucleonic isospin asymmetry, the traditionally defined nuclear symmetry energy, how to express the total energy of strange dense matter in terms of the relevant asymmetries, and how to determine their coefficients (generalized symmetry energies) have attracted some attention in recent years, see, e.g., Refs. \cite{Prov,Beda,Bedn,Yang}. Interestingly, the presence of strangeness in cold matter may affect whether isospin symmetry is preserved or broken, and brings in new terms in its ground state EOS. Moreover, new methods may be needed to extract the symmetry energy of strange matter from terrestrial nuclear experiments, astrophysical observations, and nuclear theories \cite{Yang}. Certainly, it is an interesting direction to be further explored. 

In $u+d+s$ quark matter, the up-down quark asymmetry can be directly related to the neutron-proton asymmetry in ANM according to the nucleon constituent quark model. Indeed, the symmetry energy of quark matter has been used in describing the QM EOS and investigating its effects on the properties 
of massive NSs that may have a quark core, see, e.g., Refs. \cite{Toro2006,Pag2010,Shao2012,Chu2013,Chu2015,Liu2016,Chu2017,Wu19}. The density dependence of $u$-quark and $d$-quark masses, as well as their mass-splitting, are closely related to the isospin asymmetry and symmetry energy of QM.
Moreover, the isospin chemical potential in QM and that in nucleonic matter can also be directly related, facilitating a unified and consistent description of neutron-rich matter from low to high densities through a possible phase transition. In the remaining parts of this review, however, we focus on the symmetry energy of nucleonic matter without addressing any of the interesting issues mentioned above in this section.

\section{Siemens' $\rho^{2/3}$ scaling for nuclear symmetry energy of uniform nucleonic matter}
In the late 1960s and early 1970s, Brueckner, Coon, and Dabrowski \cite{bru64,bru68}, and independently Siemens \cite{Siemens1971}, pioneered the study of the density dependence of the nuclear symmetry energy within Brueckner theory by solving the Bethe--Goldstone equation with the best nucleon-nucleon interactions available at the time. In particular, Siemens \cite{Siemens1971} found that 
\begin{quote}
{\bf ``The total symmetry energy $E_{\rm sym}(\rho)$ is proportional to $k_F^2$; this is of course the form of the kinetic part of $E_{\rm sym}(\rho)$, and since the potential energy is nearly quadratic in the single-particle momentum, it is not surprising that all of $E_{\rm sym}(\rho)$ scales quite well with $\rho^{2/3}$."}
\end{quote}
We refer to this observation as \textit{Siemens' $\rho^{2/3}$ scaling} of the symmetry energy near saturation density. 

The physical origin of this scaling can be traced to the nearly quadratic momentum dependence of the single-particle potential in Brueckner theory. In Brueckner--Bethe--Goldstone theory, the single-particle potential in SNM is 
\begin{equation}
U(k) = \sum_{k' \leq k_F} \langle kk' | G | kk' \rangle,
\end{equation}
where $G$ is the in-medium interaction. Numerical studies show that $U(k)$ is approximately quadratic in $k$ for $k \lesssim k_F$ \cite{Siemens1971,bru64,bru68,Dab73}, i.e.,
\begin{equation}
U(k,\rho) \simeq U_0(\rho) + \alpha(\rho)\, k^2.
\end{equation}
Thus the potential energy contribution to the symmetry energy scales as $k_F^2 \propto \rho^{2/3}$ if $\alpha(\rho)$ is weakly density dependent, consistent with the kinetic term. Near the Fermi surface, the single-particle spectrum is well approximated by the effective-mass form
\begin{equation}
\varepsilon(k) \simeq \frac{\hbar^2 k^2}{2 m_0^*(\rho)} + U_0(\rho).
\end{equation}
According to the general expression for the symmetry energy (e.g., Eq.~(\ref{Esymexp2})), if $m_0^*(\rho)$ and $U_{\rm sym,1}(\rho)$ vary slowly near $\rho_0$, then one obtains
\begin{equation}\label{Escaling}
E_{\rm{sym}}(\rho)\propto k_F^2 \propto \rho^{2/3}, \qquad (\rho \approx \rho_0).
\end{equation}
\begin{figure*}[ht]
\begin{center}
\resizebox{0.9\textwidth}{!}{
  \includegraphics[width=15cm,height=6cm]{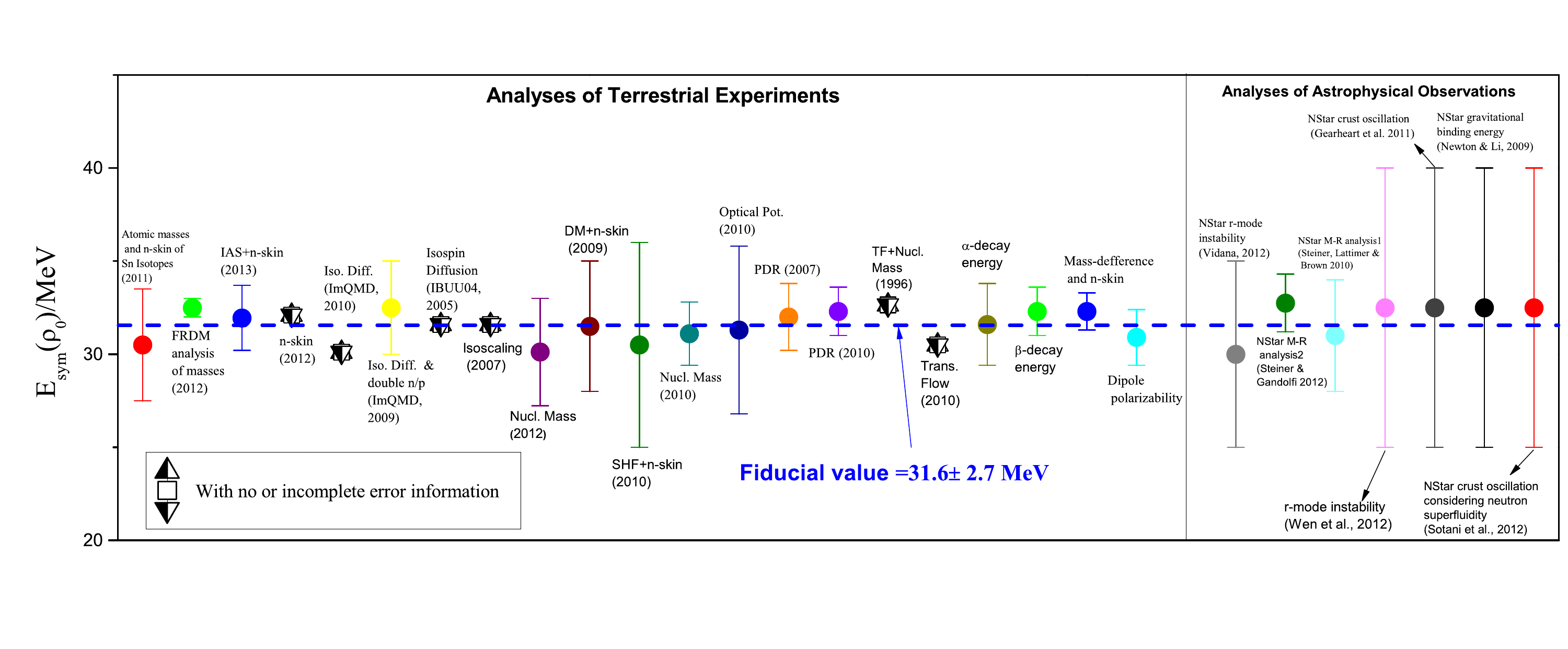}
  }
  \resizebox{0.9\textwidth}{!}{
  \includegraphics[width=15cm,height=6cm]{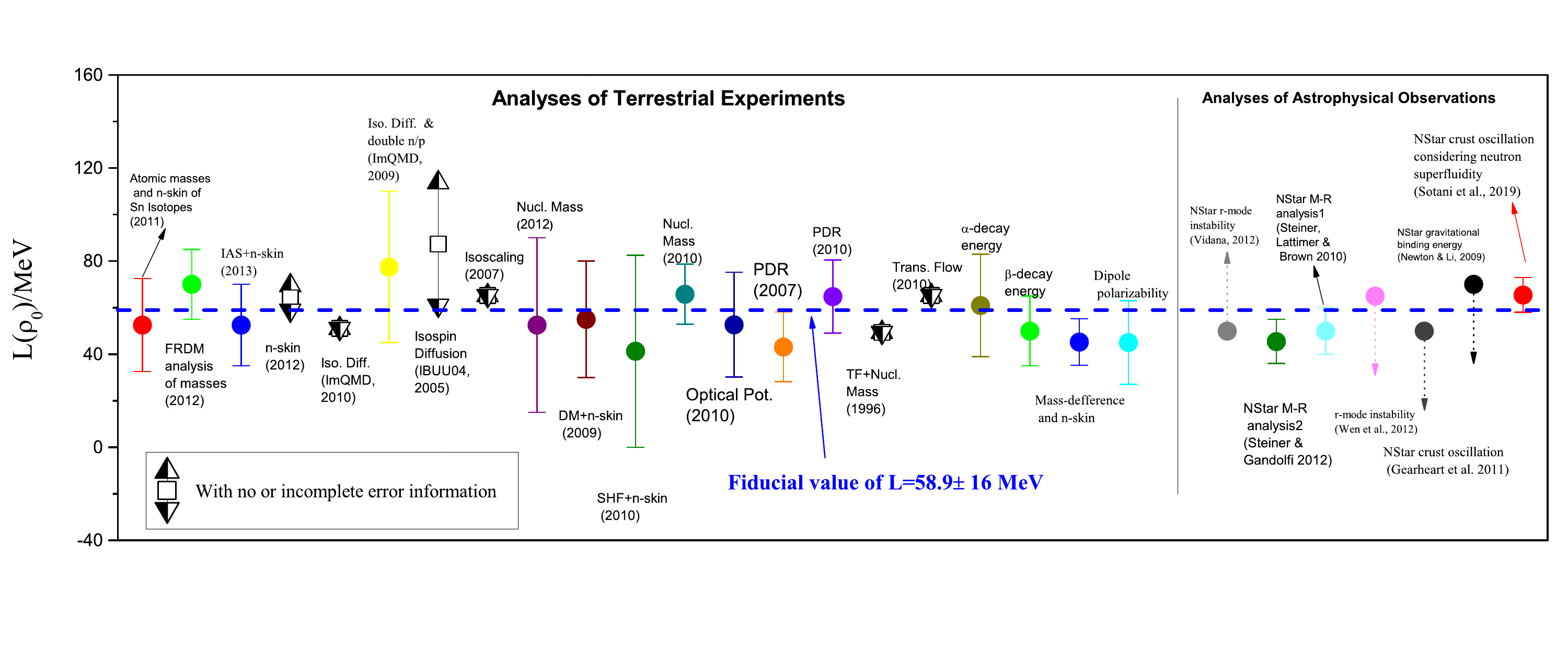}
  }
  \vspace{-0.6cm}
  \caption{(color online) The magnitude $E_{\rm{sym}}(\rho_0)$ (upper window) and slope parameter $L$ (lower window) of nuclear symmetry energy extracted by the community from analyzing various terrestrial nuclear experiments and neutron star observations up to 2013 \cite{LiBA13}}.\label{Esym0L}
\end{center}
\end{figure*}

\begin{figure*}[h!]
\centering
\includegraphics[height=10.cm]{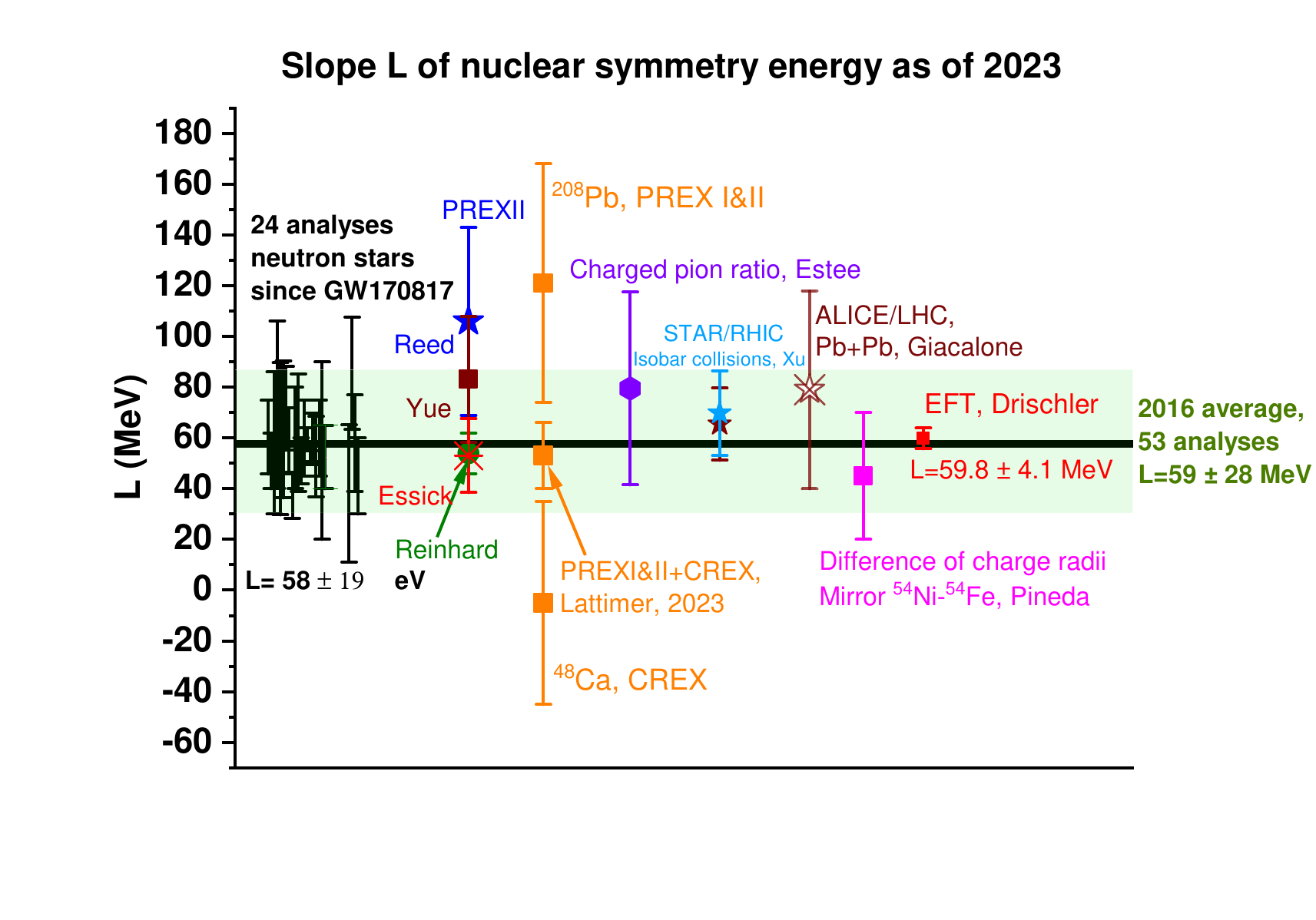}
\vspace{-1.cm}
\caption{(Color Online). Updated constraints on the slope parameter $L$ of symmetry energy up to the year 2023 including analyses of
several recent terrestrial experiments and NS observables since GW170817 in comparison with earlier systematics and the chiral EFT ($\chi$EFT) prediction. Starting from the left are the $L$ values from (1) 24 independent analyses of neutron star observables carried out by various groups between 2017 and 2021, they give an average $L\approx58\pm 19$\,MeV (thick horizontal black line)\,\cite{LCXZ2021}; (2) the original analysis of the PREX-II data\,\cite{Adh21} by Reed et al.\,\cite{Reed:2021nqk} and 3 independent analyses of PREX-II data together with different combinations of terrestrial and/or astrophysical data by Reinhard et al.\,\cite{Reinhard:2021utv,Reinhard2}, Essick et al.\,\cite{Essick21} and Yue et al.\,\cite{Yue:2021yfx}, respectively; (3) liquid drop model analyses using separately the PREX-I and II data together, CREX data only \cite{crex}, and the combination of all data assuming they are equally reliable by Lattimer\,\cite{Lattimer:2023rpe}; (4) charged pion ration in heavy-ion reactions at RIKEN by Estee et al.\,\cite{SpiRIT:2021gtq}; (5) the ratio of average transverse momentum (sky blue  star) and the ratio of charged particle multiplicities (black star) in isobar collisions  $^{96}$Zr+$^{96}$Zr and $^{96}$Ru+$^{96}$Ru) from STAR/RHIC experiments analyzed by Xu et al.\,\cite{Xu:2022ikx}; (6) using neutron-skin thickness of $^{208}$Pb inferred by Giacalone et al. from $^{208}$Pb + $^{208}$Pb collisions measured by the ALICE/LHC Collaboration\,\cite{Giacalone:2023cet}, (7) the difference of charge radii of the mirror pair $^{54}$Ni-$^{54}$Fe by Pineda et al.\,\cite{mirror}; (8) the $\chi$EFT prediction by Dirschler et al.\,\cite{Ohio20}. The horizontal band covering $L\approx59\pm 28$ MeV is the 2016 average of 53 earlier analyses of various data, mostly from terrestrial nuclear experiments\,\cite{Oer17,LiBA13}. Figure modified from those in Refs.\,\cite{Zhang:2022sep,Cai:2025nxn}}\label{L2023}
\end{figure*}

Within Landau-Migdal Fermi-liquid theory \cite{Migdal,BaymPethick}, the nuclear symmetry energy can be expressed in terms of the
isovector Landau parameter $F_0'$, which quantifies the difference in the amplitudes of proton-neutron and proton-proton forward scatterings \cite{Bentz}. In particular, the symmetry energy near saturation density is proportional to 
the isovector effective Fermi energy $E_F^\ast = \hbar^2 k_F^2 / 2m_0^\ast$ multiplied by a factor involving $F_0'$ \cite{Bentz}, 
namely
\begin{equation}
E_{\rm sym}(\rho) \simeq \frac{1}{3} E_F^\ast \, (1 + F_0'),
\end{equation}
when higher-order Landau parameters vary slowly near $\rho_0$. 
Thus, Siemens' $\rho^{2/3}$ scaling naturally emerges if both $m_0^\ast$ and $F_0'$ change only weakly with density
around $\rho_0$. This conclusion is consistent with those based on the Breuckner theory and/or the HVH theorem discussed above. 

In short, Siemens' $\rho^{2/3}$ scaling is valid under the condition that (1) the nucleon isoscalar potential depends on its momentum quadratically at all densities, (2) nucleon isovector potential/interaction is weakly density dependent. Under these conditions, the density dependence (not the magnitude) of symmetry energy is dominated by the kinetic contribution of quasi-nucleons of a constant isoscalar effective mass $m^*_0$ through the Fermi energy $E_F^\ast\propto \rho^{2/3}$. Given the widely different predictions for the density dependence of nuclear symmetry energy using various nuclear interactions within many different nuclear many-body theories, it is useful to use Siemens' $\rho^{2/3}$ scaling as a benchmark. In particular, it is interesting to know if there is any empirical evidence supporting this scaling and to examine if/when this scaling is broken by what physics ingredients compared to the conditions mentioned above.  

\section{Empirical evidence for the $\rho^{2/3}$ scaling of nuclear symmetry energy around $\rho_0$}
Siemens' $\rho^{2/3}$scaling remains a useful baseline and pedagogical reference in modern symmetry-energy studies. Here we examine empirical evidence for the $\rho^{2/3}$ scaling.
Near the saturation density $\rho_0$, the symmetry energy $E_{\rm{sym}}(\rho )$ can be expanded according to 
\begin{eqnarray}
E_{\rm{sym}}(\rho)&=&E_{\rm{sym}}(\rho_0)+L(\frac{\rho-\rho_0}{3\rho_0})+\frac{K_{\rm{sym}}}{2}(\frac{\rho-\rho_0}{3\rho_0})^2\nonumber\\
&+&\frac{J_{\rm{sym}}}{6}(\frac{\rho-\rho_0}{3\rho_0})^3+\mathcal{O}(\frac{\rho-\rho_0}{3\rho_0})^4\label{Esympara}
\end{eqnarray}
in terms of its magnitude
$E_{\rm{sym}}(\rho_0)$, slope $L$, curvature $K_{\rm{sym}}$ and skewness $J_{\rm{sym}}$ at $\rho_0$.
If Siemens' $\rho^{2/3}$ scaling is valid, then
one expects quantitatively
\begin{eqnarray}
L(\rho_0)&=&2E_{\rm{sym}}(\rho_0)\approx 56\sim 68~ {\rm{MeV}},\\\nonumber 
K_{\rm{sym}}(\rho_0)&=&-2E_{\rm{sym}}(\rho_0)\approx -(56\sim 68)~ {\rm{MeV}}, \\\nonumber 
E_{\rm{sym}}(2\rho_0)&=&1.58E_{\rm{sym}}(\rho_0)\approx 44\sim 54~ {\rm{MeV}},\\\nonumber 
J_{\rm{sym}}(\rho_0)&=&8E_{\rm{sym}}(\rho_0)\approx 224\sim 272~ {\rm{MeV}}\nonumber  
\end{eqnarray}
with the fiducial value of $E_{\rm{sym}}(\rho_0)\approx 31\pm 3$ MeV based on mostly analyses of various nuclear experimental data especially atomic masses since around 1965 \cite{Oer17,LiBA13,Cam65}.
In the following, we examine these expectations. In fact, Siemens' $\rho^{2/3}$ scaling finds strong support from both modern microscopic nuclear many-body theories and empirical constraints accumulated from analyzing many terrestrial nuclear experiments and observations of neutron stars.
\begin{enumerate}
\item {\bf Evidence for the $L(\rho_0)=2E_{\rm{sym}}(\rho_0)$ relationship:}
As shown in Fig. \ref{Esym0L}, a survey of 29 independent analyses done before 2013 found the fiducial value of $E_{\rm{sym}}(\rho_0)=31.6\pm 2.7$ MeV and $L=58.9\pm 16$ MeV\,\cite{LiBA13}, respectively. These values were slightly modified to $E_{\rm{sym}}(\rho_0)=31.7 \pm 3.2$~MeV and $L = 58.7 \pm 28.1$~MeV in the 2016 survey of 53 analyses \cite{Oer17}. These fiducial values are supported by some of the state-of-the-art nuclear many-body theories. For example, in a 2020 calculation within the $\chi$EFT (chiral effective field theory), the $E_{\rm{sym}}(\rho_0)$ and $L$ were predicted to be $E_{\rm{sym}}(\rho_0)=31.7 \pm 1.1$~MeV and $L = 59.8 \pm 4.1$~MeV \cite{Ohio20}, respectively, in perfect agreement with the fiducial values.\\

Many recent analyses involving the latest available data from both nuclear experiments and observations of neutron stars have found similar results. For example, in 2020 $E_{\rm{sym}}(\rho_0)=31.35 \pm 2.08$~MeV and $L = 59.57 \pm 10.06$~MeV \cite{ZhangYX20}, and in 2022 $E_{\rm{sym}}(\rho_0)=33.3 \pm 1.3$~MeV and $L = 59.6 \pm 22.1$~MeV \cite{Lynch22} were extracted independently, from combined data of nuclear reactions and structures as well as neutron stars. They are all consistent with the fiducial value of $E_{\rm{sym}}(\rho_0)\approx 31\pm 3$ MeV. In fact, the latter has been widely used in calibrating models and analyzing data in both nuclear physics and astrophysics. For the slope parameter $L$, however, while its fiducial value remains around $L\approx 60$ MeV with few well-known exceptions (e.g., PREX I\&II and CREX), its error bars from different analyses are still relatively large as shown in Fig. \ref{L2023}. In this 2023 update of $L$ systematics, new results from analyzing several recent terrestrial experiments and neutron stars since GW170817 are compared with earlier systematics and the $\chi$EFT prediction. Interestingly, within the error bars of both $L$ and $E_{\rm{sym}}(\rho_0)$, the relationship $L(\rho_0)=2E_{\rm{sym}}(\rho_0)$ is obviously a very good approximation.\\

It is very interesting to note that the relation $L(\rho_0)=2E_{\rm{sym}}(\rho_0)$ is also supported by the universal EOS ($E_{\rm{UG}}$)
of unitary gas (UG) interacting via pairwise $s$-waves with infinite scattering length but zero effective range.
Extensive theoretical and experimental studies of cold atoms, see, e.g., Refs.\,\cite{Zwi15,Ku12,Zu13,Endres13} for reviews,
have provided reliable information about the $E_{\rm{UG}}$. The latter constrains the PNM EOS, thus providing 
additional constraints on the $E_{\rm{sym}}(\rho)$ at sub-saturation densities. Indeed, it was conjectured that the $E_{\rm{UG}}$ provides the lower
boundary of the EOS of PNM ($E_{\rm{PNM}}$) \cite{Tews17}, namely, $E_{\rm{PNM}}(\rho)\geq E_{\rm{UG}}(\rho)$ where $E_{\rm{UG}}(\rho)=\xi E_F(\rho)$ with $\xi$ being the Bertsch parameter \cite{Zwi15,Ku12,Zu13,Endres13}.
Consequently, the lower boundaries of $E_{\rm{sym}}(\rho_0)$ and $L(\rho)$ were predicted respectively to be \cite{Tews17,NBZ}
\begin{align}\label{relbj}
E_{\rm{sym}}(\rho_0)\geq& \frac{E_{\rm{UG}}^0}{3u^{1/3}}(u+2)+\frac{K_{n}}{18}(u-1)^2\nonumber\\
  &+\frac{J_n}{81}(u-1)^3-E_0(\rho_0).
\end{align}
and 
\begin{equation}\label{relbd3}
L\geq\frac{2E_{\rm{UG}}^0}{u^{1/3}}-\frac{K_{n}}{3}(u-1)-\frac{J_n}{18}(u-1)^2
\end{equation}
where $E_{\rm{UG}}^0=E_{\rm{UG}}(\rho_0)$, $K_n=K_{\rm{sym}}+K_0$, $J_n=J_{\rm{sym}}+J_0$ and $u=\rho/\rho_0$. 
Setting $u=1$, one obtains readily $L(\rho_0)=2E_{\rm{sym}}(\rho_0)$ on the lower boundary of $E_{\rm{sym}}(\rho)$ constrained by the universal EOS 
of unitary gas.

\vspace{-0.8cm}
\begin{figure}[ht]
\begin{center}
\hspace*{-0.6cm}
\resizebox{0.55\textwidth}{!}{
  \includegraphics[width=7cm]{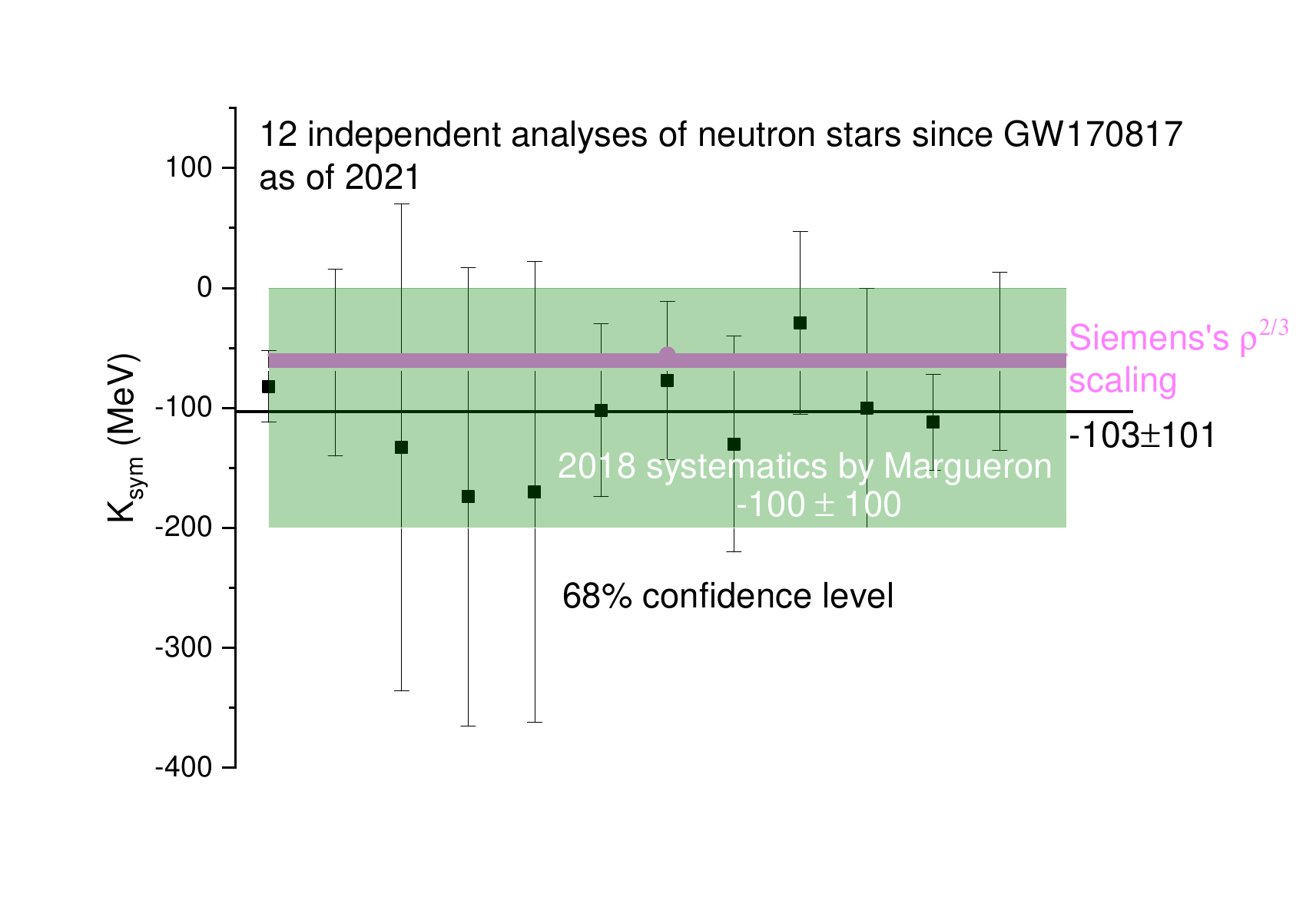}
  }
\vspace{-0.8cm}
\caption{{The curvature parameter $K_{\rm{sym}}$ at 68\% confidence level from 12 independent analyses of neutron stars (listed in Ref. \cite{LCXZ2021}). 
The black line marked by $-103\pm 101$ MeV is the average value of $K_{\rm{sym}}$ from these 12 analyses. The green band within $-100\pm 100$ MeV is from the 2018 survey by Margueron et al. based on a metamodel of nuclear EOS constrained by nuclear and astrophysical data available at the time ~\cite{Margueron18}. The pink band within $-(56-68)$ MeV is from Siemens' $\rho^{2/3}$ scaling.}}\label{Ksym}
\end{center}
\end{figure}

\item {\bf Evidence for the $K_{\rm{sym}}(\rho_0)=-2E_{\rm{sym}}(\rho_0)$ relationship:}
While there are many efforts reported in the literature about extracting the density dependence of nuclear symmetry energy, especially its slope $L$ at $\rho_0$, only a few of them have explicitly reported their $K_{\rm{sym}}(\rho_0)$ values. Shown in Figure~\ref{Ksym} is a comparison of $K_{\rm{sym}}$ values from 12 independent analyses of neutron star observables after GW170817 with respect to (1) the $K_{\rm{sym}}\approx -100\pm 100$ MeV (green band) from the 2018 systematics by Margueron~et~al.~\cite{Margueron18} from analyzing meta-model EOS predictions under the constraints of both terrestrial experiments and astrophysical observations available at the time and (2) $K_{\rm{sym}}\approx -(56-68)$ MeV (pink band) from Siemens' $\rho^{2/3}$ scaling.
The average of the 12 analyses of neutron star data is about $K_{\rm{sym}}\approx -103\pm 101$ MeV at 68\% confidence level. Obviously, within the still very large error bars, Siemens' $\rho^{2/3}$ scaling is consistent with both the 2018 systematics and the 12 analyses of neutron star data after GW170817.\\

We notice that some efforts have also been made to extract $K_{\rm{sym}}(\rho_0)$ from heavy-ion reactions, see, e.g., Ref. \cite{Cozma25} for a recent review. The results still have large error bars compatible with those from analyzing neutron star data and depend somewhat on the transport models used in analyzing the experimental data. For example, in very good agreement with Siemens' $\rho^{2/3}$ scaling, analyses of the ASY-EOS data from GSI lead to $E_{\rm{sym}}(\rho_0)=34$ MeV, $L=72\pm13$ MeV, and $K_{\rm{sym}}(\rho_0)=-(40\sim 70)$ MeV \cite{Cozma25,Rus16} within a low-intermediate energy version of the UrQMD model \cite{urqmd}. 
While analyses of the same data found $E_{\rm{sym}}(\rho_0)=31.6$ MeV, $L=85\pm 22(\rm{exp})\pm20(\rm{th})\pm 12(\rm{sys})$ MeV and $K_{\rm{sym}}=96\pm315(\rm{exp})\pm170(\rm{th})\pm166(\rm{sys})$ MeV within the T$\ddot{\rm{u}}$bingen version of the QMD model T$\ddot{\rm{u}}$QMD \cite{Cozma18}. Generally speaking, these results from heavy-ion reactions, especially the $L$ values inferred and the $E_{\rm{sym}}(\rho_0)$ used (selected {\it apriori} to be consistent with the fiducial value), are all consistent with Siemens' $\rho^{2/3}$ scaling. Obviously, the large uncertainties of $K_{\rm{sym}}$ prevent us from making a stronger conclusion. Nevertheless, it is interesting to note here that better constraining the $K_{\rm{sym}}$ parameter is a main goal of the recently finished new ASY-EOS-II experiments, and the associated error bars are expected to be significantly reduced \cite{NewASY}.\\

\begin{figure}[!hpbt]
\centering
  \resizebox{1.0\textwidth}{!}{
\includegraphics[]{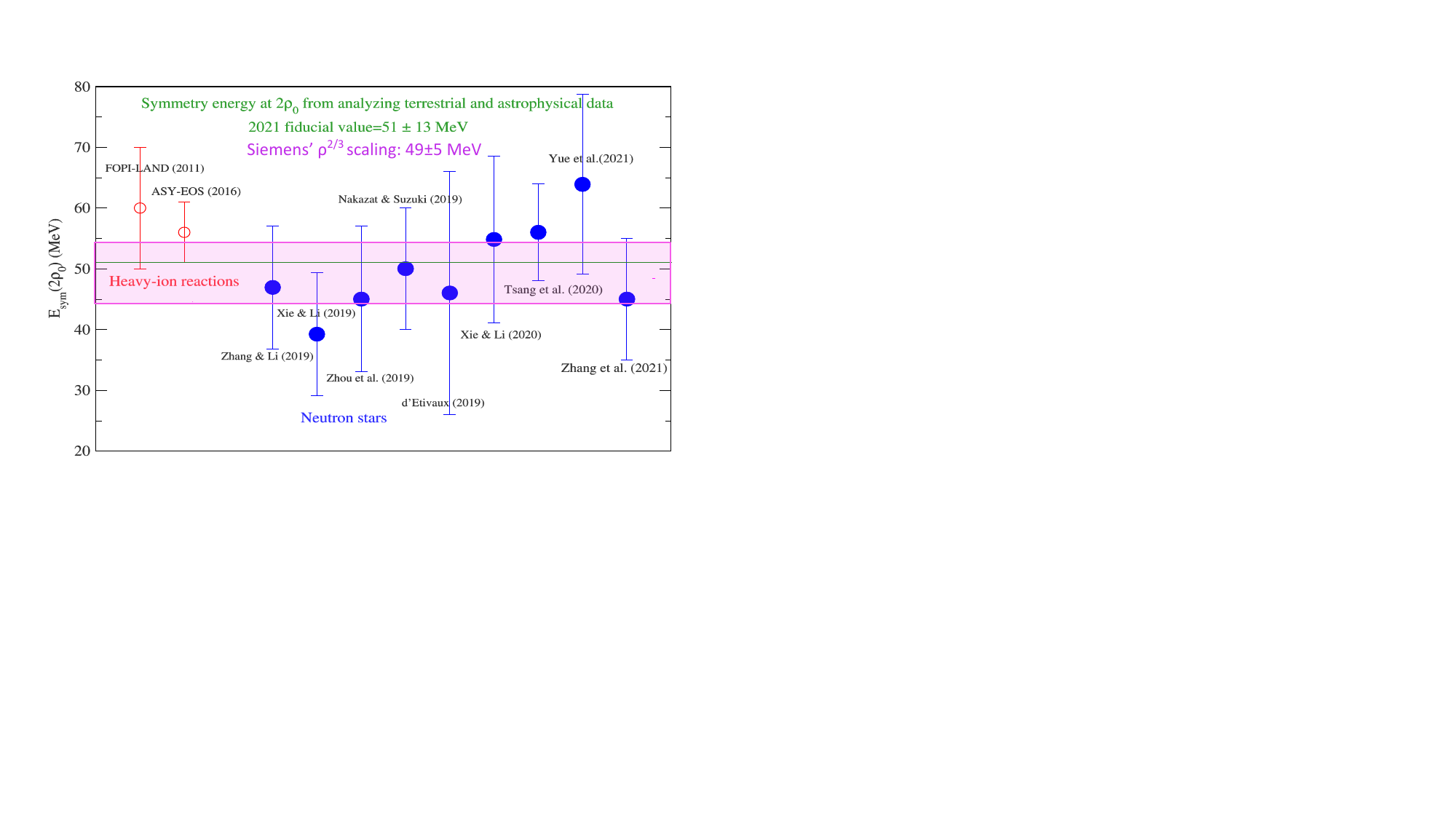}
}
\vspace{-4.6cm}
\caption{Nuclear symmetry energy at twice the saturation density of nuclear matter. The two heavy-ion results (red) are  
deduced from the FOPI-LAND \cite{Rus11} and the ASY-EOS \cite{Rus16} results from analyzing relative flows and yields of light mirror nuclei as well as neutrons and protons in heavy-ion collisions. The nine analyses of neutron star observations after GW170817 (blue) are from (1) (Zhang and Li 2019) directly inverting observed NS radii, tidal deformability, and~maximum mass in the high-density EOS space~\cite{Zhang18,Zhang19epj,Zhang19apj}, (2) (Xie and Li 2019) a Bayesian inference from the radii of canonical NSs observed by using Chandra X-rays and gravitational waves from GW170817~\cite{Xie19}, (3) (Zhou~et~al. 2019) analyses of NS radii, tidal deformability, and~maximum mass within an extended Skyrme--Hartree--Fock approach (eSHF)~\cite{YZhou19a}, (4) (Nakazato and Suzuki 2019) analyzing cooling timescales of protoneutron stars, as well as the radius and tidal deformability of GW170817~\cite{Nakazato19}, (5) (d'Etivaux~et~al. 2019) a Bayesian inference directly from the X-ray data of seven quiescent low-mass X-ray binaries in globular clusters~\cite{France1}, (6) (Xie and Li 2020) a Bayesian inference from the radii of NSs observed by NICER and LIGO/VIRGO~\cite{Xie20}, (7) (Tsang~et~al. 2020) Bayesian analyses of tidal deformation of canonical NSs from LIGO/VIRGO~\cite{Tsang20}, (8) (Yue~et~al. 2021) eSHF analyses of tidal deformation from GW170817 and radii from NICER~\cite{Yue:2021yfx}, and (9) (Zhang~et~al. 2021) Skyrme--Hartree--Fock predictions with interaction parameters constrained by heavy-ion reaction experiments, the neutron skin of heavy nuclei, as well as the tidal deformation and radii of neutron stars from LIGO/VIRGO~\cite{ZhangYX20}.}
\label{Esym-survey}
\end{figure}
\begin{figure*}[ht]
\begin{center}
\resizebox{0.9\textwidth}{!}{
  \includegraphics[]{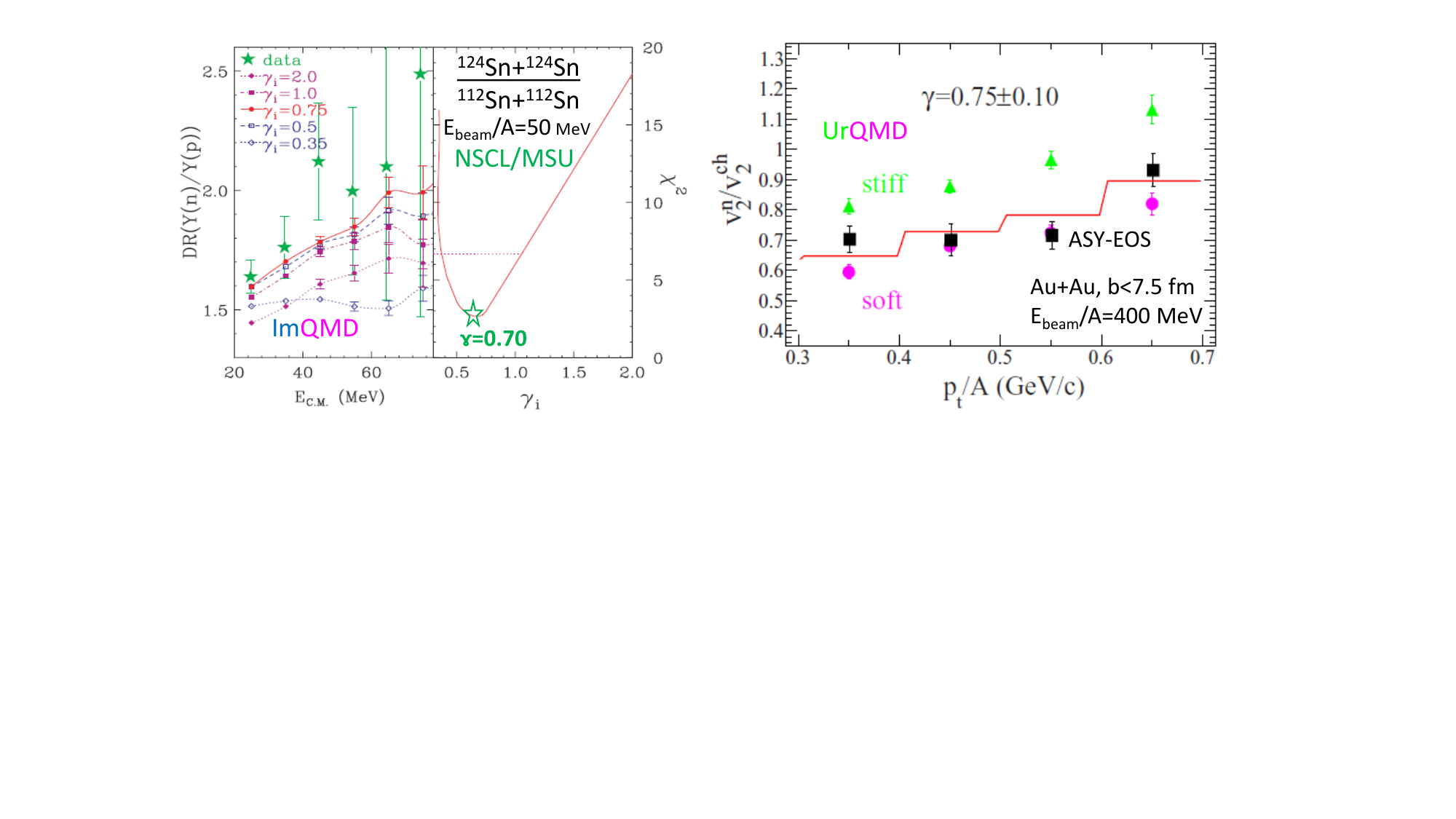}
  }
  \vspace{-4.2cm}
  \caption{Left panel of the left window: Comparison of experimental double neutron-proton ratios (star symbols) for the reaction of $^{124}$Sn+$^{124}$Sn over $^{112}$Sn+$^{112}$Sn at $E_{beam}/A=50$ MeV, as a function of nucleon center-of-mass energy, to ImQMD calculations (lines)
with different density dependencies of the symmetry energy by varying the $\gamma_i$ parameter (the same as $\gamma$ in Eq. (\ref{esym}). The right panel is a plot of $\chi^2$ as a function of $\gamma_i$ of the analyses leading to the most probable $\gamma\approx 0.7$. The figure is taken from Ref. \cite{Tsa09}. Right Window: Analyses of the elliptic flow ratio of neutrons over all charged particles for central collisions of $^{197}$Au+$^{197}$Au at 400 MeV/nucleon as a function of the transverse momentum per
nucleon $p_t$/A by the ASY-EOS Collaboration \cite{Rus16}. The black squares represent the
experimental data; the green triangles and purple circles represent the
UrQMD predictions for stiff ($\gamma$ = 1.5) and soft ($\gamma$ = 0.5) power-law
exponents of the potential term of symmetry energy, respectively. The solid line is the
result of a linear interpolation between the predictions, weighted
according to the experimental errors of the included four bins in
$p_t$/A and leading to the indicated $\gamma= 0.75 \pm 0.10$. The figure is taken from Ref. \cite{Rus16}.
}\label{Tsang-Russ}.
\end{center}
\end{figure*}
\item {\bf Evidence for the $E_{\rm{sym}}(2\rho_0)=1.58E_{\rm{sym}}(\rho_0)$ relationship:} Nuclear symmetry energy $E_{\rm{sym}}(2\rho_0)$ at twice the saturation density is a useful benchmark in exploring the EOS of dense neutron-rich matter. In particular, nuclear pressure around this density is most relevant for determining the radii of canonical neutron stars. Theoretical predictions below this density 
are relatively reliable but become gradually more diverse as the density becomes higher. Moreover, heavy-ion reactions involving radioactive beams of about 400 MeV/nucleon to be available at several facilities can produce dense neutron-rich matter around this density, providing a laboratory testing ground of nuclear symmetry energy around $2\rho_0$. As mentioned earlier, Siemens' $\rho^{2/3}$ scaling predicts that $E_{\rm{sym}}(2\rho_0)\approx 44\sim 54~ {\rm{MeV}}$ as indicated by the pink band in Fig. \ref{Esym-survey}. Surprisingly, it is in very good agreement with the 2021 fiducial value of $E_{\rm{sym}}(2\rho_0)=51\pm 13$ MeV from surveying the analyses of neutron star radii, tidal deformability and masses after GW170817 as well as the two heavy-ion reaction experiments indicated.\\

Since rather different assumptions and methods were used in analyzing the different types of nuclear laboratory and neutron star data, it is difficult to estimate rigorously the associated error bar for the fiducial value of $E_{\rm{sym}}(2\rho_0)$ as the individual errors from different analyses of even using the same observational data are often different and have different natures. Nevertheless, the general agreement between the $E_{\rm{sym}}(2\rho_0)$ based on Siemens' $\rho^{2/3}$ scaling and its fiducial value is rather encouraging. It provide at least a 
useful reference for comparisons. For example, an upper bound of $E_{\rm{sym}}(2\rho_0) \leq 53.2$ MeV was derived in Ref. \cite{Meng-PKU} by studying the radii of neutron drops using the state-of-the-art nuclear energy density functional theories. It is also interesting to note that (1) Quantum Monte Carlo calculations using local interactions derived from $\chi$EFT up to next-to-next-to-leading order predicted a value of $E_{\rm{sym}}(2\rho_0) = 46 \pm 4$ MeV \cite{Diego}, (2) $\chi$EFT predicted $E_{\rm{sym}}(2\rho_0)\approx 45\pm 3$ MeV \cite{Ohio20}, (3) a Relativistic BHF theory in full Dirac space predicted $E_{\rm{sym}}(2\rho_0)=51.6$ MeV \cite{BHF-full} while (4) a conventional Relativistic BHF theory predicted $E_{\rm{sym}}(2\rho_0)\approx 53$ MeV \cite{RBHF}.
These representative predictions of microscopic nuclear many-body theories are all consistent with the fiducial value of $E_{\rm{sym}}(2\rho_0)$ and the expectation based on Siemens' $\rho^{2/3}$ scaling given all the uncertainties associated with them.\\

\item {\bf Evidence for Siemens' $\rho^{2/3}$ scaling from directly fitting the accumulated $E_{\rm{sym}}(\rho)$ data around $\rho_0$:}
There are strong experimental evidences for Siemens' $\rho^{2/3}$ scaling from analyzing directly observables of heavy-ion reaction data at various beam energies.
There are also evidences from directly fitting the accumulated $E_{\rm{sym}}(\rho)$ data around but mostly below $\rho_0$ in analyzing heavy-ion reactions, nuclear masses, neutron-skins, isotope analog states, electrical dipole polarizability, etc. In the following, we examine a few examples. \\
\begin{figure*}[ht]
\vspace{-1.cm}
\begin{center}
\resizebox{0.65\textwidth}{!}{
  \includegraphics[]{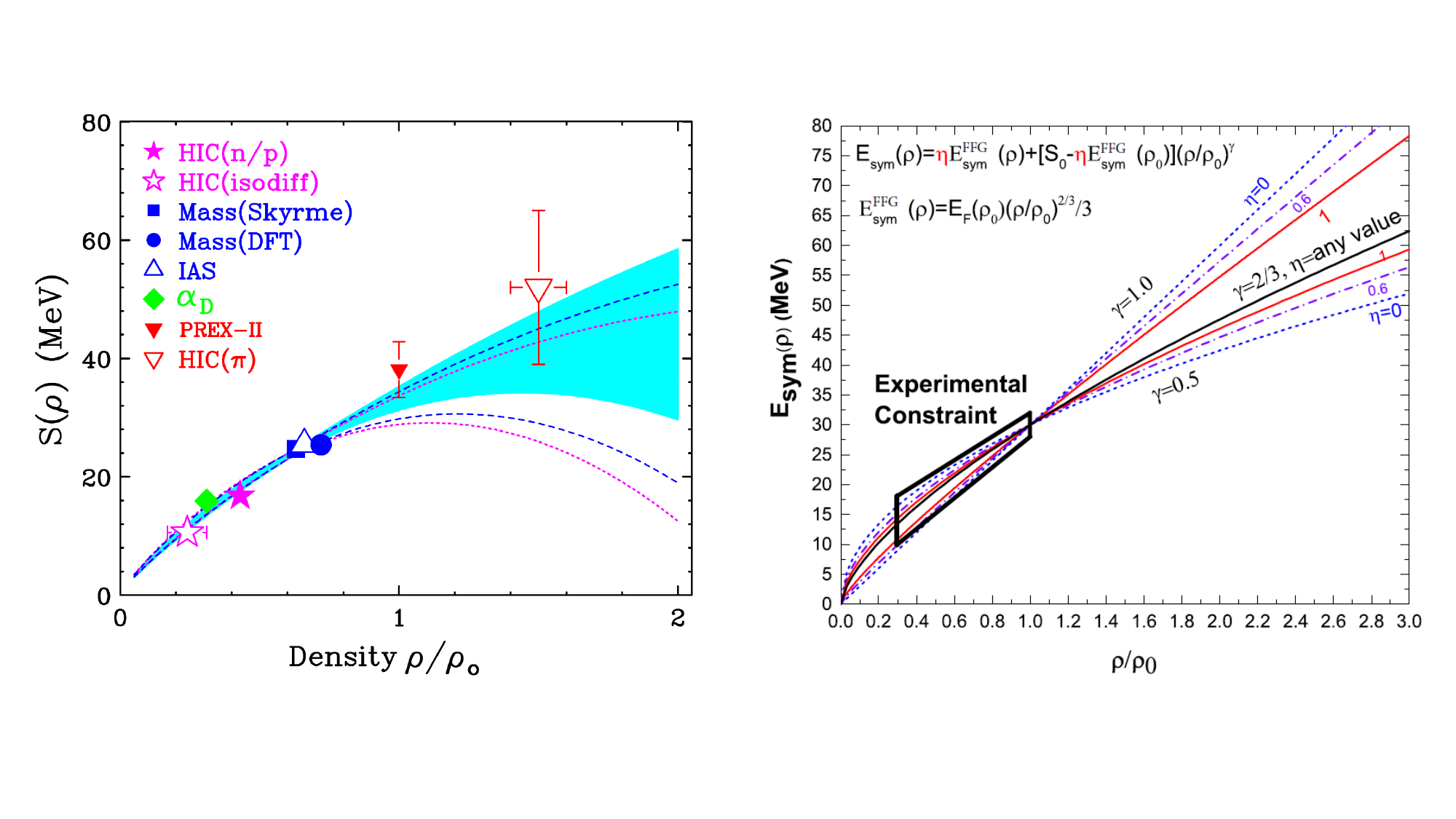}
  }
    \vspace{-1cm}
  \caption{Left: Fitting the accumulated symmetry energy data at subsaturation densities with Eq. (\ref{esym}). Left panel: examples of nuclear symmetry energy extracted from the indicated observables within approaches appropriate for the relevant densities probed \cite{Lynch22}. Right panel: The think box indicates the experimental bound including those shown in the left panel. The various curves are fits to the upper, middle and lower experimental boundaries with Eq. (\ref{esym}) using the parameters indicated \cite{LiG15}.}\label{Li+Lynch}
\end{center}
\end{figure*}
Using two parameters $\eta$ and $\gamma$ to vary separately the kinetic contribution and the density dependence of the potential part,
nuclear symmetry energy can be parameterized as
\begin{equation}\label{esym}
E_{sym}(\rho)=\eta\cdot E_{sym}^{kin}(\textrm{FG})(\rho)+[S_0-\eta\cdot E_{sym}^{kin}(\textrm{FG})(\rho_0)](\frac{\rho}{\rho_0})^{\gamma}
\end{equation}
where $E_{sym}^{kin}(\textrm{FG})(\rho_0)\approx 12$ MeV is the kinetic symmetry energy of a free Fermi gas (FG) at $\rho_0$, with $S_0\equiv E_{\rm sym}(\rho_0)$ fixed at a specific value. Normally $\eta=1$ but can be reduced if one considers effects of short-range correlations \cite{CXu11,Vid11,Xulili,Hen15b,Cai15a,Cai-RMF,Car12,Car14,LiG15,PPNP-Li}. The above parameterization is under the assumption that the SRC can modify the magnitude at $\rho_0$ but not the $\rho^{2/3}$ dependence of kinetic symmetry energy $E_{sym}^{kin}(\textrm{FG})(\rho)=E_{sym}^{kin}(\textrm{FG})(\rho_0)(\rho/\rho_0)^{2/3}$. 
The corresponding $L$ is
\begin{equation}\label{LL}
L=\frac{9}{5}(2^{2/3}-1)E_F(\rho_0)(2/3-\gamma)\eta+3\gamma S_0.
\end{equation}
{\bf Example-1:} As shown in the left window of Fig. \ref{Tsang-Russ}, Tsang et al. \cite{Tsa09} compared their experimental double neutron/proton ratios (star symbols) for the reaction of $^{124}$Sn+$^{124}$Sn over $^{112}$Sn+$^{112}$Sn at $E_{beam}/A=50$ MeV done at MSU, as a function of nucleon center-of-mass energy, with ImQMD calculations (lines) using different density dependencies of the symmetry energy by varying the $\gamma_i$ parameter (the same as $\gamma$ in Eq. (\ref{esym})) using $\eta=1$, $S_0=30.1$ MeV and $E_{sym}^{kin}(\textrm{FG})(\rho_0)=12.5$ MeV. The $\chi^2$ as a function of $\gamma_i$ of their analyses led them to the most probable $\gamma\approx 0.7$ in perfect agreement with the $\rho^{2/3}$ scaling of symmetry energy. \\

{\bf Example-2:} Shown in the right window of Fig.~\ref{Tsang-Russ} is an analysis of the elliptic flow ratio of neutrons over all charged particles for central collisions of $^{197}$Au+$^{197}$Au at 400 MeV/nucleon as a function of the transverse momentum per nucleon $p_{t}/A$ by the ASY-EOS Collaboration \cite{Rus16}.
The ASY-EOS experiment has some improvements to the earlier FOPI-LAND experiments and some of the data are complementary \cite{Rus16,Rus11}. The data were analyzed within the UrQMD model using $\eta=1$, $S_0=34$ MeV and $E_{sym}^{kin}(\textrm{FG})(\rho_0)=12$ MeV with $\gamma =0.5$ and $\gamma =1.5$ corresponding to a soft and a stiff density dependence of symmetry energy, respectively. The red solid line is the result of a linear interpolation between the predictions, weighted according to the experimental errors of the included four bins in $p_{t}/A$. It led them to conclude that $\gamma =0.75 \pm 0.10$ which is slightly larger but still consistent with the $\rho^{2/3}$ scaling within the reported error bar. This result was then used to construct a constraining band for the symmetry energy as a function of density. Quantitatively, they extracted $L = 72\pm 13$ MeV, $K_{sym}$ in the range of $-70$ to $-40$ MeV and $E_{\rm{sym}}(2\rho_0)=56\pm 5$ MeV as mentioned earlier. \\

\begin{figure*}[ht]
\begin{center}
  \includegraphics[width=8cm]{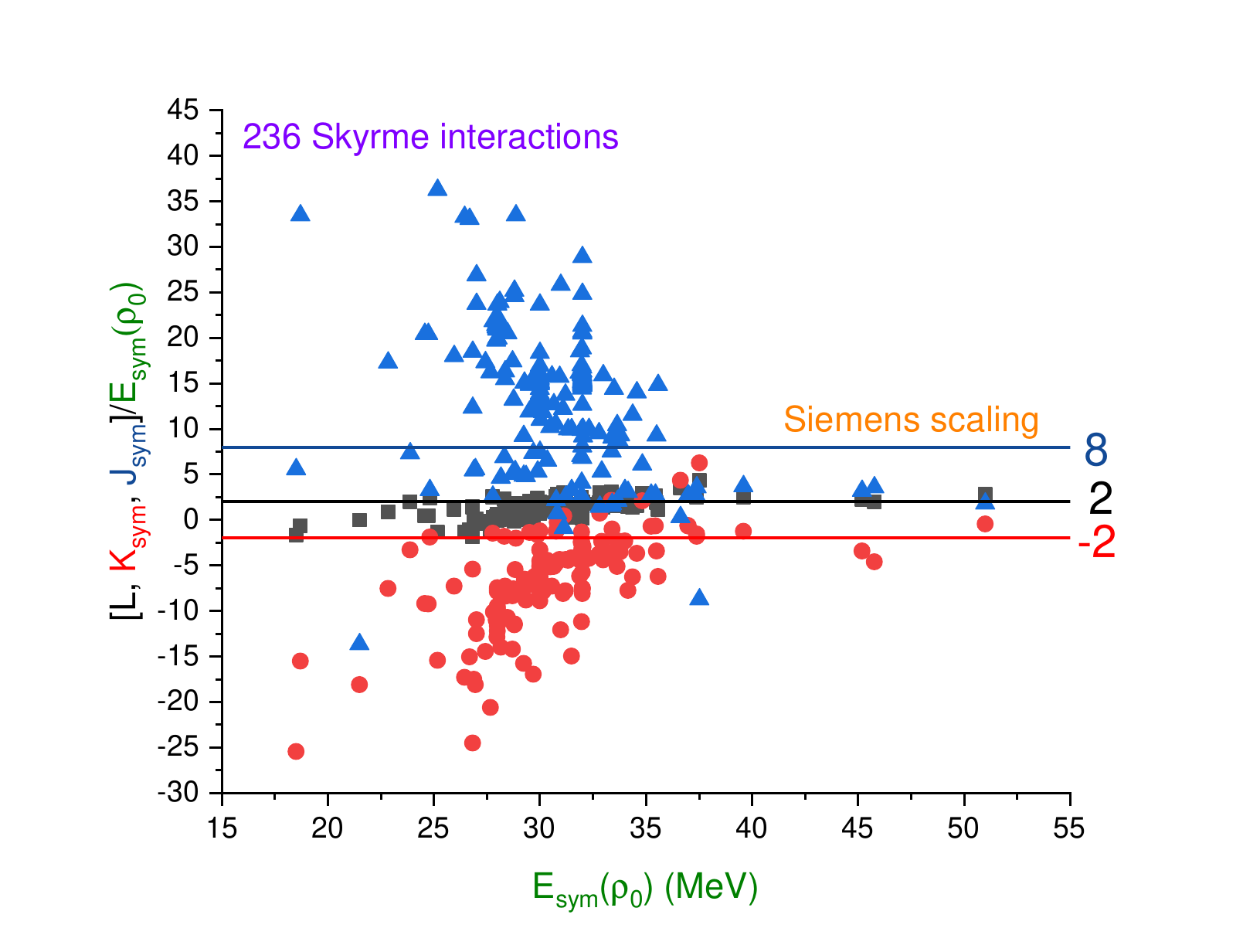}
\includegraphics[width=8cm]{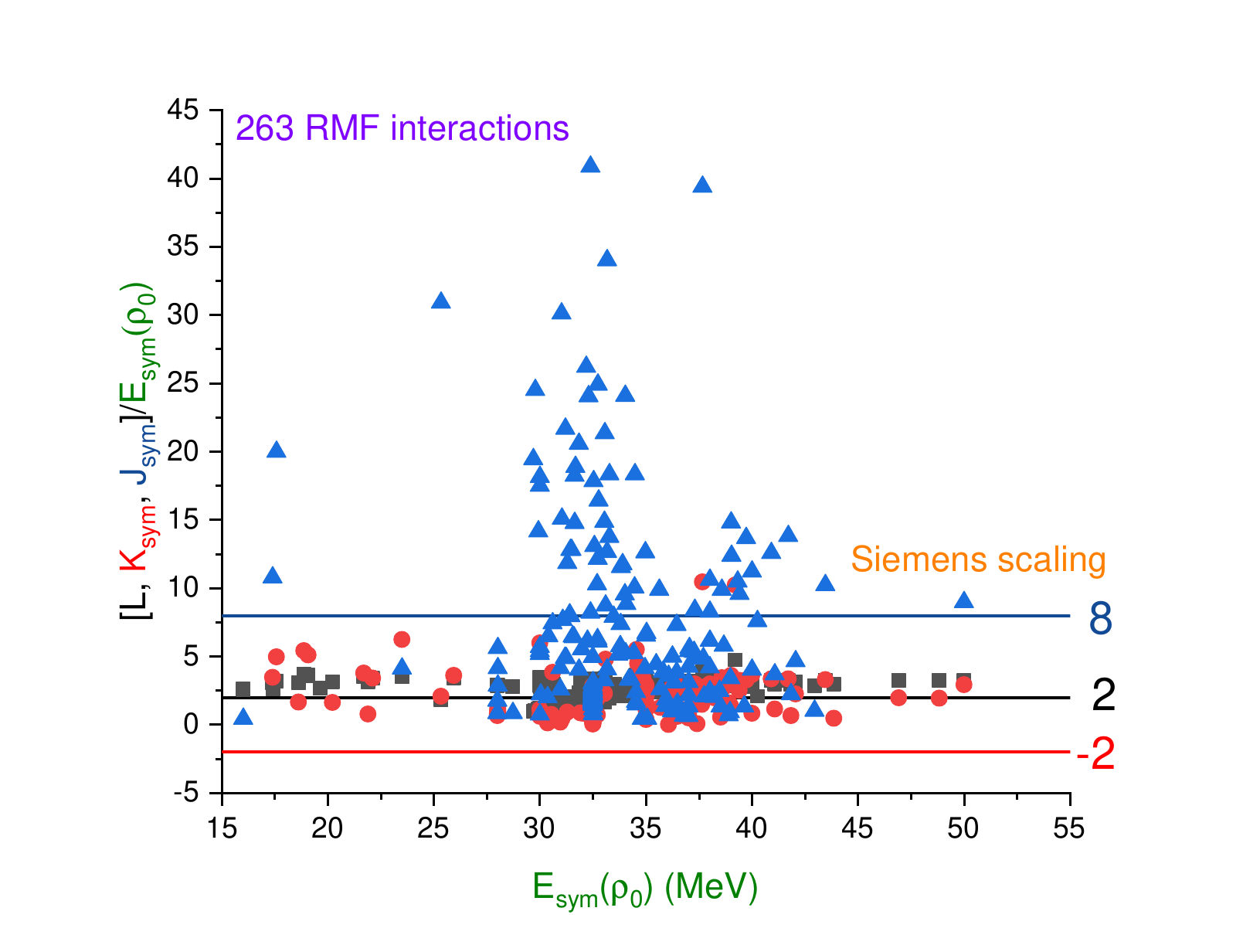}
  \caption{Extracted ratios of $L$ (black), $K_{\rm sym}$ (red) and $J_{\rm sym}$ (blue) over $E_{\rm sym}(\rho_0)$ as a function of the latter for 236 SHF (left) and 263 RMF (right) EDFs using the tabulated predictions in Refs. \cite{Dut12,Dut14}, respectively.}\label{SHF-RMF}
\end{center}
\end{figure*}

It is known that central $^{197}$Au+$^{197}$Au collisions at 400 MeV/nucleon can reach a maximum central density of about $2.0\rho_0$ depending mostly on the incompressibility and skewness of symmetric nuclear matter used in the simulations, see, e.g., Ref. \cite{LiXie25}. While in a typical central Sn+Sn reactions at $E_{beam}/A=50$ MeV, the maximum compression is about $1.4\rho_0$ \cite{LCK08}. Comparing the above results about $E_{\rm{sym}}(\rho)$ from Sn+Sn at $E_{beam}/A=50$ MeV and Au+Au at $E_{beam}/A=400$ MeV, while they are consistent there is an interesting indication that the symmetry energy becomes stiffer and starts deviating from Siemens' $\rho^{2/3}$ scaling as the density goes far above $\rho_0$. As we shall discuss in detail, this is expected. Moreover, investigating how quickly and under what conditions the $\rho^{2/3}$ scaling breaks down may help us understand the fundamental physics underlying the high-density behavior of nuclear symmetry energy.\\

{\bf Example-3:} The results discussed above are from two particular reactions using several observables, probably probing the symmetry energy at somewhat different densities mostly around and above $\rho_0$. There are also known observables from low-energy nuclear reactions and studies of nuclear structure properties. Shown in the left window of Fig. \ref{Li+Lynch} are examples of nuclear symmetry energy extracted from the indicated observables within approaches that are considered appropriate for the relevant densities probed by the observables. A fit of these data directly provides $S_0$=$(33.3\pm1.3)~\mathrm{MeV}$, $L$= $(59.6\pm22.1)~\mathrm{MeV}$ and $K_\mathrm{sym}$ =$(-180\pm96)~\mathrm{MeV}$ \cite{Lynch22}. These results are in good agreement with the systematics discussed earlier. These data all fall in the experimental bounding box of the right panel originally given in Ref. \cite{Tsa09}. The upper, middle, and lower experimental boundaries were fitted with Eq. (\ref{esym}) using the parameters indicated and $S_0=30$ MeV in Ref. \cite{LiG15}. The best description of the curve passing through the middle of the experimental constraining box is $E_{\rm{sym}}(\rho)\propto (\rho/\rho_0)^{2/3}$ independent of the $\eta$ parameter. These results together provide additional evidence supporting Siemens' $\rho^{2/3}$ scaling for nuclear symmetry energy around $\rho_0$.\\

\item {\bf Comparisons with predictions of SHF, RMF, and GHF energy density functionals:} Nuclear energy density functionals (EDFs) have been very important in studying properties of nuclei, nuclear reactions, and neutron stars. In particular, there are now over 500 Skyrme Hartree-Fock (SHF) and Relativistic Mean-Field (RMF) EDFs widely used in the literature with diverse predictions, see, e.g., surveys in Refs. \cite{Dut12,Dut14}, besides at least 11 Gogny-Hartree-Fock (GHF) EDFs \cite{Rios-G}. Thus, one may wonder how the predictions of these EDFs compare with the expectations based on Siemens' $\rho^{2/3}$ scaling for $E_{\rm sym}(\rho)$. To answer this question, shown in Fig. \ref{SHF-RMF} are the extracted ratios of $L$ (black), $K_{\rm sym}$ (red), and $J_{\rm sym}$ (blue) over $E_{\rm sym}(\rho_0)$ as a function of the latter for 236 SHF (left) and 263 RMF (right) EDFs using the tabulated predictions in Refs. \cite{Dut12,Dut14}, respectively. The horizontal lines with different colors are expectations for the indicated ratios based on Siemens' $\rho^{2/3}$ scaling. \\

It is interesting to see that for many calculations in both classes of EDFs, the $L/E_{\rm sym}(\rho_0)$ ratio agrees quite well with the $\rho^{2/3}$ scaling. However, for the high-order parameters, very few calculations, especially in the case of RMF EDFs, are consistent with the $\rho^{2/3}$ scaling. This is probably not very surprising. On one hand, as we discussed earlier, $K_{\rm sym}$ is experimentally very poorly constrained within a broad range, and there is essentially no constraint on $J_{\rm sym}$ 
at all. As we shall discuss later, neutron star radius measurements provide some constraints on $K_{\rm sym}$ but not $J_{\rm sym}$ as the latter characterizes nuclear symmetry energy around $(3-4)\rho_0$ while the radii of neutron stars are determined by nuclear pressure around $2\rho_0$. On the other hand, as we noticed earlier, Siemens' $\rho^{2/3}$ scaling is based on non-relativistic theories. It is thus expected to be broken for parameters characterizing the high-density $E_{\rm{sym}}(\rho)$. \\

\begin{figure}[ht]
\begin{center}
\vspace{-1.cm}
  \includegraphics[width=8cm]{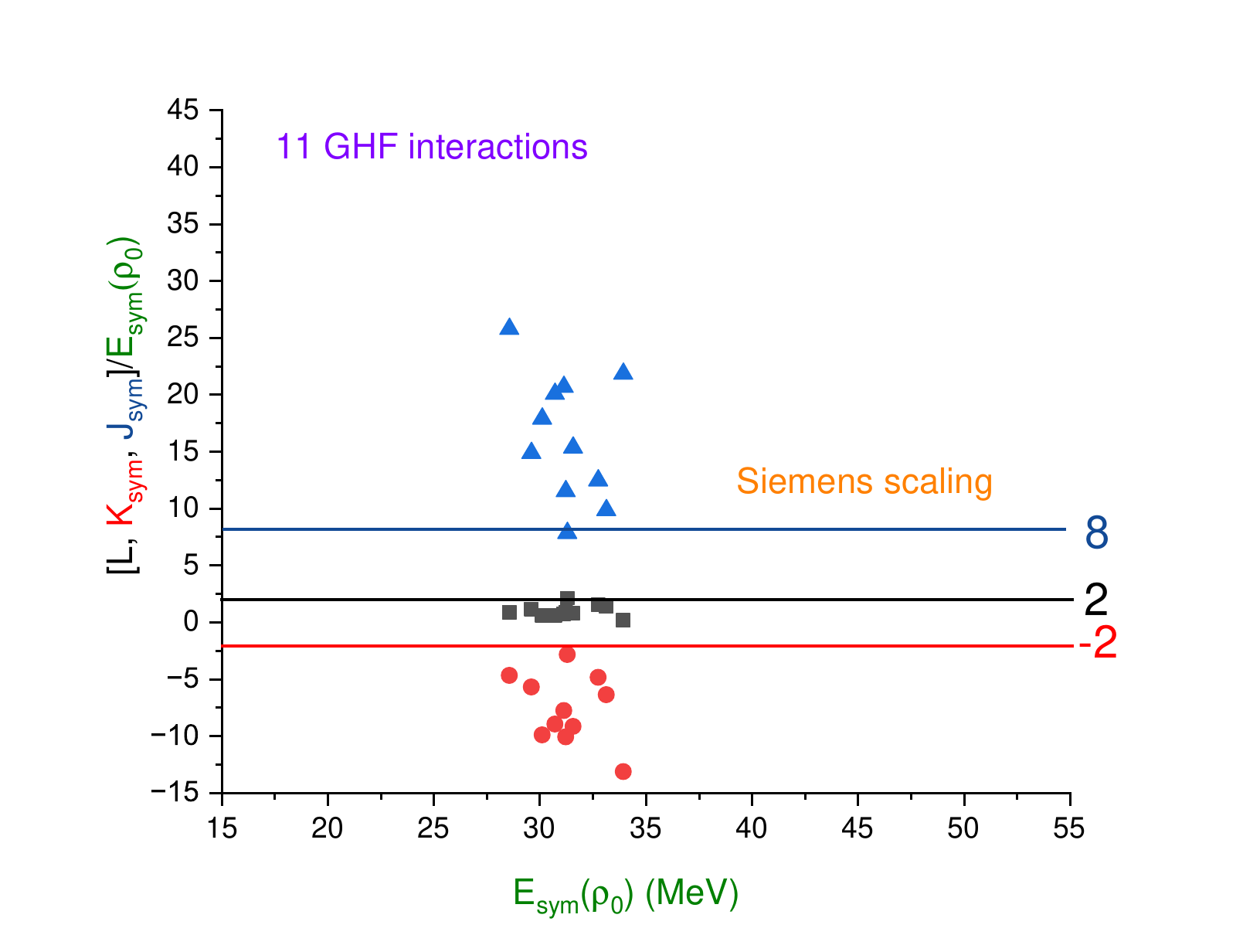}
  \caption{Extracted ratios of $L$ (black), $K_{\rm sym}$ (red) and $J_{\rm sym}$ (blue) over $E_{\rm sym}(\rho_0)$ as a function of the latter for 11 GHF EDFs using the tabulated predictions in Ref. \cite{Rios-G}.}\label{GHF-s}
\end{center}
\end{figure}

Although not as many GHF EDFs as SHF or RMF EDFs are available, as it is more difficult to handle finite-range interactions and the resulting momentum-dependence of single nucleon potentials, it is also interesting to compare GHF predictions on the symmetry energy with Siemens's scaling. Shown in Fig. \ref{GHF-s} are the extracted ratios of $L$ (black), $K_{\rm sym}$ (red) and $J_{\rm sym}$ (blue) over $E_{\rm sym}(\rho_0)$ as a function of the latter for 11 GHF EDFs using the tabulated predictions in Ref. \cite{Rios-G}. Very similar to the situation with the SHF and RMF EDFs, the $L/E_{\rm sym}(\rho_0)$ ratio is close to Siemens's scaling while the other two ratios are rather far from the expectations based on Siemens's scaling. We also note that the predicted $E_{\rm sym}(\rho_0)$ values are all very close to 32 MeV without the broad scattering as in the cases with the SHF and RMF EDFs.

To this end, it is worth mentioning that the density dependence of the symmetry energy in these mean-field type EDFs follows from their chosen functional dependence and thus leads to correlations among the nuclear matter parameters as broadly noticed in the literature, see, e.g., Refs. \cite{Chen07,Chen09,Tews17,NBZ,Mondal17,Holt18,Magno}. 

\end{enumerate}
\begin{figure}
\begin{center}
\includegraphics[scale=0.45]{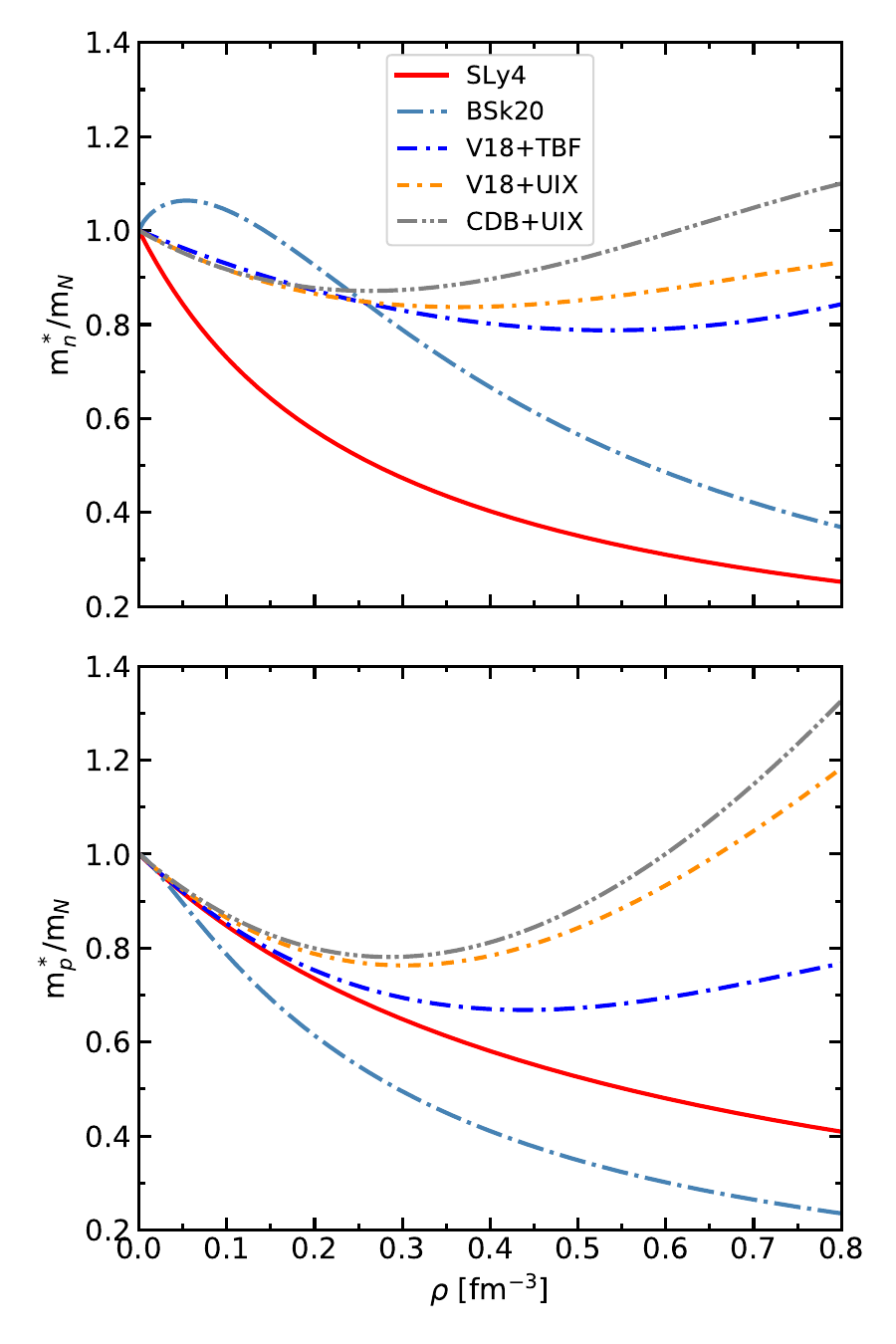} 
\caption{Nucleon effective masses computed using Skyrme interaction SLy4, the modified Skyrme interaction BSk20, and Brueckner-Hartree-Fock results with V18+TBF, V18+UIX, and CDB+UIX interactions of Ref. \cite{Baldo2014}, as functions of density in asymmetric nuclear matter with proton fraction $Y_p=0.1$. Upper panel: neutron effective mass; lower panel: proton effective mass. Taken from Ref. \cite{Duan}.}
\label{emass-france}
\end{center}
\end{figure}

\section{Neutron-proton effective mass splitting from theories, experiments and the $\rho^{2/3}$ scaling for $E_{\rm sym}(\rho)$}
The Lane potential is momentum dependent through both its isoscalar and isovector parts. 
Once an on-shell energy-momentum dispersion relation $E(k)$ or $k(E)$ is obtained from solving the equation $\rm{E}_J=k^2/2m+U_{J} (\rho,\delta,k,\rm{E})$, an equivalent single-particle potential depending on either momentum or energy can be obtained. The total effective mass $M^{\ast}_{J}=m^{\ast}_{J}$ of nucleon $J=$ n or p (similarly, m or M is used interchangeably for the free mass of nucleons in the literature)
\begin{eqnarray}\label{em1}
\frac{m^{*}_{J}}{m}&=&1-\frac{d U_{J}(\rho,\delta,k(E),E)}{dE}\Bigg|_{E(k_{\rm{F}}^{J})}\\ \nonumber
&=&\left[1+\frac{m}{k_{\rm{F}}^J}\frac{d U_{J}(\rho,\delta,k,E(k))}{d k}\Bigg|_{k_{\rm{F}}^J}\right]^{-1}
\end{eqnarray}
then characterizes either the momentum or energy dependence of the single-nucleon potential. Of course, this is the non-relativistic nucleon effective mass. 
We emphasize that there are different kinds of effective masses. Here, we focus on the non-relativistic nucleon effective masses. In relativistic models, nucleons obtain effective masses from very different sources. Nucleon effective masses in relativistic models can be different from their vacuum masses due to interactions that are not explicitly momentum-dependent. For a comparison and definitions of different nucleon effective masses, we refer the reader to a comprehensive review in Ref. \cite{PPNP-Li}.

In fact, effective masses of neutrons and protons in neutron-rich matter have been studied extensively over a long time while many interesting issues remain to be resolved. They are expected to be different and should depend on the density and isospin asymmetry of the medium. They can be described in terms of the isoscalar $M^*_{\rm{s}}$ and isovector $M^*_{\rm{v}}$ nucleon effective mass. While much progress has been made and led to $m^*_0=M^*_{\rm{s}}\approx 0.7m$ where $m$ is nucleon mass in vacuum, due to our poor knowledge about the momentum dependence of isovector interactions, the isovector nucleon effective mass has not been well constrained yet. 

For example, shown in the upper (lower) window of Fig. \ref{emass-france} is neutron (proton) effective masses computed using the Skyrme-Hartree-Fock with the SLy4 and the modified Skyrme interaction BSk20 in Ref. \cite{Duan} in comparsion with the Brueckner-Hartree-Fock (BHF) results with the V18+TBF, V18+UIX, and CDB+UIX (two-body + three-body) interactions of Ref. \cite{Baldo2014}, as functions of density in asymmetric nuclear matter with a proton fraction $Y_p=0.1$. It is seen that neutrons are more massive than protons with the BSK20, while it is the opposite with the SLY4 at all densities. Similarly, results of the BHF calculations also depend on the interaction used and vary strongly with density. Such behaviors are typical in the literature, see, e.g., Ref. \cite{PPNP-Li} for a review. Clearly, based on these typical calculations one can not conclude whether neutrons or protons have higher effective masses and how they may evolve with density and isospin asymmetry of the medium.
\begin{figure}[htb]
\begin{center}
\hspace{-1.5cm}
    \resizebox{0.65\textwidth}{!}{
  \includegraphics{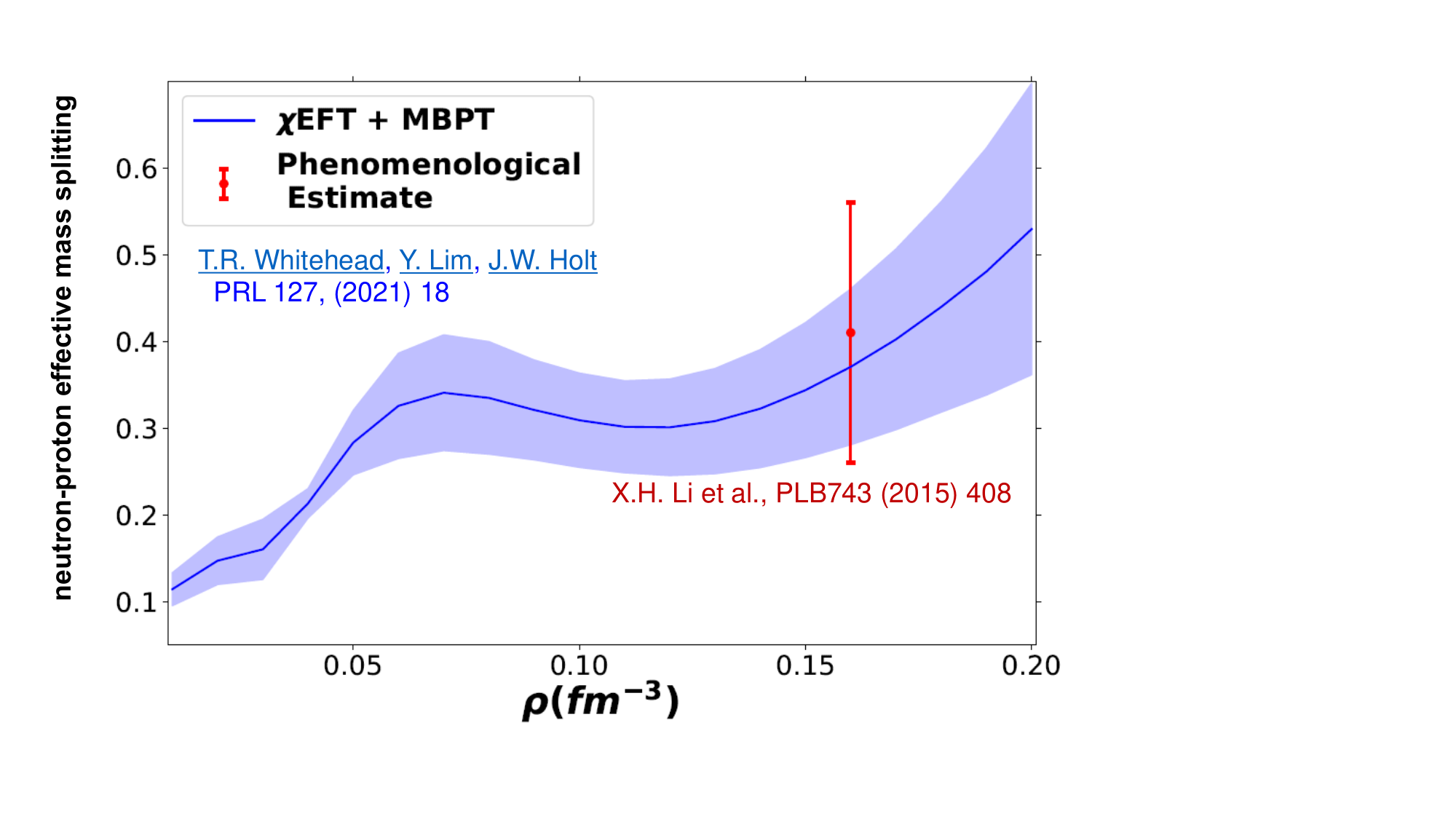}
  }
\vspace{-0.9cm}
\caption{A prediction for the density dependence of neutron-proton effective mass splitting $m^*_{\rm{n-p}}/\delta$ \cite{Whitehead} with a Many-Body Perturbative Theory (MBPT) using $\chi$EFT forces in comparison with the empirical value (red) extracted from nucleon-nucleus scattering data \cite{XHLi}. The figure is modified from a plot provided to the present author by Dr. T.R. Whitehead in 2021.
}\label{emass2}
\end{center}
\end{figure}

\subsection{\bf Neutron-proton effective mass splitting based on many-body perturbation theory using $\chi$EFT forces}
One way to measure the isovector nucleon effective mass is to use the neutron-proton effective mass splitting defined as 
\begin{equation}
m^*_{\rm{n-p}}(\rho,\delta)=(m_{\rm n}^*-m_{\rm p}^*)/m~ {\rm{at}}~ k_F. 
\end{equation}
Diverse predictions on $m^*_{\rm{n-p}}(\rho,\delta)$ exist in the literature. As an example, shown in Fig. \ref{emass2} is a prediction of many-body perturbation theory (MBPT) using $\chi$EFT forces for the density dependence of neutron-proton effective mass splitting \cite{Whitehead} in comparison with the empirical value at $\rho_0$ extracted from nucleon-nucleus scattering data \cite{XHLi}. While both still have large uncertainties, they are in very good agreement within the error bands. 
\begin{figure*}[htb]
\begin{center}
    \resizebox{0.8\textwidth}{!}{
  \includegraphics{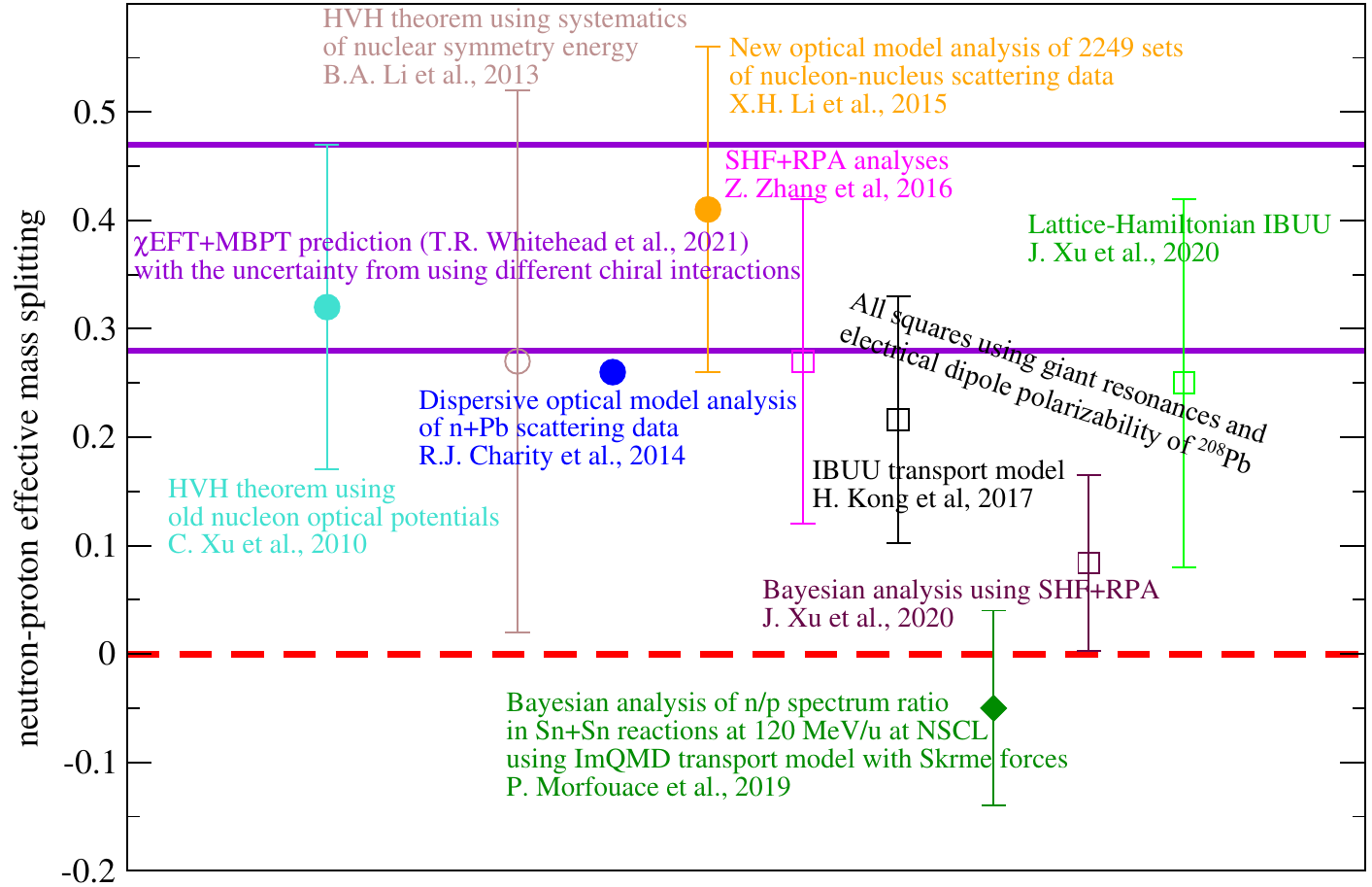}
  }
\caption{A survey of neutron-proton effective mass splitting $m^*_{\rm{n-p}}/\delta$ at $\rho_0$ from analyzing nuclear reaction and structure experiments in comparison with the latest chiral effective field theory prediction (its 68\% confidence range is indicated by the horizontal violet lines). Taken from Ref. \cite{Li-Italy}. Similar plots, possibly with additional data and/or calculations, can also be found in Refs. \cite{SWang,Yang25}.}\label{emass}
\end{center}
\end{figure*}
\subsection{\bf Neutron-proton effective mass splitting from experiments}
As shown in Fig. \ref{emass}, results from analyzing various other data indicated in the plot scatter broadly. Nevertheless, most of the results from the indicated analyses are consistent with the 
$\chi$EFT+MBPT prediction within their 68\% error bars. In particular, all earlier analyses of nucleon-nucleus scattering data \cite{XuC10,XHLi,Bob}, the analysis based on the HVH theorem using the 2013 systematics of nuclear symmetry energy \cite{LiBA13} as well as both the static and dynamical model analyses of the isoscalar and isovector giant resonances and the electrical dipole polarizability of $^{208}$Pb \cite{ZZ16,Kong17,Xu20,Xu20b} indicate surely a positive $m^*_{\rm{n-p}}(\rho_0,\delta)$. We notice that the four open squares are from using the same data set but different approaches. They give qualitatively consistent but quantitatively appreciably different $m^*_{\rm{n-p}}(\rho_0,\delta)$ values. 

One interesting exception is the result of $m^*_{\rm{n-p}}(\rho_0,\delta)=(-0.05\pm 0.09)\delta$ from a Bayesian analysis of the n/p spectrum ratios in several Sn+Sn reactions at 120 MeV/nucleon at MSU using an ImQMD transport model for nuclear reactions with Skyrme forces \cite{MSU}. More recently, using the ImQMD model again a value of $m^*_{\rm{n-p}}(\rho_0,\delta)=(-0.07^{+0.07}_{-0.06})\delta$ was extracted from analyzing the directed and elliptic flows in several Sn+Sn reactions at a beam energy of 270 MeV/A at RIKEN by the S$\pi$RIT Collaboration \cite{CTsang}. Their data have provided new challenges for some other transport models and triggered some non-traditional thoughts. For instance, within the IBUU11 transport code using momentum-dependent isoscalar and isovector potentials consistent with optical model potentials \cite{Xu15}, relative effects of the $E_{\rm{sym}}(\rho)$ and $m^*_n-m^*_p$ on the n/p ratio of free nucleons and those in light clusters were investigated. It was found that the $m^*_n-m^*_p$ has a relatively stronger effect than the $E_{\rm{sym}}(\rho)$. Indeed, the assumption of $m^*_n\leq m^*_p$ leads to a higher n/p ratio in both $^{124}$Sn+$^{124}$Sn and $^{112}$Sn+$^{112}$Sn collisions at 50 and 120 MeV/A beam energies. However, results of IBUU11 calculations using the $E_{\rm{sym}}(\rho)$ within its current uncertainty range and a positive $m^*_n-m^*_p$ consistent with the nucleon isovector optical potential from analyzing nucleon-nucleus scattering data are all too low compared to the NSCL/MSU double n/p ratio data \cite{Kong15}. 

Instead, assuming no momentum dependence at all (consequently $m^*_n=m^*_p=m$), a covariance analysis of the NSCL/MSU n/p double ratio data within BUU by adjusting the $\eta$ and $\gamma$ values in parameterizing the $E_{\rm{sym}}(\rho)$ according to Eq. (\ref{esym}) found the most probably values of 
$\eta=-0.30(1\pm18.53\%)$ corresponding to a reduced kinetic symmetry energy at $\rho_0$ to be $E^{\rm{kin}}_{\rm{sym}}(\rho_0)=-(3.8\pm0.7)$ MeV and an enhanced/stiffened potential symmetry energy with $\gamma=0.80(1\pm 5.98\%)$ \cite{Hen15b,LiG15}. Such results are consistent with earlier findings that the pre-equilibrium n/p ratio depends sensitively on the stiffness of the isovector potential \cite{Li97}. As we shall discuss in the next Section, while keep the magnitude of $E_{\rm{sym}}(\rho)$ at $\rho_0$ fixed and its density dependence within its current uncertainty range, SRC can reduce the kinetic but increase the potential contribution to $E_{\rm{sym}}(\rho)$. Analyses of data from several SRC experiments and microscopic nuclear many-body calculations incorporating tensor force and SRC effects have found evidence that the $E^{\rm{kin}}_{\rm{sym}}(\rho_0)$ is much less than 12 MeV for FFG. Indeed, several studies have shown that the $E^{\rm{kin}}_{\rm{sym}}(\rho_0)$ can be negative, as we shall elaborate in the next Section. Then, to keep the total $E^{\rm{kin}}_{\rm{sym}}(\rho)$ fixed, the potential symmetry energy (the corresponding isovector nucleon potential will then be more repulsive/attractive for neutrons/protons, leading to their more/less emissions in heavy-ion collisions) has to be increased accordingly. Consequently, the required stiffness parameter $\gamma$ of total symmetry energy is also increased to $\gamma\approx 0.8$ compared to the fiducial value of about $0.7$ as we discussed earlier.

To this end, it is necessary to emphasize that the nucleon effective mass is a fundamental quantity characterizing the propagation of a nucleon in a nuclear medium, accounting approximately for effects due to the space-time non-locality of the effective nuclear interactions \cite{Jaminon:1989wj,Jeukenne:1976uy,Sjoberg:1976tq,vanDalen:2005ns}. The magnitude and sign of $m_{np}^*(\rho)$ in neutron-rich matter have essential consequences for cosmology, astrophysics, and nuclear physics through influencing, e.g., the equilibrium neutron to proton ratio in the early universe and primordial nucleosynthesis \cite{Steigman:2005uz}, properties of mirror nuclei \cite{Nolen:1969ms}, and locations of neutron and proton drip-lines \cite{Woods:1997cs}.
Given the discrepancies of the results from analyzing various experiments discussed above and diverse model predictions, obviously, more accurate data from different approaches will be very useful. In this regard, it is encouraging to note the approved FRIB Experiment Proposal 23058 to ``Measuring the isospin dependence of the nucleon effective mass at supersaturation density" using $^{56,70}$Ni+$^{58,64}$Ni reactions at 175 MeV/A \cite{KBrown}.

\subsection{\bf Neutron-proton effective mass splitting from $\rho^{2/3}$ scaling of $E_{\rm sym}(\rho)$ within HVH theorem}
The discrepancies discussed above certainly call for more efforts both theoretically and experimentally to pin down the neutron-proton effective mass splitting at least around $\rho_0$ first. For this purpose, we make some comments below about the physics underlying the $m^*_{\rm{n-p}}(\rho_0,\delta)$ at the mean-field level. 

In terms of the isoscalar and isovector nucleon potentials, the $m^*_{\rm{n-p}}(\rho,\delta)$
can be written as \cite{PPNP-Li}
\begin{equation}\label{mnp-src}
m^*_{\rm{n-p}}(\rho,\delta)\approx 2\delta\frac{m}{k_{\rm{F}}}\left[-\frac{dU_{\rm{sym},1}}{dk}-\frac{k_{\rm{F}}}{3}\frac{d^2U_0}{dk^2}+\frac{1}{3}\frac{dU_0}{dk}\right]_{k_{\rm{F}}}\left(\frac{m^*_0}{m}\right)^2.
\end{equation}
According to the HVH theorem discussed earlier, the $m^*_{\rm{n-p}}$ at $\rho_0$ is related to $E_{\rm{sym}}(\rho_0)$ and $L(\rho_0)$ via\,\cite{LiBA13}
\begin{equation}
m^*_{\rm{n-p}}(\rho_0,\delta)\approx\delta\cdot\frac{\displaystyle3E_{\rm{sym}}(\rho_0)-L(\rho_0)-3^{-1}({m}/{m^*_0})E_{\rm{F}}(\rho_0)}{\displaystyle
E_{\rm{F}}(\rho_0)\left({m}/{m_0^*}\right)^2}.\label{mnp}
\end{equation}
Therefore, the $m_{\rm n}^*$ is equal to, larger or smaller than the $m_{\rm p}^*$ depends on the symmetry energy and its slope. For example, with the empirical values of \es0=31 MeV, $m_0^*/m=0.7$ and $E_{\rm{F}}(\rho_0)=36$ MeV, a positive $m^*_{\rm{n-p}}(\rho_0,\delta)$ implies that the slope L should be less than $76$ MeV. Moreover, using $L=2$\es0according to Siemens' $\rho^{2/3}$ scaling for $E_{\rm sym}(\rho)$, one expects $m^*_{\rm{n-p}}(\rho_0,\delta)\approx (0.148-0.229)\delta$ with \es0=$31\pm 3$ MeV. Interestingly, it is consistent with the $\chi$EFT+MBPT prediction and the phenomenological value from optical model analyses of nucleon-nucleus scattering. 

\subsection{\bf Neutron-proton effective mass splitting from SHF}
Since many SHF EDFs have been used in analyzing various experimental data, it is necessary to have a closer look at what these SHF EDFs predict for the 
$m^*_{\rm{n-p}}(\rho_0,\delta)$. Since not all publications have given detailed information about their predictions for $m^*_{\rm{n-p}}(\rho_0,\delta)$, we evaluate their predictions for this quantity in two ways here. Firstly, using their predictions for $E_{\rm{sym}}(\rho_0)$, $m_0^*/m$ and $L$, we can obtain their HVH value for $m^*_{\rm{n-p}}(\rho_0,\delta)$ using the equation (\ref{mnp}). Secondly, for those published explicitly their $M^*_{\rm{s}}$ and $ M^*_{\rm{v}}$ values, we can easily calculate their corresponding $m^*_{\rm{n-p}}(\rho_0,\delta)$ according to the following well established relationship \cite{PPNP-Li}
\begin{equation}\label{sign0}
\frac{m^*_{\rm{n}}-m^*_{\rm{p}}}{m^*_{\rm{n}}m^*_{\rm{p}}}= 2\delta\frac{M^*_{\rm{s}}-M^*_{\rm{v}}}{M^*_{\rm{s}}M^*_{\rm{v}}}.
\end{equation} 
Thus, $m^*_{\rm{n-p}} > 0$ if $M^*_{\rm{s}} > M^*_{\rm{v}}$. Moreover, assuming \[(m^*_{\rm{n}}/m)(m^*_{\rm{p}}/m)\approx (M^*_{\rm{s}}/M)^2\] valid when $\delta$ is small, one has 
\begin{equation}\label{mnp2}
m^*_{\rm{n-p}}\approx 2\delta\left(\frac{M^*_{\rm{s}}}{M}\right)^2\left[\frac{M}{M^*_{\rm{v}}}-\frac{M}{M^*_{\rm{s}}}\right]. 
\end{equation}
A comparison of the results from these two approaches will indicate whether these EDFs satisfy the HVH theorem regarding the nucleon isovector effective mass. 
\begin{figure}[htb]
\begin{center}
\vspace{-0.3cm}
    \resizebox{0.45\textwidth}{!}{
  \includegraphics{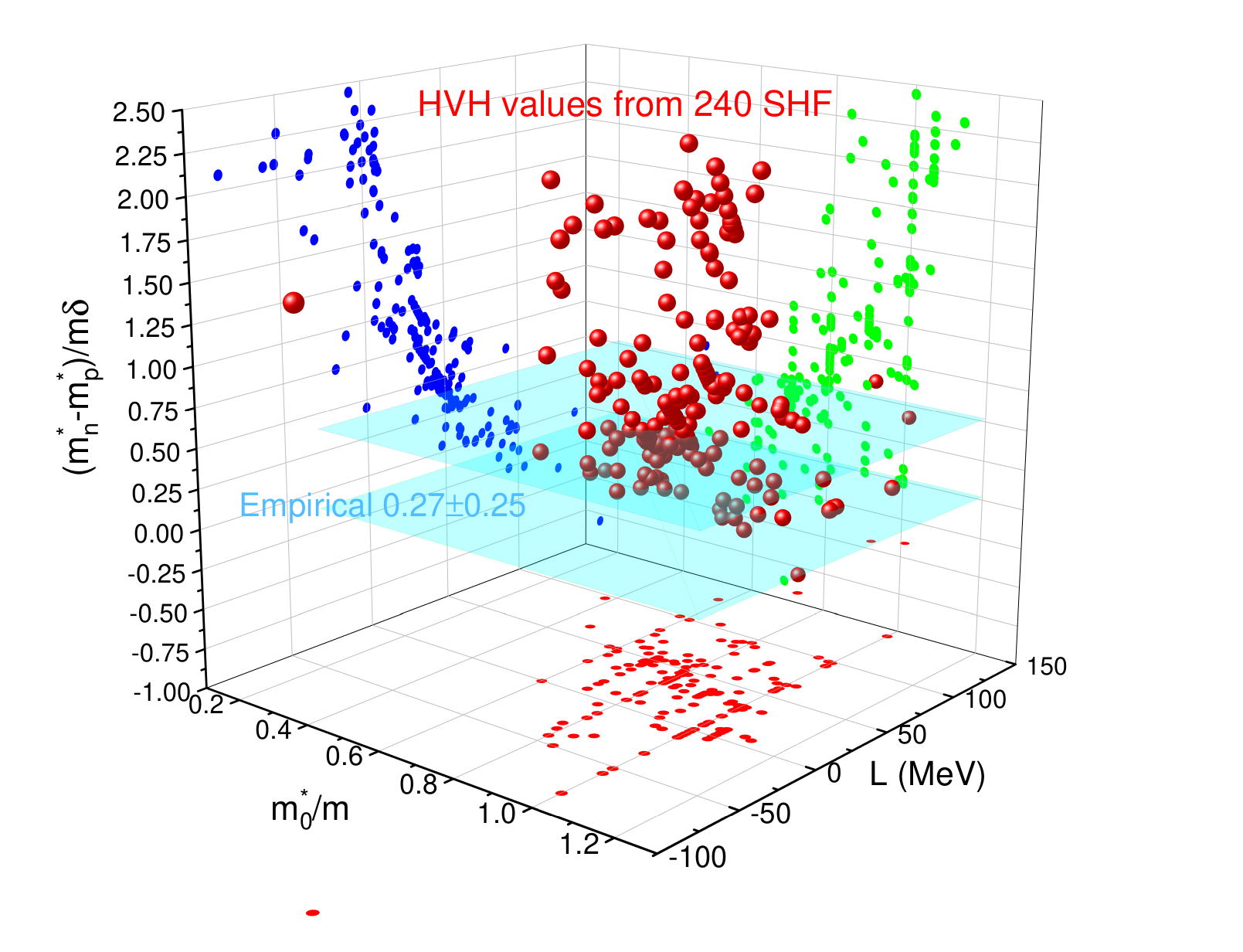}
  }
\caption{Values (red balls) of the neutron-proton effective mass splitting evaluated using EOS parameters predicted by 240 SHF EDFs \cite{Dut12} according to the HVH theorem of Eq. (\ref{mnp}) in comparison with phenomenological boundaries (cyan planes) estimated using the 2013 fiducial values (shown in Fig. \ref{Esym0L}) of nuclear symmetry energy parameters derived in Ref. \cite{LiBA13}. The red (bottom), green (back), and blue (left) points are projections of the red balls to the three 2-dimensional planes indicated by the relevant axes, respectively.
}\label{emass-SHF}
\end{center}
\end{figure}

Shown in Fig. \ref{emass-SHF} are the HVH value (red balls) for $m^*_{\rm{n-p}}(\rho_0,\delta)$ from using the EOS parameters predicted by 240 SHF EDFs \cite{Dut12} according to the HVH theorem of Eq. (\ref{mnp}). They are compared with the phenomenological boundaries estimated using the 2013 fiducial values of nuclear symmetry energy parameters in Ref. \cite{LiBA13}. Projections of them onto the three 2-dimensional planes illustrate more clearly individual effects of $L$ and $m^*_0$ as well as their correlations inherent in the 240 SHF predictions. Obviously, the predictions are rather diverse. Nevertheless, most of them predict $m^*_{\rm{n-p}}(\rho_0,\delta)\gt 0$. However, only few fall into the indicated phenomenological constraining region of $m^*_{\rm{n-p}}(\rho_0,\delta)/\delta=0.27\pm 0.25$. 

\begin{figure}[ht]
\vspace{-0.4cm}
\begin{center}
  \includegraphics[width=8.cm]{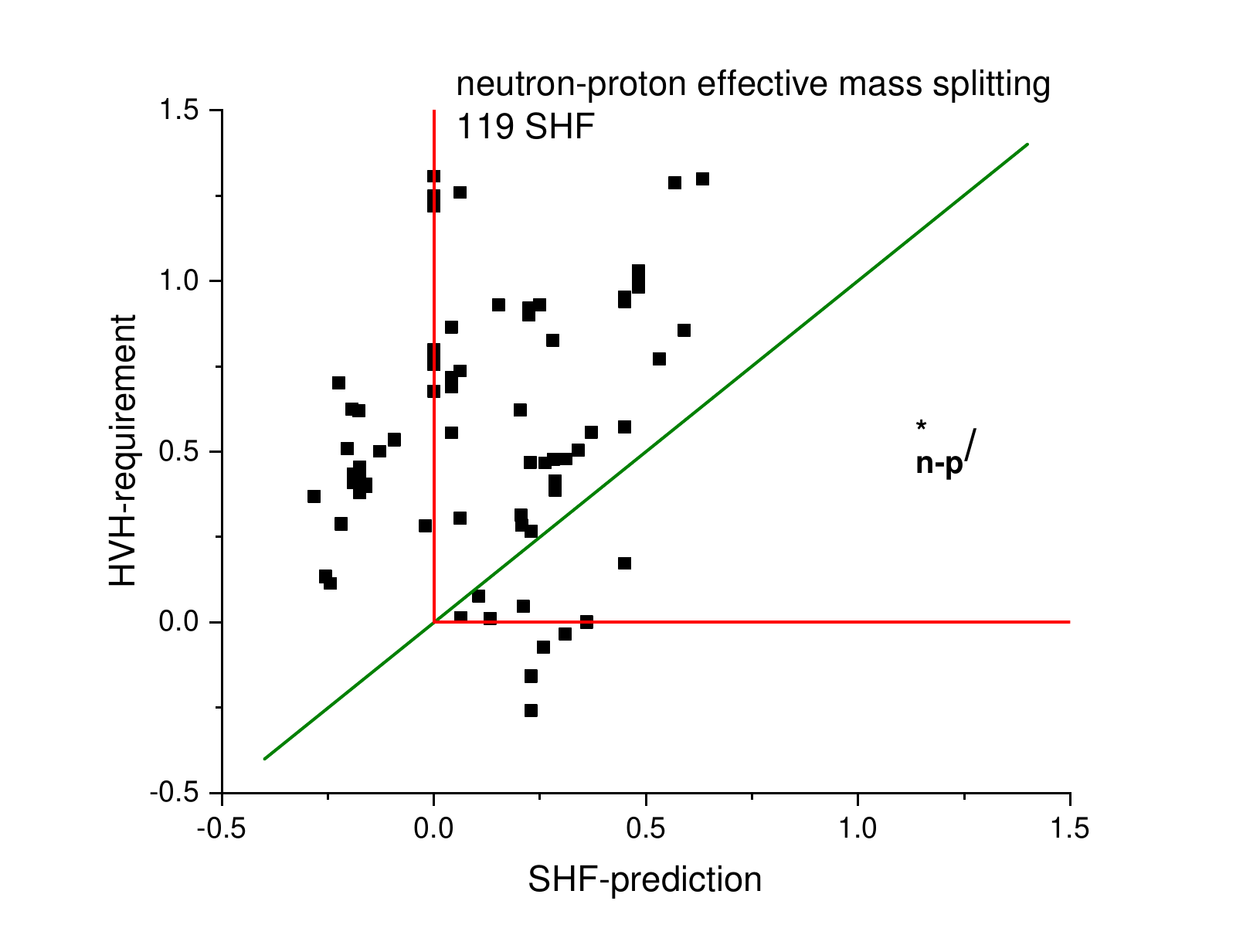}
  \caption{Comparing the HVH values of the neutron-proton effective mass splitting evaluated using Eq. (\ref{mnp}) and that using directly 
the reported $M^*_{\rm{s}}$ and $ M^*_{\rm{v}}$ values according to Eq. (\ref{mnp2}) for 119 SHF EDFs studied in Ref. \cite{Wang23}. If they are equal (satisfying the HVH theorem), they should fall on the green line. }\label{HVH-SHF}
\end{center}
\end{figure}
Shown in Fig. \ref{HVH-SHF} is a comparison of the HVH values (vertical) of the neutron-proton effective mass splitting evaluated using Eq. (\ref{mnp}) and that (horizontal) using directly the reported $M^*_{\rm{s}}$ and $ M^*_{\rm{v}}$ values according to Eq. (\ref{mnp2}) for 119 SHF EDFs studied in Ref. \cite{Wang23}. If they are equal (satisfying the HVH theorem), they should fall on the green line along the diagonal. Not surprisingly and consistent with the results shown in Fig. \ref{emass-SHF}, few of them fall exactly on the green line. Again, both approaches predict consistently positive $m^*_{\rm{n-p}}(\rho_0,\delta)$ values with most (above and on the right of the red lines) of the 119 SHF EDFs. It would be interesting to examine directly using other approaches how well these SHF EDFs satisfy the HVH theorem of Eq. (\ref{HVH}). Since it is well known that these SHF EDFs normally give different binding energies and may saturate at different densities, our finding here is not surprising.

\subsection{\bf Neutron-proton effective mass splitting as a surrogate of isospin-quartic symmetry energy}
It is very challenging to determine accurately the $m^*_{\rm{n-p}}$ because technically the isovector quantities are generally very small compared to the isoscalar ones, besides some poorly known physics reasons. However, isovector properties of nuclear matter are fundamentally important and have broad applications in exploring properties of rare isotopes and neutron stars. In the context of this brief review, it is thus interesting to note that the quartic symmetry energy $E_{\rm{sym},4}(\rho)$ is completely determined by the isoscalar and isovector nucleon effective masses according to\,\cite{PuJ17}
\begin{equation}\label{Esym4ofmsmv}
E_{\text{sym},4}(\rho)=\frac{\hbar^2}{162M}\left(\frac{3\pi^2\rho}{2}\right)^{{2}/{3}}\left[\frac{3M}{M_{\rm{v}}^*(\rho)}-\frac{2M}{M_{\rm{s}}^*(\rho)}\right].
\end{equation}
It is seen that a larger $E_{\text{sym},4}(\rho)$ would require a very small $M^*_{\rm{v}}$ but larger $M^*_{\rm{s}}$\,\cite{PuJ17}.
Indeed, it was shown numerically that the $E_{\text{sym},4}(\rho_0)$ strongly correlates positively with $M_{\rm{s}}^*$ but negatively with $M_{\rm{v}}^*$.
Given the large uncertainties of the $M_{\rm{s}}^*$ and $M_{\rm{v}}^*$ even at $\rho_0$, it is not surprising that the $E_{\text{sym},4}(\rho)$ from various models in the literature are still very different \cite{Cai15a}. However, it has been known for a long time that a small variation of $E_{\text{sym},4}(\rho)$ may lead to a big change in the crust-core transition density and pressure in neutron stars, see, e.g., Refs. \cite{Baym1971,Baym2,Arp72,Xu09,Cai12,Seif14,Gon17,Seif25}. The presence of $E_{\text{sym},4}(\rho)$ can also strongly modify the critical density for the direct Urca process, which leads to faster cooling of NSs \cite{Andrew}.
This is mainly because the incompressibility $K_{\mu}$ of uniform NS outer core at $\beta$-equilibrium vanishes at the crust-core transition point \cite{Lat}, indicating the onset of a dynamical instability by forming clusters. More specifically, $K_{\mu}$ can be written as \cite{Xu09,Lat,Kubis}
\begin{eqnarray}\label{ther5}
K_\mu &=& 2
\rho \frac{\partial E_b(\rho,x_p)}{\partial \rho} + \rho^2
\frac{\partial^2 E_b(\rho,x_p)}{\partial \rho^2}\nonumber\\ 
&-&\left(\frac{\partial^2 E_b(\rho,x_p)}{\partial \rho
\partial x_p}\rho\right)^2/\frac{\partial^2 E_b(\rho,x_p)}{\partial x_p^2}
\end{eqnarray}
where $E_b(\rho,x_p)=E(\rho,\delta)$ is the energy per baryon in ANM and $x_p=(1-\delta)/2$ is the proton fraction.
Because it involves the second-order derivatives of nuclear EOS with respect to both $\delta$ and $x_p$, for a recent review with examples, see, e.g., Ref. \cite{LiBA19}, a small variation of $E_{\text{sym},4}(\rho)$ can make big differences in the resulting crust-core transition properties by setting $K_{\mu}=0$. Thus, knowing more accurately the $m^*_{\rm{n-p}}(\rho,\delta)$ will help better constrain the $E_{\text{sym},4}(\rho)$. 

\begin{figure}[ht]
\vspace{-0.3cm}
\begin{center}
\includegraphics[width=8.cm]{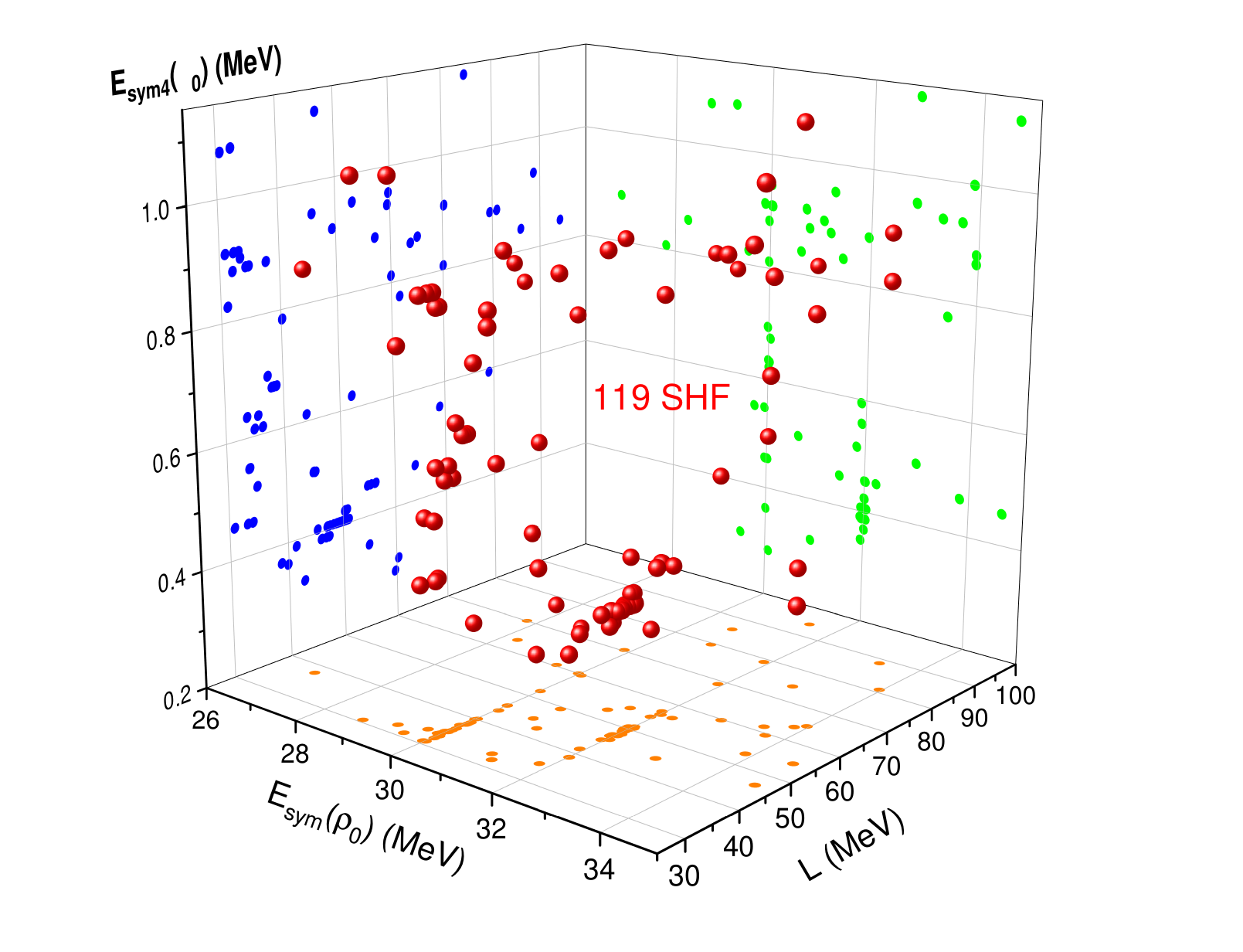}
  \caption{The quartic symmetry energy $E_{\rm{sym,4}}(\rho_0)$ (red balls) according to Eq. (\ref{Esym4ofmsmv}) versus the $E_{\rm{sym}}(\rho_0)$ and $L$ values of the same 119 SHF EDFs studied in Ref. \cite{Wang23} as in Fig. \ref{HVH-SHF}. Their projections to the 2D planes are given in blue, yellow, and green dots, respectively.}\label{Esym4SHF}
\end{center}
\end{figure}

\begin{figure}[ht]
\vspace{-0.5cm}
\begin{center}
\includegraphics[width=8.cm]{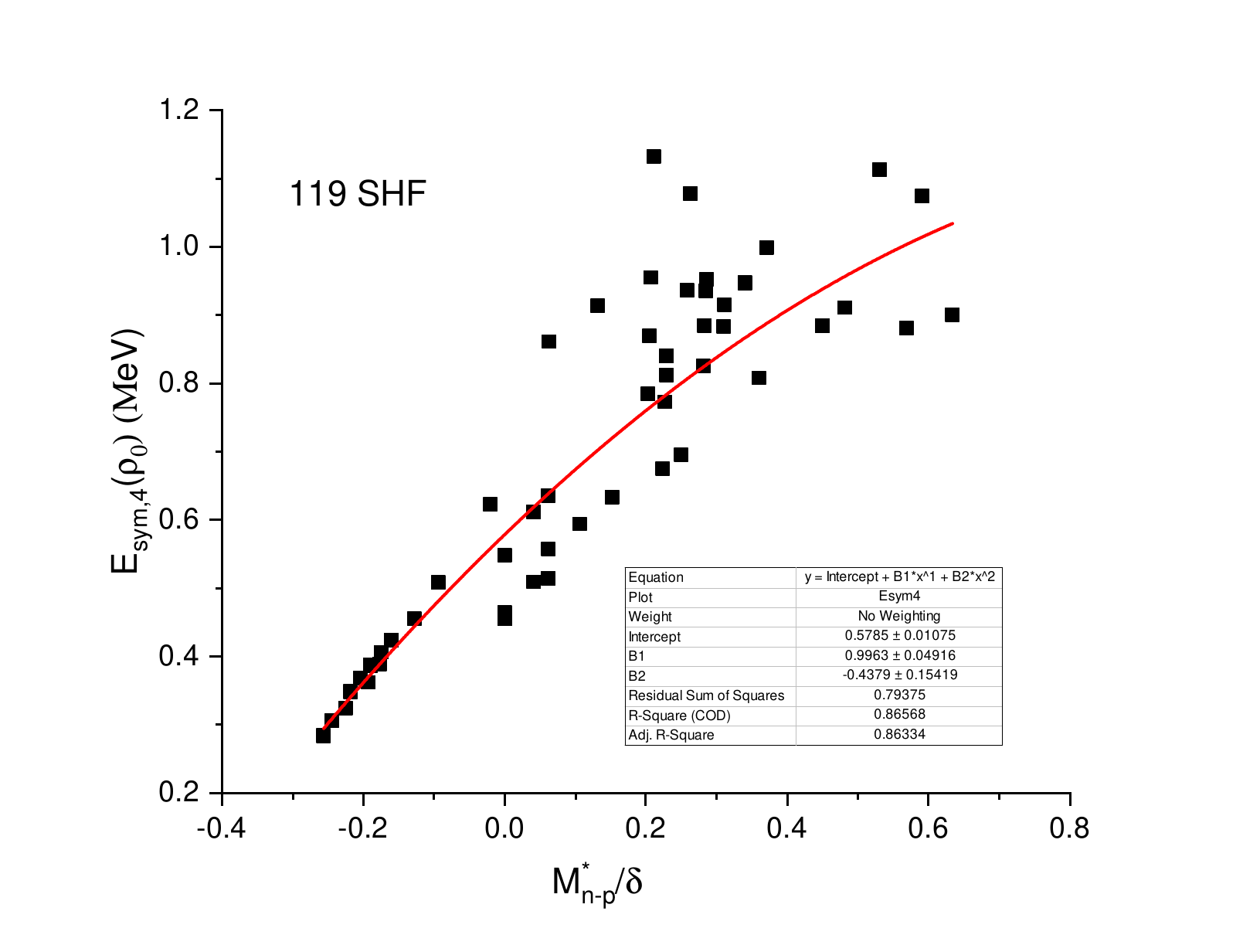}
  \caption{The quartic symmetry energy $E_{\rm{sym,4}}(\rho_0)$ versus the neutron-proton effective mass splitting 
   $m^*_{\rm{n-p}}/\delta$ at $\rho_0$ from the same 119 SHF EDFs studied in Ref. \cite{Wang23} as in Fig. \ref{HVH-SHF}. The red curve is a polynomial fit with the parameters listed in the inset.}\label{Esym4D}
\end{center}
\end{figure}

As shown in Eq.\,(\ref{sign0}), $(M^*_{\rm{n}}-M^*_{\rm{p}})\propto (M^*_{\rm{s}}-M^*_{\rm{v}})$, the $E_{\rm{sym},4}(\rho)$ is thus closely related to the 
$m^*_{\rm{n-p}}(\rho,\delta)$. As a numerical example, using $M^{*}_{\rm{s}}/m = 0.82\pm0.08$ and $M^{*}_{\rm{v}}/m = 0.69\pm0.02$ predicted by the $\chi$EFT calculations in Refs. \cite{Hol13,Hol16}, one finds $E_{\text{sym},4}(\rho_0)=0.868\pm 0.142$ MeV at $\rho_0=0.16 \pm 0.02$ fm$^{-3}$. This sets a useful reference for comparing with predictions of SHF EDFs. 

Shown in Fig. \ref{Esym4SHF} are values of $E_{\rm{sym,4}}(\rho_0)$ according to Eq. (\ref{Esym4ofmsmv}) versus the $E_{\rm{sym}}(\rho_0)$ and $L$ values of the same 119 SHF EDFs used in obtaining the results shown in Fig. \ref{HVH-SHF}. They scatter in the range of 0.2 to 1.5 MeV. They are relatively small compared to the $E_{\rm{sym}}(\rho_0)$ and are thus hard to determine both experimentally and theoretically, given the current status of the field. To our best knowledge, there is presently no experimental constraint on $E_{\rm{sym,4}}(\rho_0)$ from either astrophysical observations nor laboratory experiments. Interestingly, as shown in Fig. \ref{Esym4D}, the $E_{\rm{sym,4}}(\rho_0)$ (Eq. \ref{Esym4ofmsmv}) is indeed strongly correlated with the $m^*_{\rm{n-p}}/\delta$ at $\rho_0$ (Eq. \ref{mnp2}) for the same 119 SHF EDFs studied in Ref. \cite{Wang23}.
Quantitatively, they are related by 
\begin{equation}
E_{\rm{sym,4}}(\rho_0)\approx0.59+[m^*_{\rm{n-p}}/\delta]-0.44[m^*_{\rm{n-p}}/\delta]^2~({\rm{MeV}}). 
\end{equation}
Therefore, the $m^*_{\rm{n-p}}(\rho,\delta)$ may serve as a useful surrogate of $E_{\rm{sym,4}}(\rho_0)$.

Our discussions above indicate that the $m^*_{\rm{n-p}}(\rho,\delta)$ and $E_{\rm{sym,4}}(\rho_0)$ may be determined together. As illustrated in Fig. \ref{emass}, efforts to determine the $m^*_{\rm{n-p}}(\rho,\delta)$ using terrestrial nuclear experiments are currently inconclusive. Nevertheless, they are very encouraging as they at least show the investigated observables have strong potentials to help pin down the $m^*_{\rm{n-p}}(\rho,\delta)$ and subsequently the $E_{\rm{sym,4}}(\rho)$. Of course, new ideas and approaches using nuclear reactions and structures, especially involving rare isotopes, are clearly needed to make further progress in this direction.

\begin{figure*}[htb]
\centering
  \includegraphics[bb=-20 0 540 500,width=11.cm,height=10cm]{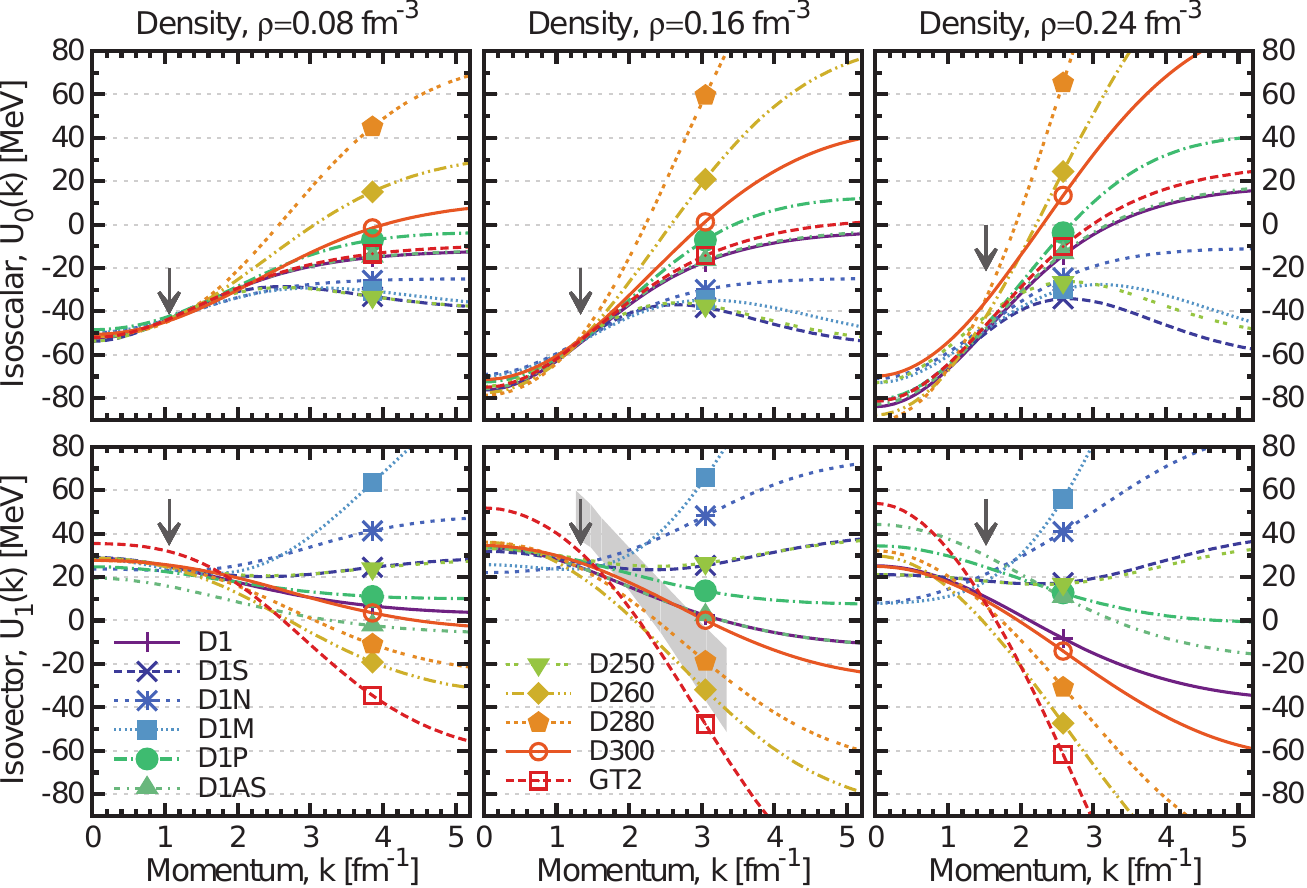}

\vspace{0.5cm}
\caption{The momentum dependence of nucleon isoscalar (upper) and isovector (lower) potentials from eleven GHF energy functionals at $\rho_0/2$, $\rho_0$ and $1.5\rho_0$, respectively. Taken from Ref. \cite{Rios-G}}
\label{CS}
\end{figure*}
\begin{figure*}[htb]
\centering
 \resizebox{0.35\textwidth}{!}{
\includegraphics[]{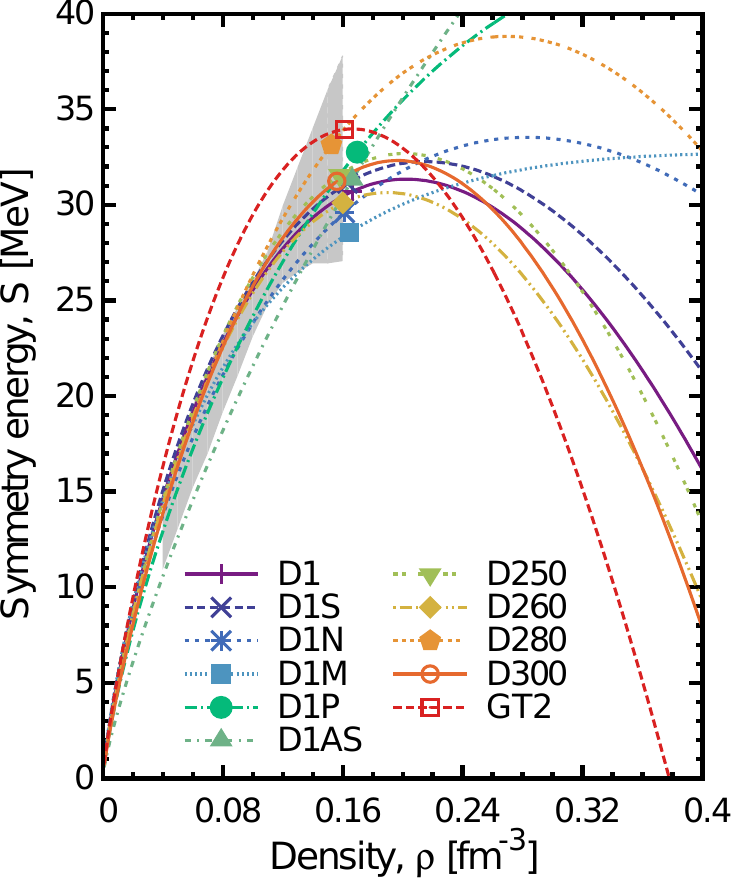}
}
\hspace{1.5cm}
 \resizebox{0.35\textwidth}{!}{
  \includegraphics[]{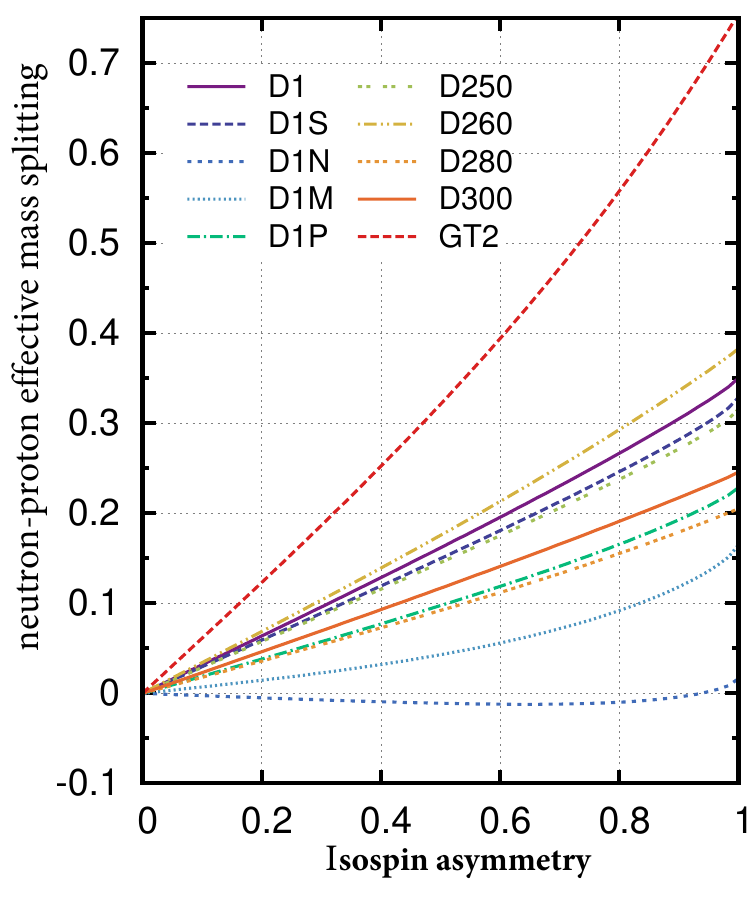}
  }

\vspace{0.5cm}
\caption{Left: The density dependence of nuclear symmetry energy. The shaded area indicates the experimental constraints at sub-saturation densities. Right: The isospin dependence of neutron-proton effective mass splitting $m^*_{\rm{n-p}}$ at $\rho_0$ (right) from eleven GHF EDFs. Taken from Ref. \cite{Rios-G}}\label{Rios-em}
\end{figure*}
\section{Expected breakdown of the $\rho^{2/3}$ scaling at high densities and its microscopic causes}
Siemens's $\rho^{2/3}$ scaling near $\rho_0$ emerges under two key assumptions: (i) the isoscalar mean-field potential is quadratic in momentum, and (ii) the isovector potential is only weakly density dependent. These assumptions are reasonable near $\rho_0$, but at supra-saturation densities and in very neutron-rich matter, they can fail for several reasons. For example, 
\begin{enumerate}
    \item \textbf{Non-quadratic momentum dependence.}  
    At high densities, the nucleon single-particle potential acquires higher-order momentum dependence due to finite-range interactions, making the quadratic approximation questionable \cite{Gale87,Wir88,Bom01}.\\

    \item \textbf{Strong density dependence of the isovector interaction.}  
    Generally, the symmetry potential can vary strongly with density \cite{Rios-G}. While in this review we focus mainly on non-relativistic approaches, it is important to point out that relativistic models, especially with contributions from $\rho$- and $\delta$-mesons or density-dependent couplings, the predicted symmetry energy may strongly deviate from $\rho^{2/3}$ scaling \cite{Thakur2022}, as demonstrated in Fig. \ref{SHF-RMF}.\\

    \item \textbf{Spin-isospin dependent tensor forces, short-range correlations, and three-body forces.}  
    These effects become increasingly important at high density, modifying nucleon effective masses and spectral functions, leading to density-dependent corrections to the symmetry energy \cite{PPNP-Li,GEB94,Eng,Sam,Muther,Lee}. In particular, the spin-isospin dependent tensor force has long been known to affect the density dependence of nuclear symmetry energy \cite{Kuo}, see, e.g., Ref. \cite{Ba85} for an earlier review and a recent one in Ref. \cite{PPNP-Li}. Because the tensor force exists mostly in the neutron-proton isosiglet channel, it can make the EOS of SNM increase faster than that of PNM as the density increases, leading to a decreasing $E_{\rm{sym}}(\rho)$ above certain suprasaturation density \cite{Pan72,WFF}. Moreover, the kinetic part of $E_{\rm{sym}}(\rho)$ can become negative \cite{CXu11,Vid11,Xulili,Hen15b} as the isospin-dependent SRC makes protons more energetic than neutrons in neutron-rich matter \cite{Hen14,Hen20a,Cai16}.\\

    \item \textbf{Relativistic self-energies and Lorentz structure.}  
    In relativistic mean-field approaches, scalar and vector self-energies evolve with density in a way that cannot be represented by a simple quadratic dependence on momentum \cite{Maslov2015}.\\

    \item \textbf{Emergence of new degrees of freedom.}  
    At higher densities, hyperons, $\Delta$ resonances, meson condensates, or a hadron-quark transition are generally expected to occur \cite{Stu}, qualitatively altering the dynamics of the system and invalidating not only the $\rho^{2/3}$ scaling of nuclear symmetry energy but all predictions based on many-body theories considering only nucleons at suprasaturation densities.\\

    \item \textbf{Large isospin splitting of nucleon effective masses.}  
    While the neutron-proton effective mass splitting is rather small around $\rho_0$ as shown in the previous section, a strong momentum dependence of the isovector potential may appear at high densities \cite{Rios-G}, which may lead to a $E_{\rm{sym}}(\rho)$ incompatible with a universal $\rho^{2/3}$ scaling.\\

    \item \textbf{In-medium modifications of nucleon quasiparticles.}  
    At high densities, the nucleon spectral functions broaden and the strength shifts away from a sharp quasiparticle peak \cite{SCGF}, breaking the simple single-particle picture assumed to derive the $\rho^{2/3}$ scaling.
\end{enumerate}
To be clearer, we illustrate below the main points mentioned above with two examples from the literature.

\subsection{\bf Example-1: Density and momentum dependences of nucleon isoscalar and isovector potentials as well as the corresponding $E_{\rm{sym}}(\rho)$ and $m^*_{\rm{n-p}}$ from Gogny-Hartree-Fock (GHF)}\label{ghf} 
Currently, eleven popular versions of Gogny-type forces are widely used in the literature mainly for studying nuclear structure successfully \cite{Gogny,Blaizot}. Applying them to studying the properties of ANMs and neutron stars has met some challenges. In particular, various attempts have been made to stiffen the Gogny-type EOSs for dense neutron-rich matter to be consistent with astrophysical observations without compromising their successes at low densities in studying nuclear structures \cite{NewG1}. Unfortunately, little success has been achieved, and some of the proposed mechanisms are under debate \cite{Rios-G,NewG2}. 

As an illustration of the diversity and importance of fixing the Gogny energy functional, shown in Fig. \ref{CS} are isoscalar and isovector nucleon potentials from the eleven GHF energy functionals \cite{Rios-G}. They vary broadly, especially at suprasaturation densities, thus affecting the high-density behavior of nuclear symmetry energy as well as the nucleon effective masses and their isospin splittings in dense neutron-rich matter. Especially at high momenta and suprasaturation densities, the isoscalar potential is not necessarily quadratic in momentum depending on the interaction used. 

The density dependence of nuclear symmetry energy is shown in the left panel of Fig. \ref{Rios-em}. Most of the predicted symmetry energy functionals are super-soft at supra-saturation densities. It is currently not clear if this is partially responsible for predicting NS maximum masses less than 2 M$_{\odot}$. While at subsaturation densities, they are all consistent with the existing constraints. Obviously, the resulting $E_{\rm{sym}}(\rho)$ does not scale with $\rho^{2/3}$ especially at suprasaturation densities, as already demonstrated in Fig. \ref{GHF-s}. The corresponding neutron-proton effective mass splitting $m^*_{\rm{n-p}}$ at $\rho_0$ is shown as a function of isospin asymmetry $\delta$ in the right panel of Fig. \ref{Rios-em}.
It is seen that with most of the interactions, the $m^*_{\rm{n-p}}$ varies almost linearly with $\delta$, especially below $\delta\leq 0.6$. Moreover, essentially all of them predict positive $m^*_{\rm{n-p}}$ values at $\rho_0$.

\begin{figure*}[htb]
\centering
\resizebox{1.0\textwidth}{!}{
\includegraphics[]{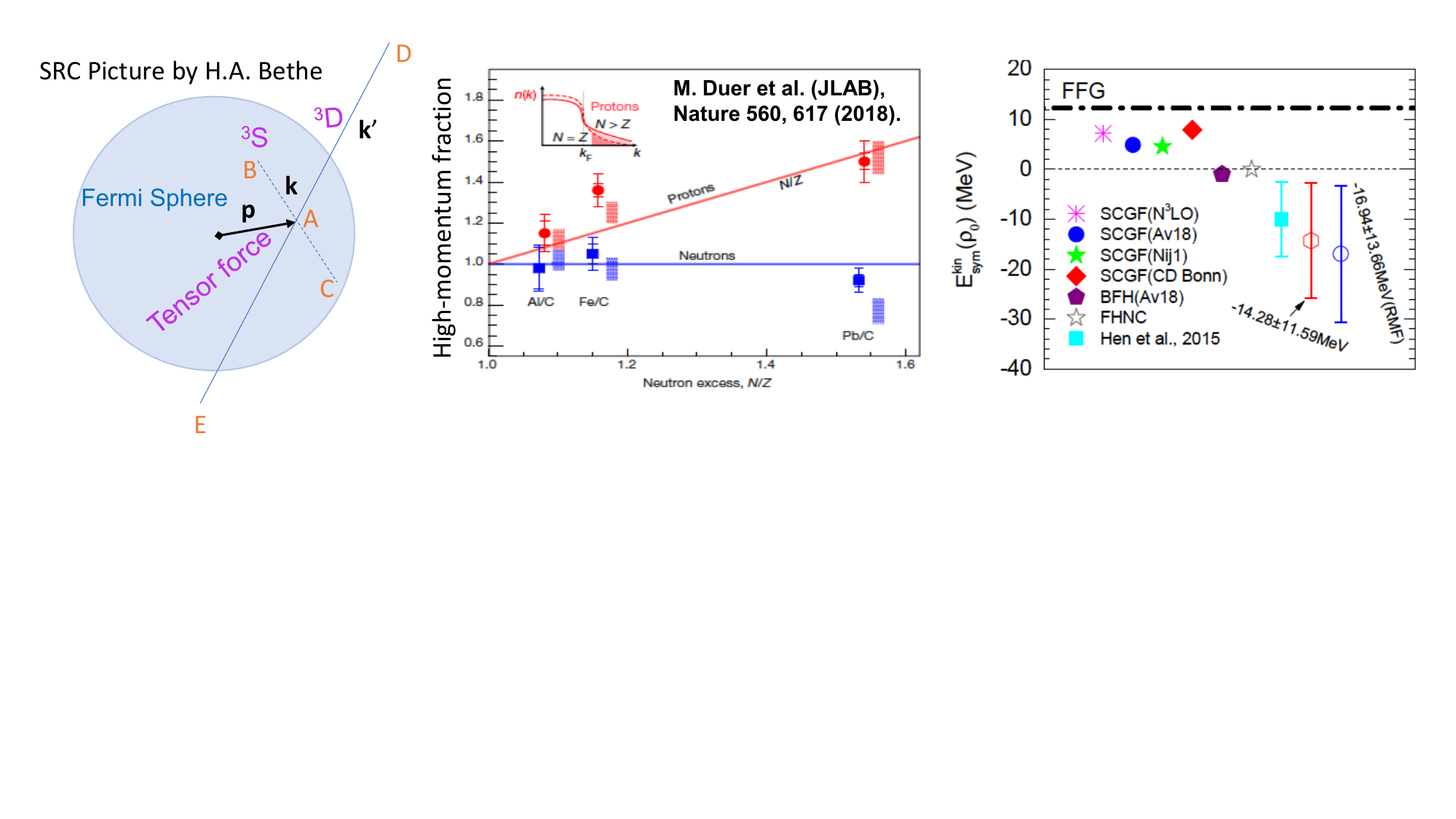}
}
\vspace{-4.5cm}
\caption{Left: A picture of SRC in momentum space by H.A. Bethe \cite{bethe71}: a neutron-proton pair initially in the $^{3}$S state with relative momentum {\bf k} and center of mass momentum {\bf p} in the Fermi sphere will scatter into a $^{3}$D state outsde the Fermi sphere by the tensor operator $\vec{S}_{12}(\Vec{r})=Y^{(2)}(\theta)f(r)$\cite{Otsuka1,Otsuka2}, leading to a HMT in the single-nucleon momentum distribution illustrated in the inset of the middle panel.} Middle: fractions of protons and neutrons in the HMT as functions of the neutron/proton ratio of the reaction system measured at JLAB \cite{Hen20a}. Right: Kinetic symmetry energy at $\rho_0$ from the listed microscopic nuclear many-body theories, including SRC effects, in comparison with the free Fermi (FFG) model prediction of about 12 MeV. The symbols with error bars are extracted from the JLAB data by using the quasi-deuteron dominance model \cite{Hen14}, assuming the SRC is dominated by the spin triplet but isospin singlet neutron-proton pairs \cite{PPNP-Li,Hen15b,Cai15a}.
\label{bethe}
\end{figure*}
\subsection{\bf Example-2: Effects of short-range correlation (SRC) on nuclear symmetry energy} 
Nucleon-nucleon SRCs in nuclei and cold nuclear matter refer to nucleon pairs that have temporally fluctuated into a high-relative-momentum state with approximately zero total center-of-mass momentum (c.m.) and a spatial separation of about 1 fm \cite{Hen14,bethe71,Mig57,Lut60,Tang,Sub08,anto88,Fra88,Ben93,Pan99,Claudio15,Ryc15,Fa17,Arr12,RMP2017}.
As illustrated in the left panel of Fig. \ref{bethe} by H.A. Bethe \cite{bethe71}, the short-range tensor interaction predominantly in the isosinglet and spin-triplet neutron-proton pairs leads to a $^{3}$S-$^{3}$D mixing. As a result, the neutron-proton pair originally at points B and C in the Fermi sea will be scattered to points E and D above the Fermi surface, forming the high momentum tail (HMT) in the single-nucleon momentum distribution (inset of the middle panel). 

Proton-nucleus and electron-nucleus scatterings as well as nucleus-nucleus reaction experiments in inverse kinematics have confirmed the above Bethe picture and revealed many interesting and fundamentally new physics, especially during the last few years, see, e.g., Refs. \cite{Fa17,Arr12,RMP2017} for reviews. In particular, studies on the isospin dependence of SRC have revealed some crucial information about the sources of nuclear symmetry energy. In neutron-rich matter, since it is easier for protons to find neutron partners to form SRC pairs that will be populating the HMT after their interactions via the tensor force, there are relatively higher fractions of protons than neutrons in the HMT as the system becomes more neutron-rich, as shown by the results of JLAB experiments \cite{Hen20a,Hen20b} in the middle panel of Fig. \ref{bethe}. This phenomenon has a direct impact on the kinetic part ($E^{kin}_{\rm sym}(\rho)$) of nuclear symmetry energy. For neutron-rich FFG without SRC, because neutrons have a higher Fermi energy, the $E^{kin}_{\rm sym}(\rho)$ is always positive and has a value of about 12 MeV at $\rho_0$ (indicated by the dash-dotted line in the right panel) as normally given in standard nuclear physics textbooks. However, with the SRC, because of the larger fraction of protons in the HMT in neutron-rich matter, the average kinetic energy of protons is higher than that of neutrons, leading to a $E^{kin}_{\rm sym}(\rho)$ much less than that for the FFG. In fact, as shown by the predictions of both microscopic nuclear many-body theories including self-consistently the SRC effects and phenomenological models, the $E^{kin}_{\rm sym}(\rho)$ is reduced compared to the FFG value and can even become negative depending on the properties of the tensor force used or the size of SRC/HMT effects incorporated, see, e.g., Ref. \cite{PPNP-Li} for a review. 

\begin{figure*}[ht]
\begin{center}
\resizebox{0.9\textwidth}{!}{
\includegraphics{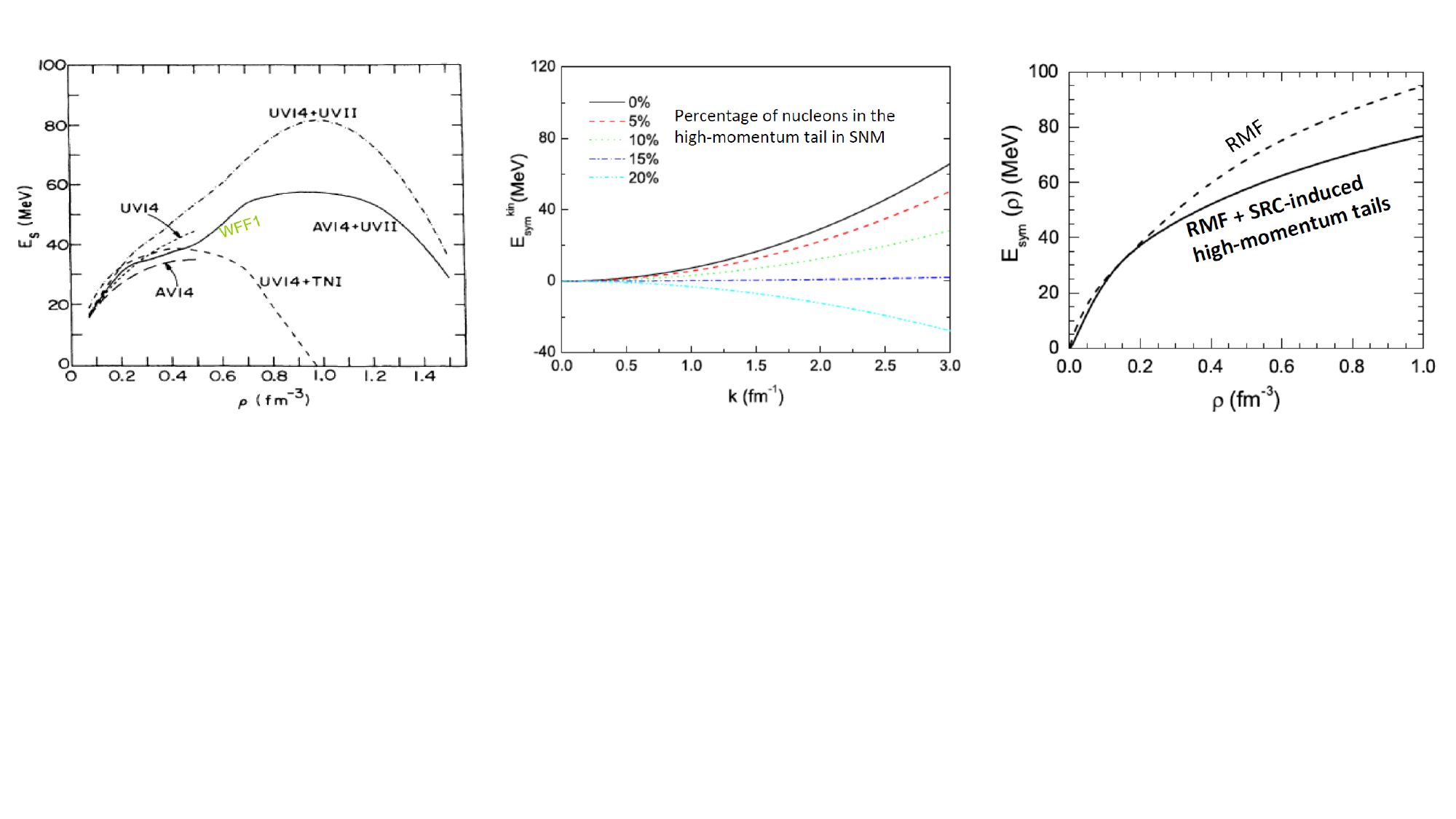}
}
\end{center}
\vspace{-4.8cm}
\caption{{\protect Left: \esym from the Variational Many-Body Theory using different two- and three-body interactions by Wiringa et al. \cite{WFF}.
Middle: The kinetic symmetry energy of correlated nucleons as a function of Fermi momentum with different fractions of high-momentum nucleons in SNM \cite{Xulili}. 
Right: \esym within the traditional RMF (dashed) or RMF incorporating the SRC-modified single-nucleon momentum distribution with a high-momentum tail \cite{Cai-RMF}.}}\label{SRC}
\end{figure*}

The tensor force and the resulting SRC/HMT also affect the potential part of nuclear symmetry energy. Essentially, all physics issues discussed above are related to the density and momentum dependence of the nucleon isovector potential $U_{\rm{sym},1}(\rho,k)$. The underlying physics is the isospin dependence of the nuclear force. For example, the Hartree (direct) term of the nucleon isovector potential 
$U_{\rm{sym},1}(\rho,k)$ at $k_F$ in the interacting Fermi gas model \cite{pre,Xu-tensor} is
\begin{equation}U_{\rm sym,1} (\rm{direct})=
\frac{\rho}{4}\int [V_{T1}\cdot f^{T1}(r_{ij})-V_{T0}\cdot f^{T0}(r_{ij})]d^3r_{ij}
\end{equation}
in terms of the isosinglet (T=0) and isotriplet (T=1) NN interactions $V_{T0}(r_{ij})$ and $V_{T1}(r_{ij})$, as well as the corresponding NN correlation functions $f^{T0}(r_{ij})$ and $f^{T1}(r_{ij})$, respectively. 
While $V_{nn}=V_{pp}=V_{np}$ in the T=1 channel due to the charge independence of NN interactions, the $V_{np}$ interactions and the associated NN correlations in the T=1 and T=0 channels are not the same due to the isospin dependence of strong interactions. It is well known that the tensor force and the resulting NN SRC in the T=0 channel (thus in SNM) is much stronger than that in the T=1 channel (thus in PNM). Consequently, since the symmetry energy is approximately the difference in nucleon specific energy in PNM and SNM, if the short-range repulsive tensor force contribution due to $\rho$ meson exchange at high densities grows faster in SNM than in PNM with increasing density, the $U_{\rm{sym},1}(\rho,k)$ at $k_F$ and the resulting $E_{\rm{sym}}(\rho)$ will decrease as the density increases. 
Therefore, the nucleon isovector potential, the resulting symmetry energy and its slope (Eqs. \ref{FKW} and \ref{Lexp2}), neutron-proton effective mass splitting (Eq. \ref{mnp-src}), and their scattering cross sections are all reflections of the isospin dependence of nuclear forces and correlations. However, several features of the SRC, such as its strength, isospin, and density dependence, as well as the shape of the resulting HMT are still poorly known \cite{Rios}. 

The isospin dependence of SRC is also being probed by using nuclear spectroscopic factors extracted from direct/transfer reactions involving rare isotopes in experiments at RIKEN, FRIB, and GSI. While much progress has been made in these studies, there are still ongoing debates about the isospin dependence of the spectroscopic factors from various analyses of different experiments \cite{Gade,Bob06,Tsang1,Tsang2,Tom}. Moreover, it is well known that the evolution of nuclear shells with neutron/proton ratio and structures of exotic nuclei depend strongly on the nuclear tensor force, see, e.g., Refs. \cite{Otsuka1,Otsuka2}. Normally, a cut-off around the nucleon-nucleon spatial separation of 1.0 fm is used in these studies on nuclear structure. At this distance critical for the SRC effects, the net strength of the tensor force is very model-dependent partially because of the cancellation between the attractive tensor force due to the $\pi$-meson exchange dominating at long distances and the repulsive one due to the $\rho$-meson exchange dominating at short distances.
Moreover, the in-medium $\rho$-meson mass may be significantly different from its free-space value \cite{Brown:1990kj,Brown:1991kk,Rapp:1997ii}. Such modifications of the $\rho$-meson mass have been found to significantly affect the strength of the tensor force, and consequently the high-density behavior of nuclear symmetry energy \cite{Xulili,Xu-tensor,Li2011}. Thus, investigating SRC effects in heavy-ion reactions may provide complementary information about the short-range behavior of the tensor force that is often being cut off in nuclear structure studies.

Two interesting effects of the tensor force on the high-density $E_{sym}(\rho)$ deserve further discussion here. These effects might be probed indirectly using heavy-ion collisions or properties of neutron stars. As mentioned earlier, when the repulsive tensor force due to the $\rho$ meson exchange in the isosinglet n-p channel dominates at high densities, the potential energy in SNM can increase faster than that in PNM where the tensor force is negligible, leading to a super-soft or even negative symmetry energy above certain densities \cite{Pan72,WFF}. This effect can be seen in the left window of Fig.\ \ref{SRC} where some variational many-body theory predictions on effects of the three-body force and/or tensor force at high densities are shown. A summary of more predictions of similar high-density behavior of symmetry energy can be found in Ref. \cite{Kut06}. The decreasing or negative symmetry energy at high densities leads to the interesting possibility of forming proton polarons \cite{Kut93,Kut94} in neutron-rich nucleonic matter, the possible need for alternative gravity theories in massive neutron stars \cite{Wen09,Wlin14,Jiang15,XTHE} or the existence of a weakly interacting light boson mediating a new force \cite{Yong13,Kra16,Feng16}. These possibilities are closely related to the fundamental degeneracy or duality between the strong-field gravity and supradense matter in compact objects \cite{gr1,gr2,gr3}.

Another interesting effect of the tensor force is in reducing the kinetic symmetry energy, as mentioned above. Based on the information extracted from laboratory experiments \cite{Hen14,Sub08,Arr12}, the percentages of nucleons in the high momentum tail are estimated to be about 25\% in SNM and (1-2) \% in PNM. Theoretical calculations predict about (10-25)\% high momentum nucleons for SNM and (1-5)\% for PNM, depending on the model and interaction used \cite{Rios,VMC,mu04,Rio09,Yin13,ZHLi}. This isospin dependence of HMT is expected to modify the kinetic symmetry energy away from the traditional $E_{\rm sym}^{\rm Kin}(\rm{FFG})\approx 12.5(\rho/\rho_0)^{2/3}$ for an uncorrelated FFG. For example, shown in the middle of  Fig.\ref{SRC} is the kinetic symmetry energy as a function of Fermi momentum with different fractions of high-momentum nucleons in SNM, while that in PNM is set to be zero. It is seen that the HMT significantly affects the kinetic symmetry energy, especially at supra-saturation densities. With about 15\% high momentum nucleons in SNM, the $E_{sym}^{kin}(\rho)$ is almost zero in a broad range of Fermi momentum. With more nucleons in the HMT, the kinetic symmetry energy becomes negative at higher densities. Moreover, as shown in the right window of Fig. \ref{SRC}, incorporating the HMT in calculating the kinetic energy and the scalar mass while maintaining the same empirical properties of SNM as well as the $E_{\rm{sym}}(\rho_0)$ and $L$ as in the original RMF, the \esym becomes more concave around $\rho_0$. Consequently, the experimentally measured curvature of the symmetry energy (or the isospin dependence of the incompressibility $K_{\tau}$) can be better reproduced, and the resulting EOS (its SNM part becomes stiffer because of the increased kinetic pressure from HMT nucleons) was found to affect appreciably properties of neutron stars \cite{Cai-RMF}. Interestingly, more recent studies incorporating HMT in RMF or other models have found many interesting SRC effects on properties of neutron stars \cite{Hong24,Gau24,Lia24,Liu24,Kou23,Rod23,Hong23,Pel22,Dutra22,Lou22,Lu22,Hong22,Lou22b,Sou20,ALi19,Liu18}, the cooling of protoneutron stars \cite{Sedrakian:2024uma} as well as observables in nuclear reactions and/or structures \cite{Yong17a,Wang17,Guo21,Bur22,Dala22,Xu25a,Xu25b}.

At this point, it is useful to make a connection between the above two examples, namely the HMT from SRC may help resolve the challenge for GHF models (lacking SRC) to support massive neutron stars above about 2M$_{\odot}$ as we noticed earlier in Section \ref{ghf}. The single-nucleon momentum distribution $n_{\v{k}}(\tau_3,\rho,\delta)$ enters the GHF energy density functional by calculating the kinetic energy and the momentum-dependent part of the potential energy. Without considering the HMT due to SRC, the Fermi step function is used for the $n_{\v{k}}(\tau_3,\rho,\delta)$ for cold nuclear matter. As mentioned above, about (15-20)\% (model dependent) nucleons are in the HMT in SNM, while only about (1-5)\% are in PNM. This relatively small fraction of nucleons in HMT may warrant a perturbative treatment of them. Since the SRC is the strongest in SNM, weighted by the phase space factor $k^2$, the (15-20)\% nucleons in the HMT above the Fermi surface will enhance the kinetic pressure of SNM at high densities without changing its EOS significantly around and below $\rho_0$. The enhanced pressure in SNM at high densities may help support massive neutron stars and generally increases the stiffness of high-density SNM EOS.

In summary of this section, Siemens's $\rho^{2/3}$ scaling is not a universal feature of dense matter; it holds only under restrictive assumptions about the momentum and density dependence of the nuclear mean field. At supra-saturation densities and in very neutron-rich systems, a variety of physical effects --- higher-order momentum dependence, density dependence of isovector potentials, isospin dependence of short-range correlations, relativistic self-energies, new degrees of freedom, and neutron-proton effective mass splitting due to the momentum dependence --- can all lead to significant deviations from Siemens's $\rho^{2/3}$ scaling. Investigations of these deviations will help us better understand the fundamental physics underlying nuclear symmetry energy. Fortunately, heavy-ion collision observables (e.g., transverse momentum dependence of various flow components of protons and light clusters, ratios of charged pions, neutron-proton differential flow, etc) are known to be sensitive to non-quadratic momentum dependence. Optical model analyses of nucleon-nucleus scattering at high energy may also reveal deviations from a simple quadratic potential. Moreover, neutron star properties (mass-radius relation, tidal deformability) may provide astrophysical tests of the high-density symmetry energy and scaling assumptions. In the following two sections, we give a few examples of signals of high-density behavior of nuclear symmetry energy from both astrophysics and nuclear physics.

\begin{figure*}[ht]
\begin{center}
 \resizebox{0.8\textwidth}{!}{
  \includegraphics[]{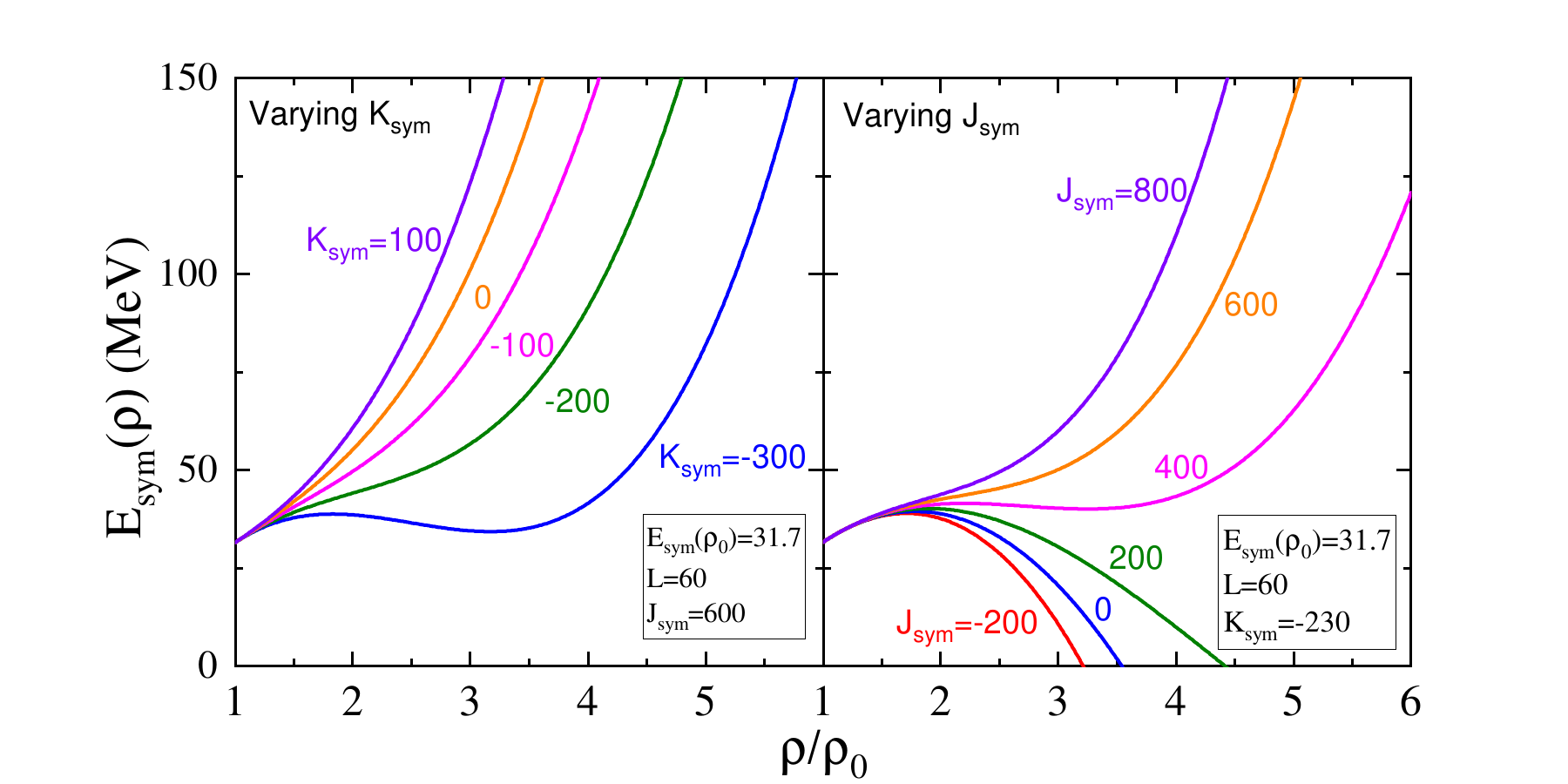}
  }
  \caption{High-density behaviors of nuclear symmetry energy by varying the curvature $K_{\rm{sym}}$ (left) or skewness $J_{\rm{sym}}$ (right) while keeping the magnitude $E_{\rm{sym}}(\rho_0)$ and slope $L$ fixed at their known most probable values indicated. Similar plots can be found in Refs. \cite{Zhang18,Zhang19epj}.}\label{examp1}
\end{center} 
\end{figure*}

\begin{figure*}[ht]
\begin{center}
  \resizebox{0.5\textwidth}{!}{
  \includegraphics[trim={18mm, 0mm, 15mm, 0mm}, clip]{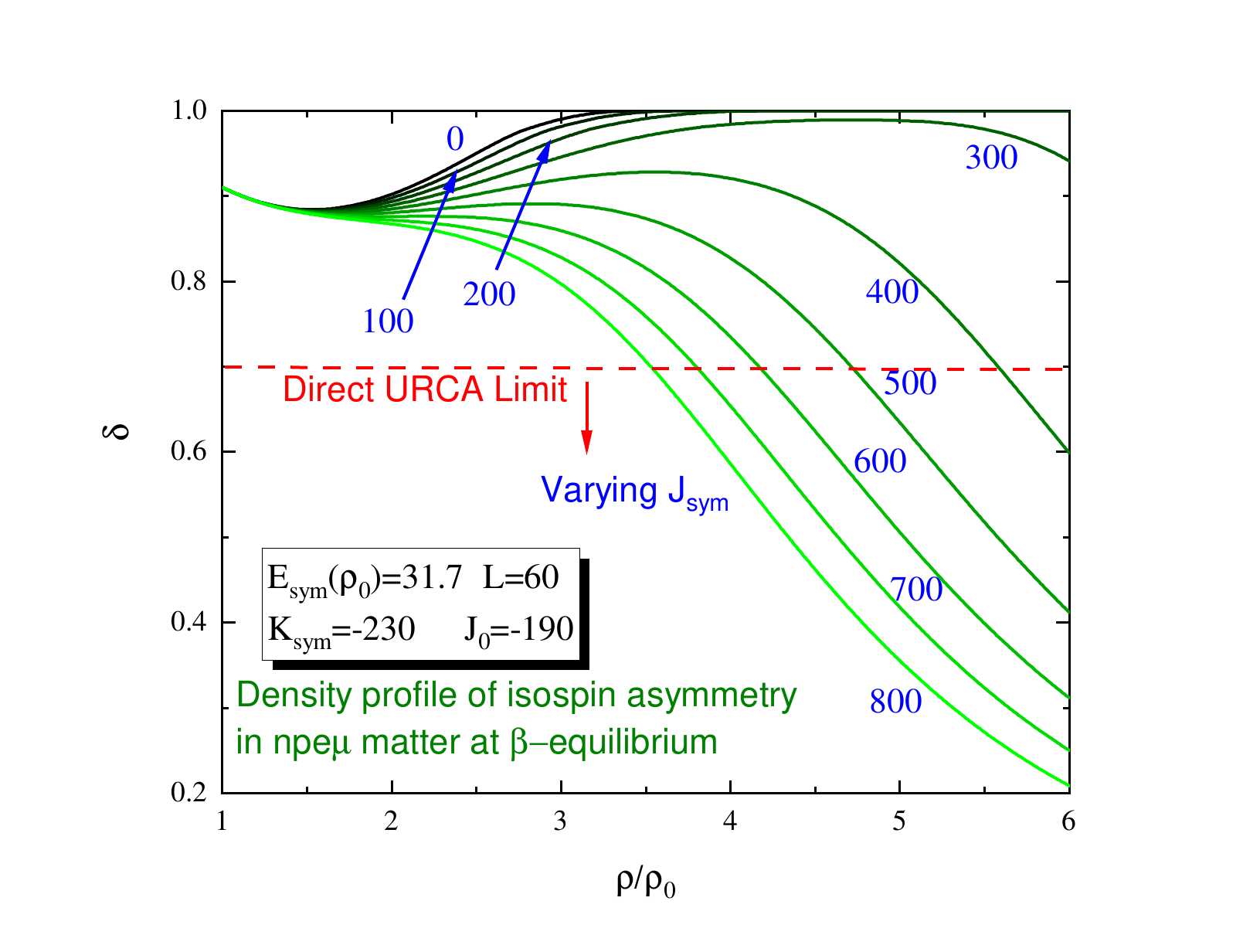}
  }
  \hspace{-0.5cm}
   \resizebox{0.5\textwidth}{!}{
  \includegraphics[trim={18mm, 0mm, 15mm, 0mm}, clip]{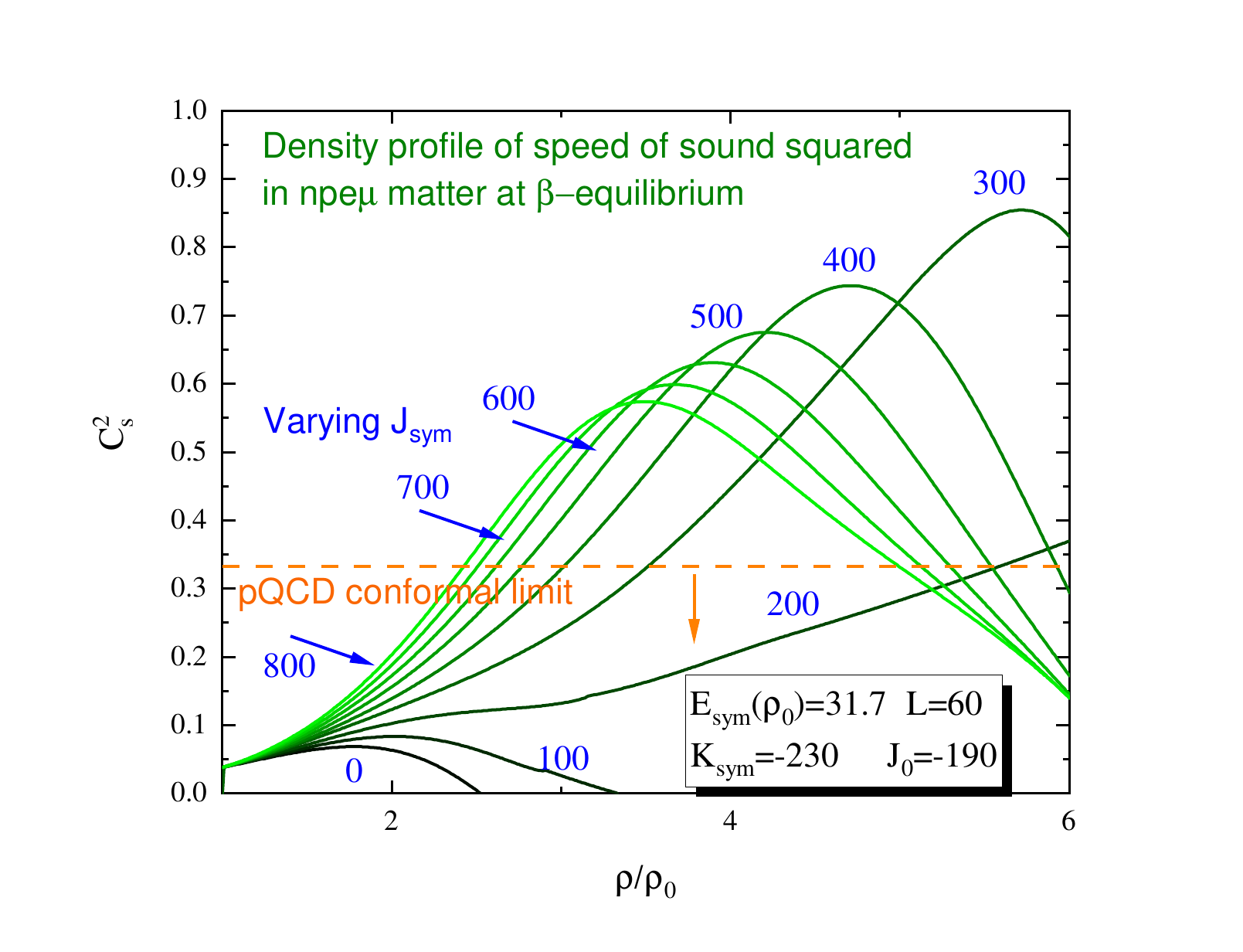}
  }
  \setlength{\abovecaptionskip}{-0.5cm}
  \caption{Density profiles of isospin asymmetry $\delta(\rho)$ (left) and the speed of sound squared $C^2_s\equiv{dP\over d\epsilon}$ (right) in the $npe\mu$ matter at $\beta-$equilibrium by varying $J_{\rm{sym}}$ as in the right window of Fig. \ref{examp1}. The red dashed line at $\delta=0.704$ in the left window is the direct URCA limit in $npe\mu$ matter. The yellow dashed line at $C^2_s=1/3$ in the right window is the conformal limit predicted by pQCD. Modified from a figure in Ref. \cite{Zhang:2022sep}.}\label{examp2}
\end{center} 
\end{figure*}

\section{High-density nuclear symmetry energy from neutron star observations}
Here, we summarize constraints on the high-density behavior of nuclear symmetry energy from analyzing both currently available and expected future high-precision neutron star radius measurements. Solving the neutron star inverse-structure problem, i.e, inferring the EOS of dense matter directly from observations, has been a long-standing goal of astrophysics \cite{Lin92}. Significant progress has been made to achieve this goal by performing both Bayesian statistical and brute-force direct inversions of neutron star observables. 

For completeness and ease of the following discussions, we first recall here the differential equation that has to be solved simultaneously with the Tolman-Oppenheimer-Volkov (TOV) equations ~\cite{Tolman1934,Oppenheimer39}
to investigate the tidal deformability $\lambda$ ~\cite{Flanagan2008,Hinderer2008,Binnington2009,Damour2009,Damour2010,Hinderer2010,Postnikov2010,Baiotti2010,Baiotti2011,Lackey2012,Pannarale2011,Damour2012,Fattoyev2013,Fattoyev2014} together with the mass-radius sequence. The $\lambda$ is related to the tidal Love number $k_2$ and radius $R$ via $\lambda=\frac{2}{3}k_2R^5$ ~\cite{Flanagan2008,Damour2009,Damour2010,Hinderer2010}.
The $k_2$ can be calculated by using the following expression~\cite{Hinderer2008,Postnikov2010}
\begin{eqnarray}\label{k2}
  k_2&=&\frac{1}{20}(\frac{R_s}{R})^5(1-\frac{R_s}{R})^2[2-y_R+(y_R-1]\frac{R_s}{R} \\\nonumber
   &\times&\{(6-3y_R+\frac{3R_s}{2R}(5y_R-8)+\frac{1}{4}(\frac{R_s}{R})^2[26  \\\nonumber
   &-&22y_R+(\frac{R_s}{R})(3y_R-2)+(\frac{R_s}{R})^2(1+y_R)])  \\\nonumber
   &+&3(1-\frac{R_s}{R})^2[2-y_R+(y_R-1)\frac{R_s}{R}]  \\\nonumber
   &\times&{\rm log}(1-\frac{R_s}{R})\}^{-1},
\end{eqnarray}
where $R_s\equiv2M$ is the Schwarzschild radius, and $y_R\equiv y(R)$ can be calculated by solving the following differential equation:
\begin{equation}\label{yr}
  r\frac{dy(r)}{dr}+y(r)^2+y(r)F(r)+r^2Q(r)=0,
\end{equation}
with
\begin{equation}\label{fr}
  F(r)=\frac{r-4\pi r^3(\varepsilon(r)-P(r))}{r-2M(r)},
\end{equation}
\begin{eqnarray}
  Q(r)&=&\frac{4\pi r(5\varepsilon(r)+9P(r)+\frac{\varepsilon(r)+P(r)}{\partial P(r)/\partial\varepsilon(r)}-\frac{6}{4\pi r^2})}{r-2M(r)}\\\nonumber
   &-&4[\frac{M(r)+4\pi r^3P(r)}{r^2(102M(r)/r)}]^2.
\end{eqnarray}
As the above equations are related to $M(r)$ and $r$, they are solved together with the TOV equations~\cite{Tolman1934,Oppenheimer39}
\begin{equation}\label{TOVp}
\frac{dP}{dr}=-\frac{G(M(r)+4\pi r^3P/c^2)(\epsilon+P/c^2)}{r(r-2GM(r)/c^2)},
\end{equation}
\begin{equation}\label{TOVm}
\frac{dM(r)}{dr}=4\pi\epsilon r^2
\end{equation}
by adopting the boundary conditions: $y(0) = 2$, $P(0)=P_c$, and $M(0) = 0$. The NS radius is determined by the condition of vanishing pressure $P(R)=0$ on its surface, and the NS mass is obtained from integrating Eq. (\ref{TOVm}).

\subsection{\bf Basic roles of nuclear symmetry energy in determining NS EOS and its observational properties}\label{Basics}
There are many comprehensive reviews in the literature on issues related to constructing the EOS of charge-neutral NS matter at $\beta$-equilibrium. In the context of this review and for ease of the following discussions, we stress that nuclear symmetry energy plays a significant role in determining the composition, EOS, and observables of NSs as well as the cooling mechanisms of protoneutron stars. For example, in the minimum model of NSs consisting of $npe\mu$ matter at $\beta$-equilibrium, the EOS can be constructed from the energy density
\begin{equation}\label{lepton-density}
  \varepsilon(\rho, \delta)=\rho [E(\rho,\delta)+M_N]+\varepsilon_l(\rho, \delta),
\end{equation}
where $M_N$ represents the average nucleon mass, and $\varepsilon_l(\rho, \delta)$ denotes the lepton energy density \cite{Oppenheimer39}. The particle densities (consequently the density profile of isospin asymmetry $\delta(\rho)$) can be obtained by solving the $\beta$-equilibrium condition $\mu_n-\mu_p=\mu_e=\mu_\mu\approx4\delta E_{\rm{sym}}(\rho)$ where $\mu_i=\partial\varepsilon(\rho,\delta)/\partial\rho_i$ and charge neutrality condition $\rho_p=\rho_e+\rho_\mu$. Before the appearance of muons, one can write the pressure analytically as \cite{Lat}
\begin{equation}\label{pre}
P(\rho,\delta)=\rho^2[\frac{dE_{\rm{SNM}}(\rho)}{d\rho}+\frac{dE_{\rm{sym}}(\rho)}{d\rho}\delta^2]
+\frac{1}{2}\delta(1-\delta)\rho E_{\rm sym}(\rho).
\end{equation}
Moreover, in this case, the density profile of isospin asymmetry $\delta(\rho)$ (or the corresponding proton fraction $x_p(\rho)$)
at $\beta$-equilibrium is determined completely by the \esym via \cite{Lat}
\begin{equation}\label{xp}
x_p(\rho)= 0.048 \left[E_{\rm sym}(\rho)/E_{\rm sym}(\rho_0)\right]^3
(\rho/\rho_0)(1-2x_p(\rho))^3.
\end{equation}
Once the density profile of isospin asymmetry $\delta(\rho)$ or $x_p(\rho)$ is obtained, the pressure becomes barotropic and can be calculated from
\begin{equation}\label{pressure}
  P(\rho)=\rho^2\frac{d\varepsilon(\rho,\delta(\rho))/\rho}{d\rho}.
\end{equation}
Similarly, the energy density $\varepsilon(\rho, \delta(\rho))\rightarrow \varepsilon(\rho)$ becomes barotropic. Finally, the resulting EOS $P(\varepsilon)$ can then be used in solving the relevant differential equations governing NS structures given above.

It is also well known that $x_p(\rho)$ is the most critical quantity determining the cooling mechanisms of protoneutron stars and the related neutrino emissions \cite{LPPH}. In the $npe\mu$ matter, the threshold $x^{DU}_p$ enabling the fast cooling through the direct URCA process (DU) is given by
\begin{equation}\label{xdu}
x^{DU}_p=1/[1+(1+x_e^{1/3})^3]
\end{equation}
with $x_e\equiv \rho_e/\rho_p$ between 1 and 0.5 leading to a $x^{DU}_p$ between 11.1\% to 14.8\%  \cite{Kla06}. 
Moreover, the high-density behavior of $E_{\rm sym}(\rho)$ and the associated $x_p(\rho)$ also determines the onset densities of various new particles, e.g., hyperons \cite{Lee96,Jiang,Ye24}, $\Delta$ resonances \cite{Dra14}, and dark matter particles \cite{NBZ25}.

\begin{figure*}[ht]
\begin{center}
\vspace{-0.4cm}
\resizebox{0.7\textwidth}{!}{
  \includegraphics{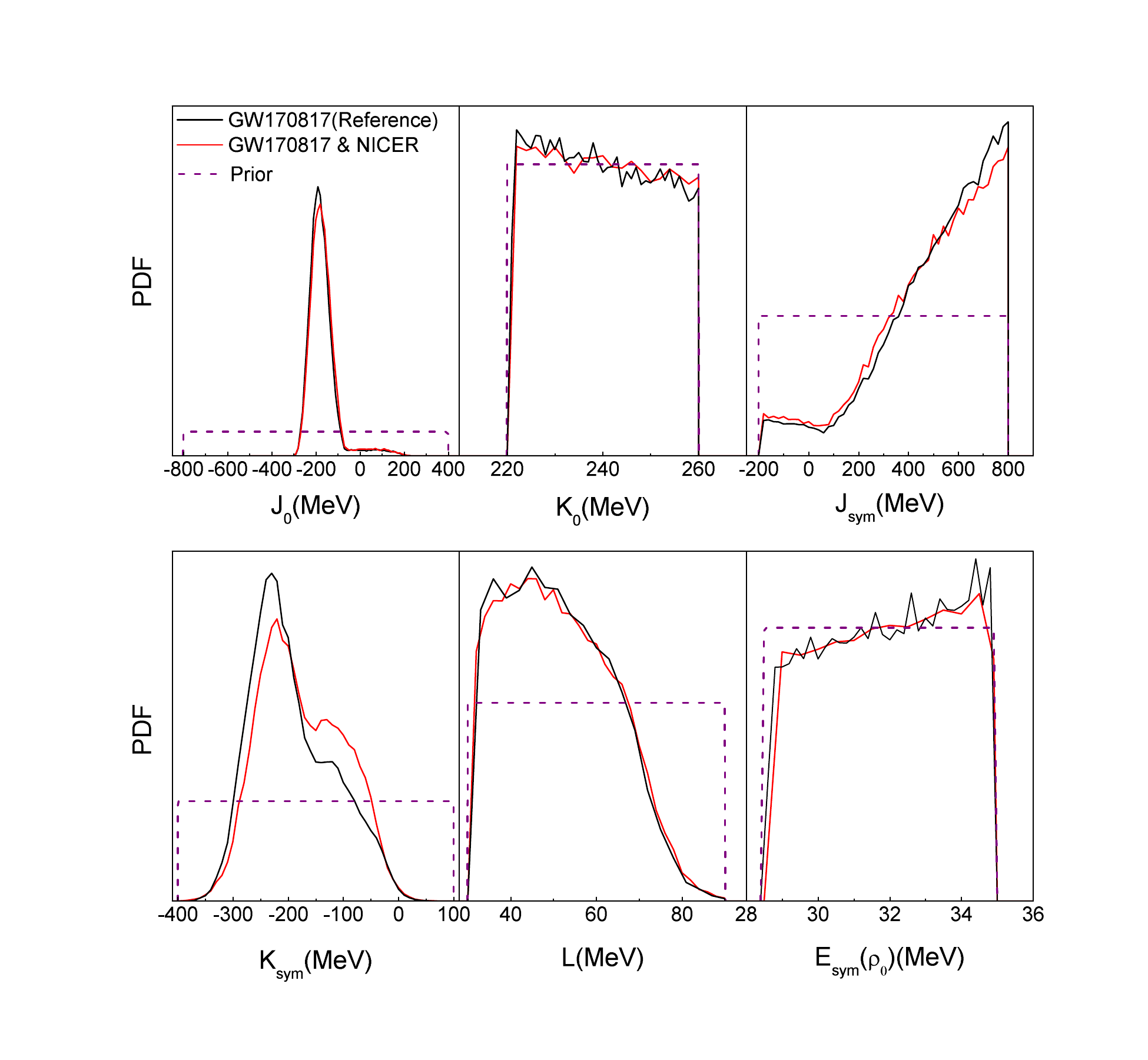}
  }
    \end{center}
    \vspace{-1cm}
    \caption{Posterior probability distribution functions of six EOS parameters from Bayesian analyses of neutron star radius data from GW170817 by LIGO/VIRGO \cite{LIGO18} and the NICER data for PSR J0030+0451 \cite{Riley19,Miller19}. Taken from Ref. \cite{Xie20}. }\label{XiePDF}
\end{figure*}

As an example, shown in Fig. \ref{examp1} are typical high-density behaviors of nuclear symmetry energy using the presently known most probable values of $E_{\rm{sym}}(\rho_0)$ and $L$ by varying the curvature $K_{\rm{sym}}$ (left) or skewness $J_{\rm{sym}}$ (right). Such density dependence of nuclear symmetry energy $E_{\rm{sym}}(\rho)$ covers the diverse model predictions at high densities while it is consistent with existing constraints around $\rho_0$. Shown in Fig. \ref{examp2} are examples of the resulting density profiles of isospin asymmetry $\delta(\rho)$ (left) and the speed of sound squared $C^2_s\equiv{dP\over d\epsilon}$ (right) in the $npe\mu$ matter at $\beta-$equilibrium with $K_{\rm{sym}}=-230$ MeV and varying $J_{\rm{sym}}$ within its prior range. As previously discussed in detail in Refs. \cite{Zhang:2022sep,Li24-PRD}, mainly because of the isospin fractionation \cite{LCK08} induced by the $E_{\rm{sym}}(\rho)\cdot \delta^2$ term in the EOS of neutron-rich matter and the approximate isospin-equilibrium condition, when the $E_{\rm{sym}}(\rho)$ is very stiff with either very large $K_{\rm{sym}}$ and/or $J_{\rm{sym}}$ the resulting $\delta$ approaches zero (symmetric nuclear matter). On the other hand, when the $E_{\rm{sym}}(\rho)$ is very soft with either very negative $K_{\rm{sym}}$ and/or $J_{\rm{sym}}$ the $\delta$ approaches 1.0 (pure neutron matter). Moreover, considering two connected regions with local densities $\rho_1$ and $\rho_2$, respectively, the isospin-equilibrium requires that \cite{shi} $E_{\rm sym}(\rho_1)\delta(\rho_1)=E_{\rm sym}(\rho_2)\delta(\rho_2)$. Thus, for a given \esym function, the local isospin asymmetries will adjust themselves according to the relative symmetry energy in those regions. Therefore, self-consistent studies considering the above physics conditions lead to physical conclusions very different from expectations based on the presumption that $\delta\approx 1$ everywhere in NSs (which is equivalent to assuming a zero symmetry energy at all densities according to Eq. (\ref{xp}) obtained from the $\beta$-equilibrium and charge neutrality condition \cite{Lat}). 

In many published studies adopting directly either parametric or non-parametric $P(\epsilon)$ relations, the physics related to $E_{\rm{sym}}(\rho)$ discussed above does not appear anywhere explicitly. This is mainly because the TOV equations are intrinsically composition-degenerate, namely, only the EOS $P(\epsilon)$ is needed as input to obtain the mass-radius sequence and/or tidal deformabilities, regardless of how the $P(\epsilon)$ relations are constructed by considering what particles or physics ingredients. How to make use of the $P(\epsilon)$ relations from these studies to learn anything about $E_{\rm{sym}}(\rho)$ has then 
become an interesting question. Unlike in some studies where NSs are assumed to be PNM, based on the formalism given above for $npe$ matter at $\beta$-equilibrium, an iterative inversion procedure (solving the inverse $\beta$-equilibrium problem) to extract the profile of isospin asymmetry $\delta(\rho)$ and the symmetry energy $E_{\rm{sym}}(\rho)$ from the combined quasi-data of $P(\epsilon)$ from NSs and heavy-ion collisions was first proposed in Ref. \cite{LiXie20}. This procedure was very recently extended in Ref. \cite{Dima} to include muons within a Bayesian approach. The approach was shown to have a remarkable accuracy, and the inferred $E_{\rm{sym}}(\rho)$ was found to strongly support the operation of the direct URCA neutrino emission process in NSs \cite{Dima}.

Interestingly, depending on the high-density behavior of $E_{\rm{sym}}(\rho)$, the resulting speed of sound squared $C^2_s(\rho)$ in $npe\mu$ matter at $\beta-$ equilibrium can have a peaked density profile. Such behavior was also found independently in Ref. \cite{Ye24} recently, and it was shown there that the appearance of hyperons sharpens the peak. Compared to the many interesting physics mechanisms for the peaked density profile of $C^2_s(\rho)$ proposed in the literature, the one illustrated in Fig. \ref{examp2} (discussed in detail in Refs. \cite{Zhang:2022sep,Ye24}) is probably the simplest one without introducing any new physics beyond the uncertain high-density behavior of nuclear symmetry energy and its impact on hyperon formation in NSs.

\begin{figure}[ht]
\hspace{-0.5cm}
  \resizebox{0.7\textwidth}{!}{
   \includegraphics{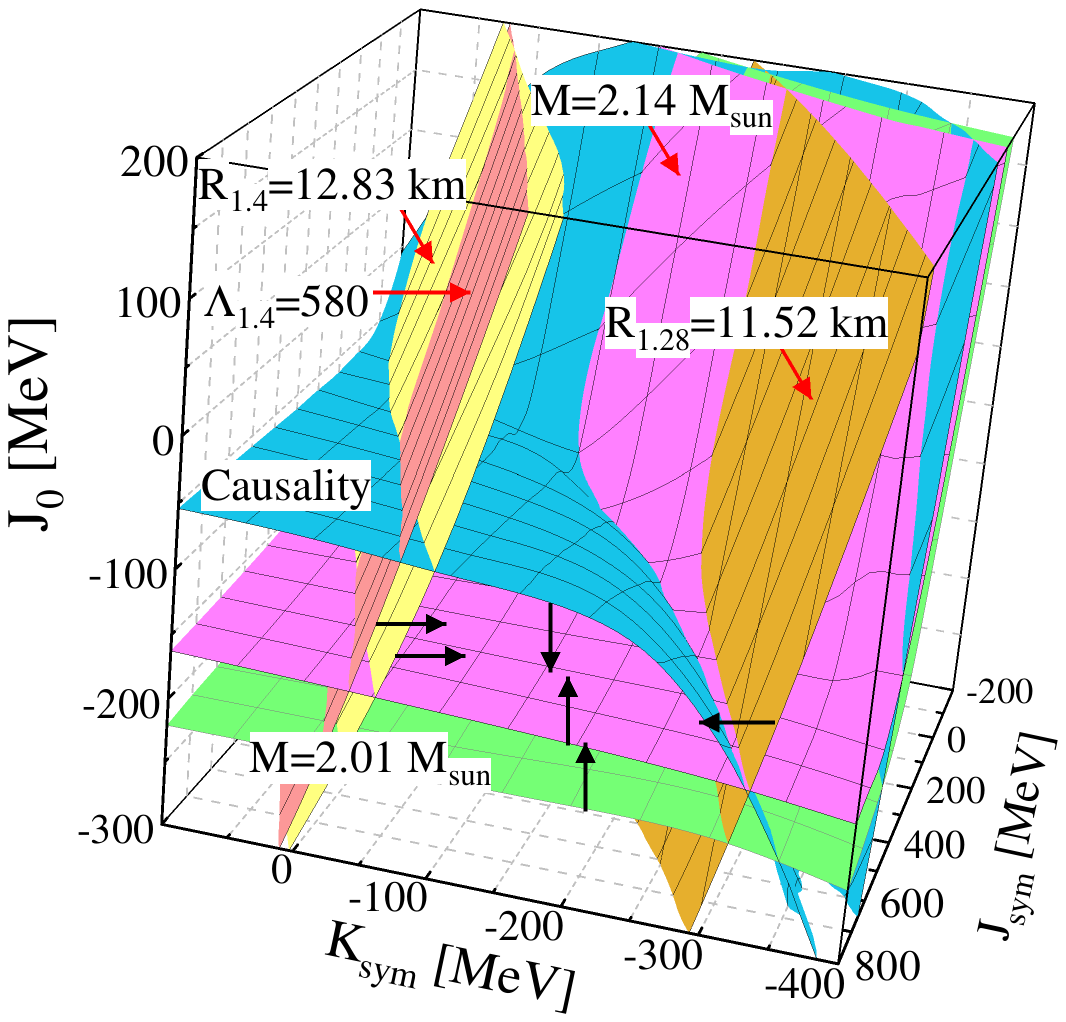}
   }
  \vspace{-1.cm}
  \caption{Constant surfaces of neutron star observables with values indicated by the red arrows and the causality condition in the 3D $K_{\rm sym}-J_{\rm sym}-J_0$ EOS parameter space: the NS maximum mass of M=2.14$M_{\odot}$ (green surface) or 2.01$M_\odot$ (pink surface), the radius of canonical NS $R_{1.4}=12.83$ km (yellow surface) or $R_{1.28}=11.52$ km (orange surface) for a NS of mass 1.28$M_{\odot}$, the dimensionless tidal deformability of canonical NS $\Lambda_{1.4}=580$ (red surface), and the causality surface (blue) on which the sound speed equals the speed of light in centers of most massive NSs supported at the point of the EOS parameter space. The red arrows point to the constant surfaces on which the specified observables have the same values, while the black arrows indicate the directions satisfying the specified observational constraints. Taken from Ref. \cite{Zhang20apj}. }
  \label{mass217}
\end{figure}

\begin{figure}[ht]
\hspace{-1.5cm}
   \resizebox{0.7\textwidth}{!}{
   \includegraphics{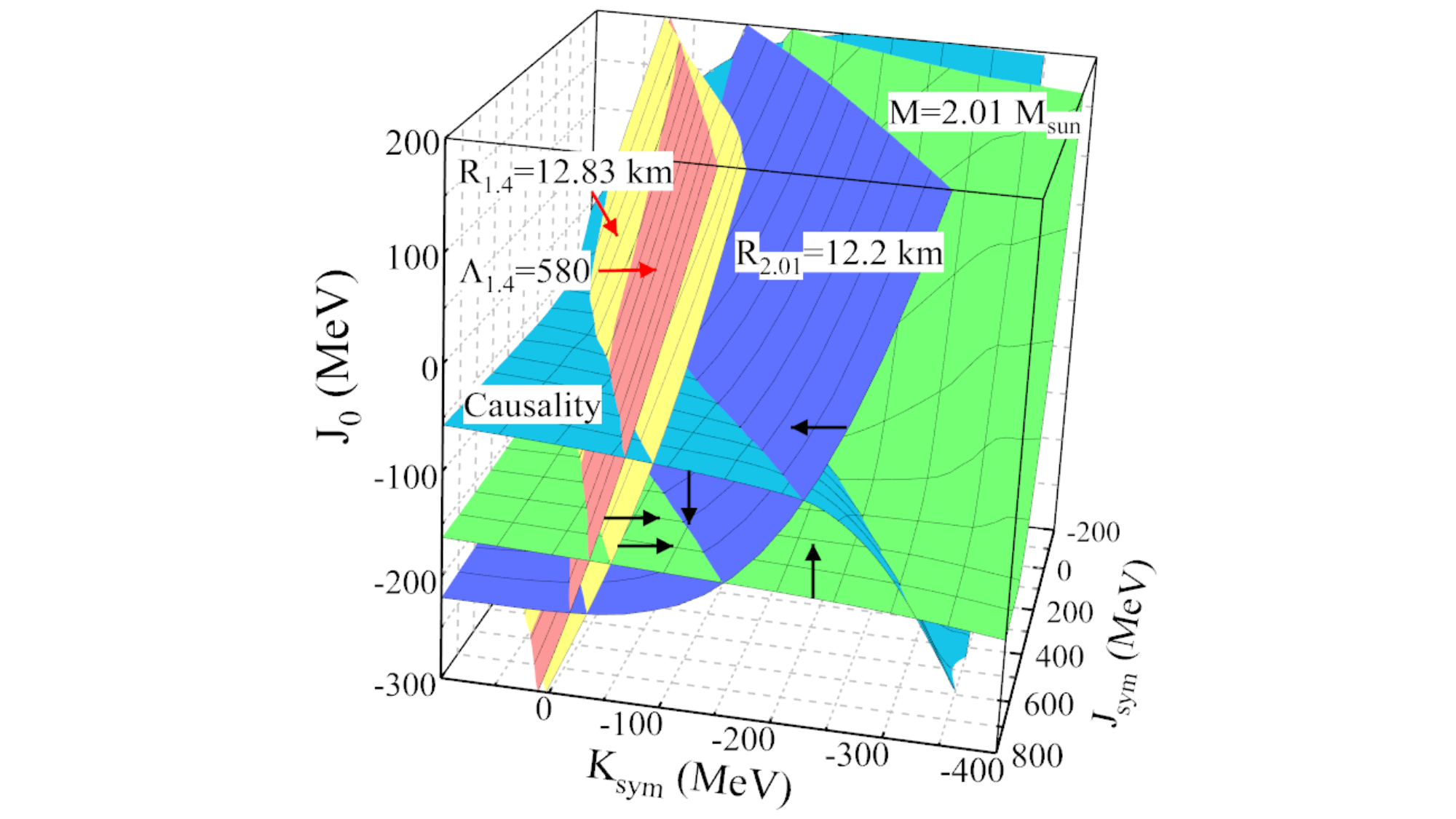}
   }
  \vspace{0.3cm}
  \caption{
 Same as in  Fig. \ref{mass217} but with the $68\%$ confidence lower-boundary $R_{2.01}=12.2$ km for the radius of PSR J0740+6620 with a mass of $2.08\pm 0.07$~$M_{\odot}$. Taken from Ref. \cite{LCXZ2021,Zhang21apj}. }\label{NICER2}
\end{figure}

\begin{figure*}
\centering
\includegraphics[width=.4\linewidth,angle=0]{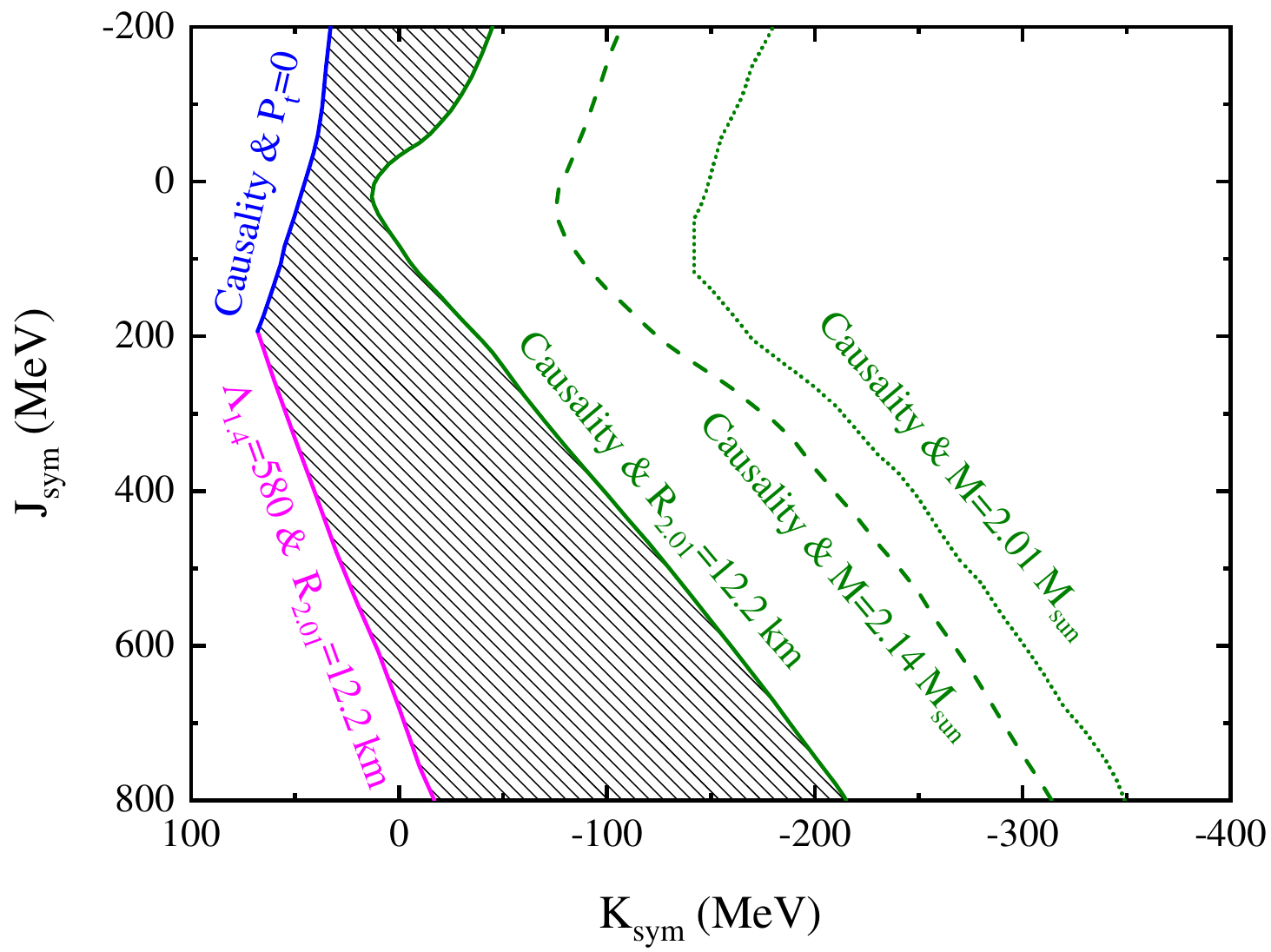}
\includegraphics[width=.4\linewidth,angle=0]{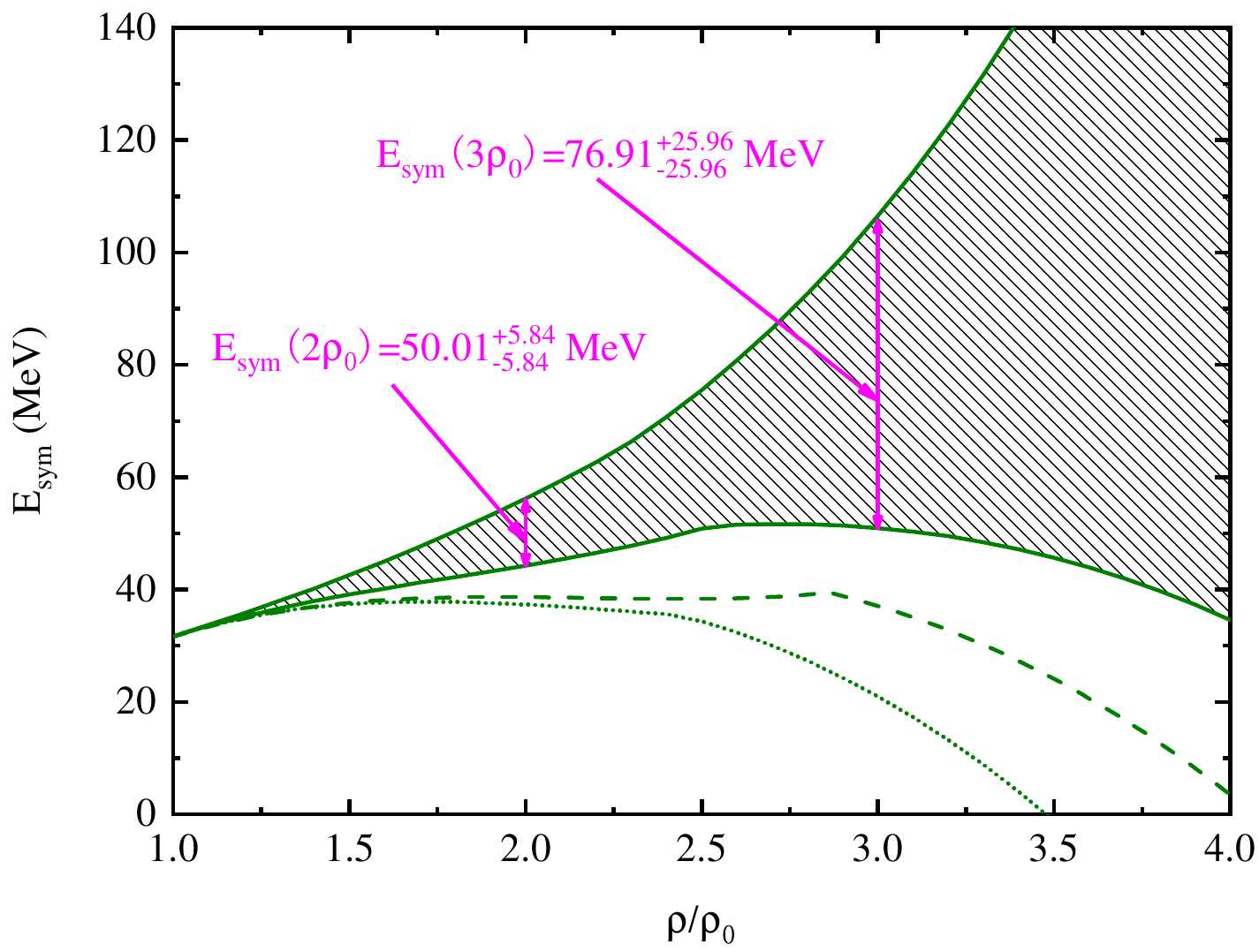}
\caption{({Left}) Projections of the indicated crosslines of constant surfaces of NS observables to the $K_{\rm sym}-J_{\rm sym}$ plane for $L=58.7$ MeV. ({Right}) The constraints on symmetry energy extracted from the combinations of $K_{\rm sym}-J_{\rm sym}$ in the left plot.~$E_{\rm sym}(2\rho_0)$ and $E_{\rm sym}(3\rho_0)$ are also extracted and~labeled. Taken from Ref. \cite{LCXZ2021,Zhang21apj}. }\label{KsymJsymL60}
\end{figure*}

\subsection{\bf Example-1: Bayesian statistical inference of high-density symmetry energy from neutron star observables}
In Bayesian analyses one seeks the posterior probability distribution functions (PDFs) of model parameters through the Markov-Chain Monte Carlo (MCMC) sampling, given a set of data and assumed prior PDFs for the model parameters. For example, shown in Figures \ref{XiePDF} are the marginalized posterior PDFs of the six EOS parameters using $R_{1.4}=11.9^{+1.4}_{-1.4}$ (GW170817) \cite{LIGO18} and the NICER data for PSR J0030+0451 \cite{Riley19,Miller19}, respectively. Results of the two independent analyses of the NICER data: (1) $M=1.34^{+0.16}_{-0.15}$ M$_{\odot}$ and $R=12.71^{+1.19}_{-1.14}$ km \cite{Riley19} and (2) $M=1.44^{+0.15}_{-0.14}$ M$_{\odot}$ and $R=13.02^{+1.24}_{-1.06}$ km \cite{Miller19}, are used as independent data sets. Besides the three parameters characterizing the $E_{\rm{sym}}(\rho)$, the $K_0$ and $J_0$ are the incompressibility and skewness in parameterizing the energy per nucleon $E_{0}(\rho)$ in SNM as 
\begin{equation}\label{E0para}
E_{0}(\rho)=E_0(\rho_0)+\frac{K_0}{2}(\frac{\rho-\rho_0}{3\rho_0})^2+\frac{J_0}{6}(\frac{\rho-\rho_0}{3\rho_0})^3.
\end{equation}
Compared to the uniform prior PDFs used, except the $J_{\mathrm{sym}}$, all other EOS parameters are significantly constrained by the radius data. It was also found in Ref. \cite{Xie19} that the saturation parameters $K_0$ and $E_{\mathrm{sym}}(\rho_0)$ are essentially not affected by the radius data compared to their prior PDFs. More quantitatively, the most probably values and 68\% confidence boundaries of $J_0$, $J_{\mathrm{sym}}$, $K_{\mathrm{sym}}$ and L were found to be $J_0=-190_{-40}^{+40}$, $J_{\mathrm{sym}}=800_{-360}^{+0}$, $K_{\mathrm{sym}}=-230_{-50}^{+90}$ and $L=39_{-0}^{+19}$, respectively. 
It is important to emphasize that the radius data of canonical neutron stars do not constrain the skewness $J_{\mathrm{sym}}$ of symmetry energy. It was shown explicitly in Ref. \cite{Xie20} that the peak of PDF($J_{\mathrm{sym}})$ at the upper boundary continues to shift with its previous range.
Using these parameters, the constraining bands on $E_0(\rho)$ and $E_{\rm{sym}}(\rho)$ can easily be reconstructed \cite{Xie19,Xie20}. We notice that many other Bayesian analyses of neutron star observations since GW170817 have extracted very similar results, see, e.g., Ref. \cite{LCXZ2021} for a review.

\subsection{\bf Example-2: Direct inference of high-density symmetry energy from inverting neutron star observables by brute force}
The likelihood function in Bayesian analyses discussed above is normally a product of several factors. Besides normally using a Gaussian function to compare the model predictions with observational data, the likelihood function also involves several filters to enforce some general physical requirements including (1) the causality condition is upheld at all densities, and (2) the generated NS EOS must be sufficiently stiff to support NSs at least as massive as the mass of the most massive NS found most recently, e.g., (1.97, 2.01, 2.14, etc) M$_{\odot}$ that has been changing in recent years. As a statistical approach in nature, the posterior PDFs from Bayesian analyses mix up the effects statistically of all physics ingredients considered. While it is possible to examine the statistical impact of some factors in Bayesian analyses, it is sometimes necessary to visualize the direct impact of individual observables on specific features of NS EOSs. For this purpose, in a series of analyses of several issues concerning neutron stars, see, e.g., Refs. \cite{Zhang18,Zhang19epj,Zhang19apj,Zhang19jpg,Zhang20apj,Zhang21apj}, the neutron star inverse-structure problem was solved by brute force in the 3-dimensional high-density EOS parameter space by fixing the low-density EOS parameters at their currently known most probable values. More specifically, usually constant values of $E_0(\rho_0)$, $K_0$, $E_{\rm sym}(\rho_0)$ and $L$ are taken within the ranges of
$E_0(\rho_0)=-15.9 \pm 0.4$ MeV, $K_0=230 \pm 20$ MeV, $E_{\rm sym}(\rho_0)=31.7\pm 3.2$ MeV and $L=58.7\pm 28.1 $, while the three high-order parameters $J_0$, $K_{\rm{sym}}$ and $J_{\rm{sym}}$ are taken as variables in the range of $-400 \leq K_{\rm{sym}} \leq 100$ MeV, $-200 \leq J_{\rm{sym}}\leq 800$ MeV, and $-800 \leq J_{0}\leq 400$ MeV, respectively. The neutron star inverse-structure problem is then solved by brute force in the 3D  $J_0-K_{\rm{sym}}-J_{\rm{sym}}$ space. 

As an example, Figure \ref{mass217} illustrates how combinations of neutron star observable and causality conditions may help constrain the $K_{\rm sym}-J_{\rm sym}-J_0$ high-density EOS parameter space within the $npe\mu$ model for neutron stars \cite{Zhang20apj}. In these calculations, $E_{\rm sym}(\rho_0)=31.7$ MeV and $L$ is set at $L=58.7$ MeV. The examples shown are the NS maximum mass of M=2.14 $M_{\odot}$ (pink surface) \cite{Cro19Mmax} or 2.01 $M_\odot$ (green surface) \cite{Antoniadis13}, the upper limit of canonical NS radius $R_{1.4}=12.83$ km at 68\% confidence level from earlier X-ray observation (yellow surface) \cite{Lattimer2014} or the lower limit $R_{1.28}=11.52$ km (orange surface) for a NS of mass 1.28 $M_{\odot}$ \cite{Riley19} for PSR J0030+0451 from NICER, the dimensionless tidal deformability $\Lambda_{1.4}=580$ (red surface) of canonical NSs involved in GW170817 \cite{LIGO18}, and the causality surface (blue) on which the speed of sound equals the speed of light at the central density of the most massive NS supported by the nuclear pressure at each point with the specific EOS there \cite{Zhang19epj}. Detailed analyses in Ref. \cite{Zhang20apj} found that the upper limit of the radius of canonical NSs from NICER observation of PSR J0030+0451 is consistent with the $R_{1.4}\leq 12.83$ km and $\Lambda_{1.4}\leq 580$ boundaries shown in Figure \ref{mass217}. 

Every point on each constant surface of a given observable represents a unique EOS. An approximately vertical surface in the $J_0$ direction, e.g., the tidal deformation or radius of canonical NSs, means that the considered observable does not depend on the $J_0$ parameter. On the other hand, an approximately flat surface, e.g., the maximum mass of M=2.14 $M_{\odot}$ in the lower-left corner, means that the high-density symmetry energy parameters $K_{\rm{sym}}$ and $J_{\rm{sym}}$ in that region have little influence on the NS maximum mass.

The lower limit $R_{1.28}\geq 11.52$ km provides a tighter constraint on the EOS parameter space than the lower limit of $\Lambda_{1.4}$ or $R_{1.4}$ previously reported. It thus sets a new constraint on the right-back corner of the high-density EOS parameter space in Fig.\ \ref{mass217}. However, the $R_{1.28}\geq11.52$ km constant surface is still outside the crossline between the constant surface of NS maximum mass of M=2.14 $M_{\odot}$ and the causality surface. Therefore, the high-density EOS parameter space surrounded by the pink (lower limit of NS maximum mass any EOS has to support), yellow (upper limit of canonical NS radius), and blue surfaces (upper limit of sound speed) shown in Figure \ref{mass217} satisfy all existing NS observational constraints and the general physics requirements. 

The cross lines of the constant surfaces of observables determine the boundaries of the 3D EOS parameter space. In particular, the crossline between the causality surface and the maximum mass of M=2.14 $M_{\odot}$ sets a boundary from the lower-right side, constraining the lower limit of the skewness parameter $J_0$ of SNM. On the other hand, the crossline between the causality surface and the upper limit of canonical NS radius $R_{1.4}\leq12.83$ km sets an upper-left boundary of the high-density EOS parameter space, limiting the maximum value of $J_0$. The knowledge about both the mass and radius of PSR J0740+6620 had a significant effect on refining the lower boundary of the EOS parameter space from the lower-right side, as shown in Fig. \ref{NICER2}, constraining the lower boundary of high-density symmetry energy. This can be seen more clearly by projecting the relevant crosslines to the 
$K_{\rm sym}-J_{\rm sym}$ plane as we shall discuss next.

\begin{figure*}[ht]
\centering
\resizebox{0.9\textwidth}{!}{
  \includegraphics{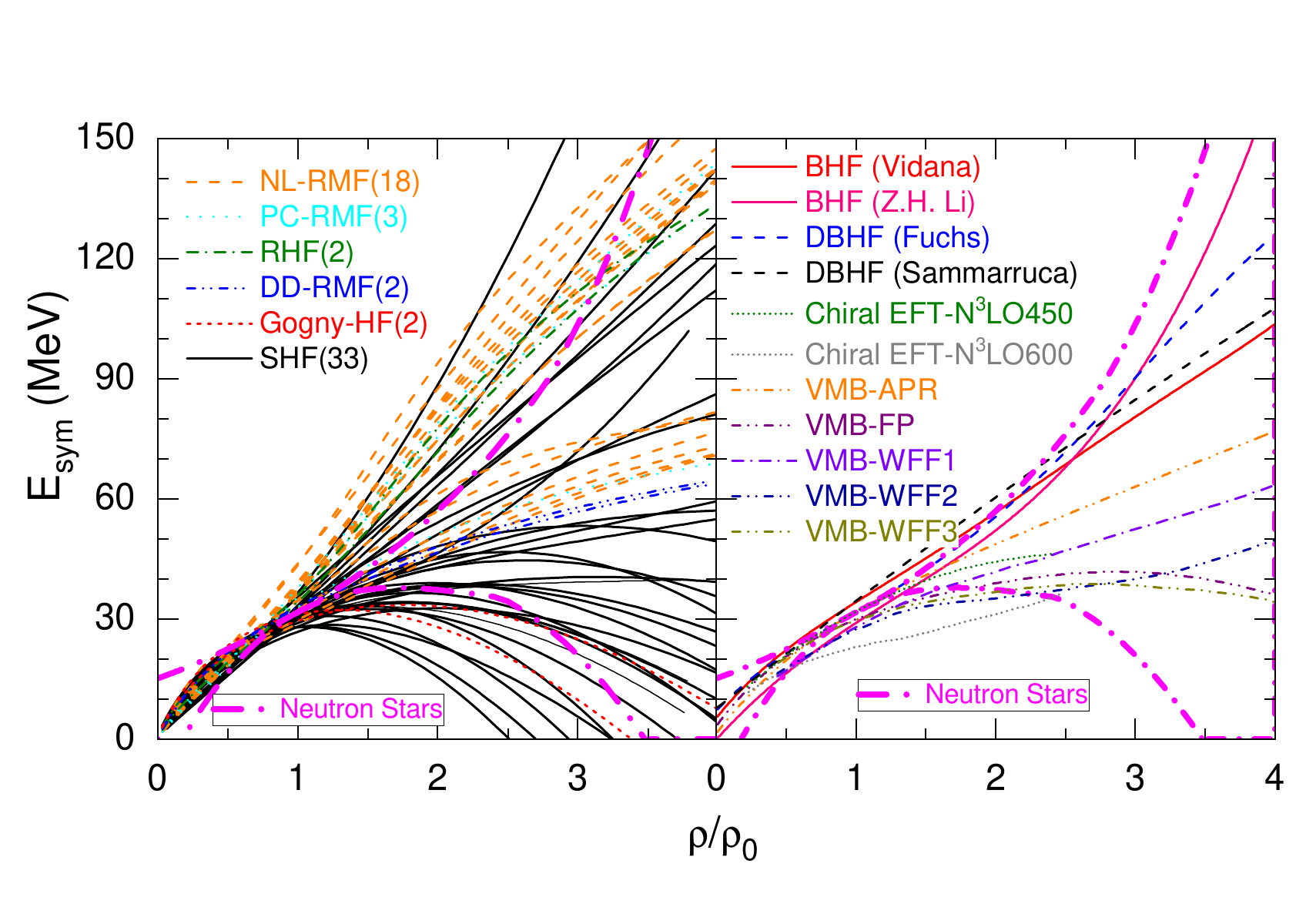}
  }
    \vspace{-0.4cm}
    \caption{Left: 60 examples of predicted $E_{\rm{sym}}(\rho )$ using 6 classes of nuclear energy density functionals. Right: predictions using 11 microscopic and/or {\it ab initio} nuclear many-body theories \cite{ChenLW14,ChenLW15}. In both panels, they are compared with the upper and lower boundaries of symmetry energy extracted from analyzing neutron star observations (dot-dashed magenta curves) \cite{Zhang19epj,Zhang19apj}. The lower boundary is the most conservative one shown in Fig. \ref{KsymJsymL60}.}\label{Esym-all}
\end{figure*}

Shown in the left window of Figure~\ref{KsymJsymL60} are projections of the indicated crosslines of constant surfaces of NS observables to the $K_{\rm sym}-J_{\rm sym}$ plane with $E_{\rm sym}(\rho_0)=31.7$ MeV and $L=58.7$ MeV. The right window shows the resulting constraints on the symmetry energy. For~quantitative comparisons with the systematics discussed earlier, the~new bounds on $E_{\rm sym}(2\rho_0)$ and $E_{\rm sym}(3\rho_0)$ were also extracted and labeled. Clearly, the radius measurement for PSR J0740+6620 had a significant impact on refining the constraint on the lower boundary of high-density symmetry energy. It is seen that the latter is sensitive to the lower limit of NS mass (i.e., the currently observed maximum mass of NSs). This feature is consistent with the findings within the extended Skyrme-Hartree-Fock (eSHF) \cite{YZhou19a} and Bayesian analyses \cite{Xie19,Xie20}. Most interestingly, however, is that both the lower and upper boundary below about $2\rho_0$ are largely independent of the maximum mass used. In particular, the lower boundary is set by the crossline of causality and $R_{2.01}=12.2$ km surfaces.
This feature was used in Ref. \cite{Zhang19epj,Zhang19apj} to constrain the symmetry energy at $2\rho_0$ to $E_{\rm{sym}}(2\rho_0)=46.9\pm10.1$ MeV as shown in Fig. \ref{Esym-survey}. 

Regarding the influence of different NS observables on constraining various high-density EOS parameters, one may naturally ask why the currently available NS observables appear to be significantly more sensitive to $J_0$ than to $J_{\rm sym}$, given that both parameters enter the density expansion at the same order, have comparable magnitudes, and NS matter is often considered to be highly isospin asymmetric. This is a reasonable question and merits a careful explanation.

First, as discussed in Section~\ref{Basics}, the widely used assumption that ``the isospin asymmetry in neutron stars is close to unity'' can substantially reduce the role played by the nuclear symmetry energy in determining NS properties. Indeed, many studies approximate NSs as PNM when addressing certain problems, since the TOV equations require only the EOS $P(\epsilon)$ as input and do not explicitly depend on the matter composition. However, when the purpose is to investigate the effects of the nuclear symmetry energy through NS observables, this assumption becomes a strong one. In particular, under $\beta$-equilibrium and charge-neutrality conditions, Eq.~(\ref{xp}) indicates that such an approximation would effectively suppress the contribution of the symmetry energy throughout the star. Quantitatively, Eq.~(\ref{xp}) requires $E_{\rm sym}(\rho)=0$ for PNM.

Second, as shown in Figs.~\ref{mass217} and \ref{NICER2} and discussed above, the parameter $J_0$ is constrained from above by the causality condition and from below by the requirement that the EOS must support at least the presently observed maximum NS mass. By contrast, the constraints on the symmetry-energy sector arise mainly from constant surfaces of the tidal polarizability measured in GW170817 and the radius of PSR~J0740+6620. These constraint surfaces are largely vertical, suggesting that they depend only weakly on $J_0$.

Third, as illustrated in Fig.~\ref{KsymJsymL60}, the projections of the relevant crosslines in Figs.~\ref{mass217} and \ref{NICER2} onto the high-density symmetry-energy plane do not close in the $J_{\rm sym}$ direction. This behavior indicates that the NS observables considered here provide limited sensitivity to this parameter. This observation is consistent with the results of Bayesian analyses shown in Fig.~\ref{XiePDF}, although statistical approaches of this type do not directly reveal the physical origin of such parameter dependencies clearly.

Finally, to the best of the present author’s knowledge, existing studies based on the latest NS observational data have not achieved substantially tighter constraints on $J_{\rm sym}$ than those presented here. As will be discussed further in Section~\ref{future}, NS radii—particularly the canonical radius $R_{1.4}$—are primarily determined by the pressure around $\sim 2\rho_0$ \cite{Lat}, whereas $J_{\rm sym}$ mainly influences the pressure of NS matter at higher densities. As a result, NS radius measurements alone, even when highly precise, are not expected to tightly constrain $J_{\rm sym}$. This conclusion is supported quantitatively by Fig.~5 of Ref.~\cite{Farr14}, which shows that $R_{1.4}$ correlates most strongly with $L(\rho \approx 1.5\rho_0)$, while $R_{1.8}$ correlates with $L(\rho \approx 2$--$3\rho_0)$ within SHF energy density functionals. Moreover, it was found there that the third term in the expansion of $L(\rho)$ in Eq.~(\ref{Lexp2}), associated with the momentum dependence of the nucleon isovector potential, provides the dominant contribution to these correlations.

\begin{figure*}[ht]
\begin{center}
 \resizebox{0.7\textwidth}{!}{
  \includegraphics[width=10cm,height=10cm]{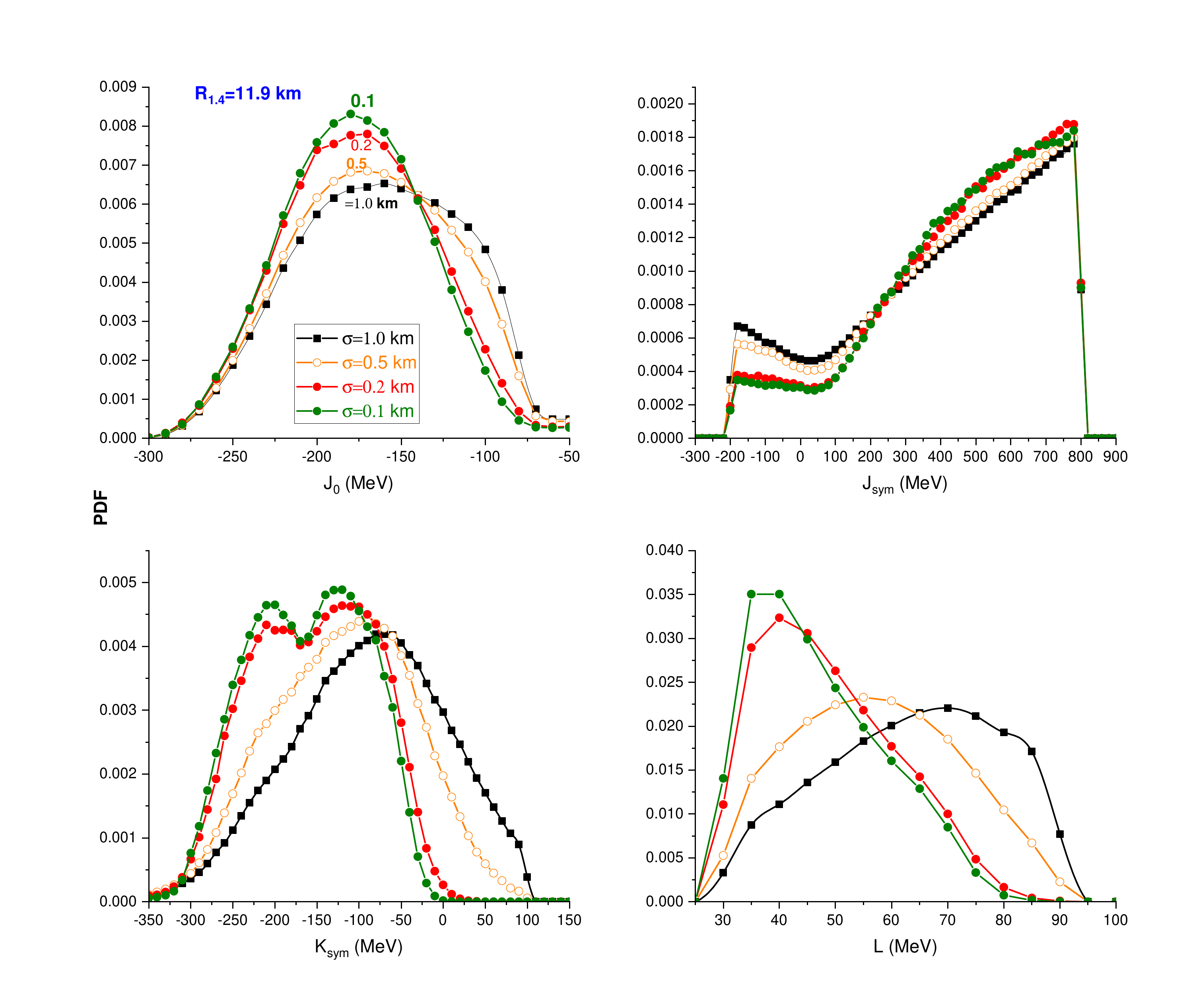}
  }
   \vspace{-0.5cm}
  \caption{Posterior PDFs of EOS parameters with $R_{1.4}=11.9$ km and a precision of $\Delta R=1.0, 0.5, 0.2$, and $0.1$ km, respectively. The figure is taken from Ref. \cite{Li24-PRD}. }\label{R14PDF}
\end{center}
\end{figure*}

\subsection{\bf Example-3: High-density symmetry energies from nuclear many-body theories in comparison with those from analyzing neutron star observables}
Because of the broadly recognized importance of knowing precisely the $E_{\rm{sym}}(\rho )$ in both astrophysics and nuclear physics, essentially all existing nuclear many-body theories using all available nuclear forces have been used to predict the $E_{\rm{sym}}(\rho )$. Shown in the left window of Fig. \ref{Esym-all} are 60 examples selected from 6 classes of over 520 phenomenological models and/or energy density functional theories (among those shown in Fig. \ref{SHF-RMF}), while the right window shows 11 examples from microscopic and/or {\it ab initio} theories \cite{ChenLW14,ChenLW15}. Mostly by design, they all agree with existing constrains available around the saturation density $\rho_0$. However, at supra-saturation densities their predictions diverge very broadly. Besides the technical challenges of solving nuclear many-body problems, the uncertain high-density behavior of nuclear symmetry energy is caused by our poor knowledge about the normally weak isospin-dependence of strong interactions and the resulting nucleon-nucleon correlations as discussed in the previous section. 

To reveal effects of the normally small isospin dependence of strong interactions, one has to use either the naturally existing neutron-rich environments inside neutron stars or create them in terrestrial nuclear laboratories with heavy-ions especially those near the neutron drip line. Constrains on the high-density nuclear symmetry energy from both astrophysical observations and terrestrial experiments will help screen the existing predictions and reveal the underlying fundamental physics of ultra-dense neutron-rich nuclear matter. For example, the solid blues lines in the two windows of Fig. \ref{Esym-all} are the upper and lower boundaries of $E_{\rm{sym}}(\rho )$ extracted from studying the radii and tidal deformability of canonical NSs as demonstrated in Fig. \ref{mass217}. Clearly, while these constraints can already exclude many of the model predictions, the constraints at densities above $2\rho_0$ are still rather loose mainly because both the radii and tidal deformability of 1.4M$_{\odot}$ NSs are mostly sensitive to the pressure at densities around $2\rho_0$. To constrain the symmetry energy significantly above $2\rho_0$, one may have to study additional messengers especially those directly from NS cores or emitted during collisions between two NSs in space or two heavy nuclei in the laboratory \cite{Xie19,Xie20}. 

\begin{figure*}[ht]
\begin{center}
\resizebox{0.8\textwidth}{!}{
  \includegraphics{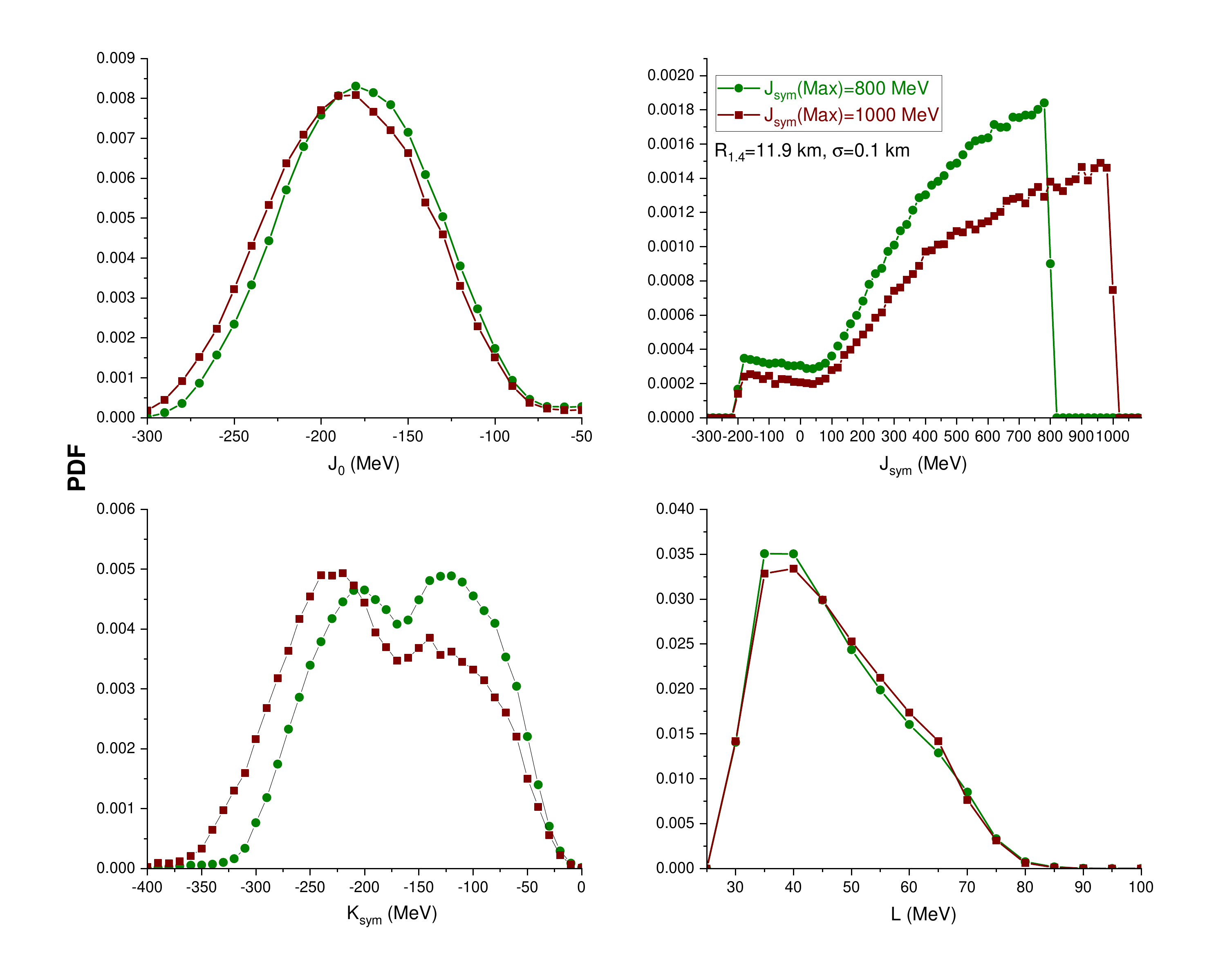}
  }
  \resizebox{0.9\textwidth}{!}{
  \includegraphics{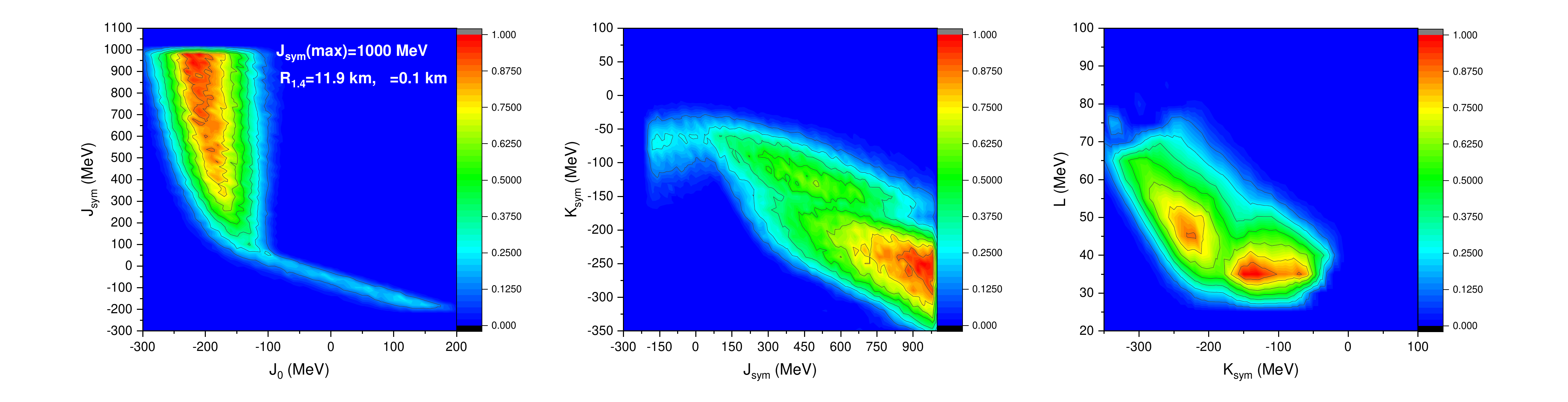}
  }
    \end{center}
    \vspace{-0.2cm}
    \caption{Upper two rows: Comparisons of the PDFs of high-density EOS parameters with $R_{1.4}=11.9$ km and a precision of $\Delta R=0.1$ km, with the maximum prior value of $J_{\rm{sym}}$ set at $800$ (green) and 1000 MeV (wine), respectively. Lower row: the pairwise correlations of the high-density EOS parameters with $J_{\rm{sym}}=1000$ MeV and $R_{1.4}=11.9 \pm 0.1$ km. The figure is taken from Ref. \cite{Li24-PRD}. }\label{fut}
    \end{figure*}
    
\subsection{\bf Example-4: High-density symmetry energies from future neutron star radius measurement to 0.1 km accuracy}\label{future}
The current accuracy in measuring the radii of neutron stars is about 1 km. For instance, LIGO/VIRGO inferred a radius of $R_{1.4}=11.9\pm 0.875$ km at 68\% confidence level for neutron stars involved in GW170817 \cite{LIGO18}. Subsequent independent analyses of the tidal polarizability of GW170817 within various models found the mean radius of a canonical NS is about $R_{1.4}=12.0\pm 1.13$ km at 68\% credibility assuming all reports are equally reliable \cite{LiBA19}. More recent NICER observations for neutron star radii generally have similar or larger errors, see, e.g., Refs. \cite{Riley19,Miller19}.
Moreover, considering together results from both gravitational wave and various X-ray observations of neutron stars, it has been found empirically that $R_{1.4}\approx R_{1.6}\approx R_{1.8}\approx R_{2.0}$ within about 1.0 km precision \cite{MR-Russia,Ayr25}. Overall, the current precision of measuring neutron star radii is about (10-15)\% of the mean radii. 

While the currently available neutron star radius data have certainly improved our understanding of the EOS and nucleon effective mass in dense neutron-rich matter, see, e.g., Ref. \cite{Prad}, more precise radius measurements may lead to major progress. For instance, to distinguish many EOSs and identify twin stars (two NSs having the same mass bust different radii) or strange stars, it is necessary to carry out the differential mass/radius measurement $dM/dR$ in the mass region $(1.2-2.0$M$_{\odot}$ \cite{Cai:2025nxn,Li24-PRD,Zhao20,Han22,Pro,Zhang25,Xavier}. 
In particular, two neutron stars may be born as twins or evolve into ones having the same mass but different compositions and radii. How and why they exist, as well as their composition, all depend on the poorly known EOS, especially its symmetry energy term. To observationally detect twin stars, one key challenge is to predict and measure the mass range $\Delta M$ in which the twins can coexist and the maximum difference $\Delta R$ of their radii. With the expected high-precision radius and mass measurements, there is a great chance to detect twin stars and understand the underlying EOS governing their properties based on both forward-modelings \cite{Zhang25} and Bayesian inference \cite{Xavier}. 

To realize the goals mentioned above, NS radii have to be measured much better than 1.0 km. Indeed, such measurements have been proposed by using the next-generation X-ray pulse profile observatories and gravitational wave detectors. For instance, the enhanced X-ray Timing and Polarimetry mission (eXTP) \cite{eXTP:2018anb,AngLi25} to be launched around 2030 is designed to measure the radius of PSR 10740+6620 to about $\pm 6\%$ accuracy, while the Advanced Telescope for High Energy Astrophysics (NewATHENA) \cite{Ath} to be launched around 2037 can measure the radius of PSR 10740+6620 to about $\pm 3\%$ accuracy \cite{Seb}.
The third-generation gravitational-wave detectors \cite{Hild:2009ns,LIGOScientific:2020zkf}, e.g., Einstein Telescope \cite{Sathyaprakash:2012jk} and Cosmic Explorer \cite{Evans:2021gyd} may measure the radii at even higher precision. For example, based on several recent analyses and simulations, see, e.g., Refs. \cite{Chatziioannou:2021tdi,Pacilio:2021jmq,Bandopadhyay:2024zrr,Finstad:2022oni,Walker:2024loo}, the planned new gravitational wave facilities can measure the $R_{1.4}$ to a precision better than 2.0\%. For example, considering only the 75 loudest events of the expected more than $3\times 10^5$  binary NS mergers in the one-year operation of a network consisting of one Cosmic Explorer and the Einstein Telescope, the radii of NSs in the mass range (1-1.97) M$_{\odot}$ will be constrained to at least $\Delta R\leq 0.2$ km at 90\% credibility \cite{Walker:2024loo}. Of course, all these new proposals have multiple science drivers. Understanding the nature of compact stars and the associated EOS of superdense neutron-rich matter is only one of them. 

Given the super-difficult work involved and the super-expensive investments needed in many aspects to get the super-precise neutron star radius data, what new physics can we learn about the EOS of supradense neutron-rich matter from the expected data? Some efforts have been made recently to answer this question, see, e.g., Refs. \cite{Li24-PRD,Li25-PRD}. In the following, we summarize the key points about extracting information regarding the high-density nuclear symmetry energy. 
Shown in Fig. \ref{R14PDF} is a comparison of the posterior PDFs of $J_0$, $J_{\rm{sym}}$, $K_{\rm{sym}}$, and $L$ from Bayesian analyses using the fiducial radius data of $R_{1.4}=11.9$ km with an imagined precision of $\Delta R=1.0, 0.5, 0.2$, and $0.1$ km, respectively. Several interesting observations can be made: 
\begin{enumerate}
    \item 
While the most probable value of the skewness $J_0$ measuring the stiffness of high-density SNM remains about the same, its precision (68\% confidence width) gets appreciably improved as the precision $\Delta R $ changes from 1.0 km to 0.1 km. \\
\item The PDFs of the symmetry energy parameters especially $L$ and $K_{\rm{sym}}$ have significant changes. In particular, the most probable $L$ shifts to smaller values and the PDF($K_{\rm{sym}}$) starts to show two peaks as the precision improves. On the other hand, the  PDF($J_{\rm{sym}}$) still peaks at the upper boundary of its uniform prior. These findings are qualitatively expected. It is known that the radii of canonical neutron stars are determined by the pressure around $2\rho_0$ to which the symmetry energy makes a major contribution. Thus, using the same mean radius $R_{1.4}=11.9$ km with better precision from 1.0 to 0.1 km will lead mainly to a more precise inference of nuclear symmetry energy around $2\rho_0$ characterized by $L$ and $K_{\rm{sym}}$. It will not improve much about the EOS at significantly higher densities characterized by $J_0$ and $J_{\rm{sym}}$. In fact, as we shall show next, the $J_{\rm{sym}}$ remains unconstrained regardless of the precision of measuring $R_{1.4}$. Nevertheless, it is interesting to see that as the $\Delta R$ decreases, because the pairwise anti-correlations of $J_{\rm{sym}}-J_0$ and $J_{\rm{sym}}-K_{\rm{sym}}$ become pronounced with better precision, the mean value of $J_{\rm{sym}}$ increases somewhat to counterbalance the decreases of $J_0$ and $K_{\rm{sym}}$. Very interestingly, a distinguished two-peak structure is revealed in the PDF of $K_{\rm{sym}}$ when the precision of NS radius measurement is better than about 0.2 km. As in many other areas of science, depending on measurements, seeing such fine structures with high-precision probes is not surprising. As explained in detail in Ref. \cite{Li24-PRD}, the two-peak structure of PDF($K_{\rm{sym}}$) is completely due to the strong anti-correlations of $K_{\rm{sym}}-J_{\rm{sym}}$ and $K_{\rm{sym}}-L$ that are only visible when the precision of radius measurements is high enough.\\
\item The response of the PDFs of symmetry energy parameters to the variation of the precision $\Delta R$ is rather asymmetric. This is rather different from the response of SNM skewness $J_0$.
It is seen that the PDF$(J_0)$ mostly narrows its width symmetrically around approximately the same most probable value. It is known that the $J_0$ is mainly determined by the maximum mass and causality requirements, as shown in the left window of Fig. \ref{mass217}, while the radii and tidal deformation are mainly controlled by the symmetry energy. Thus, the PDF$(J_0)$ is not affected much by the variation of $\Delta R$. On the other hand, a small change in radius can cause a significant change in the density profile or average density of neutron stars. Since the ToV equations governing the structure of neutron stars are highly nonlinear, the correspondence between the mass-radius sequence and the EOS is unique but nonlinear. While the Gaussian likelihood function used in Bayesian analyses is symmetric between the observational data and model prediction, it was demonstrated in Ref. \cite{Li24-PRD} that the necessary symmetry energy parameters to reproduce a neutron star radius above or below the mean value of $R_{1.4}=11.9$ km vary asymmetrically because of the nonlinear nature of the TOV equations.
\end{enumerate}
\begin{figure*}[!hpbt]
\vspace{-0.5cm}
\centering
\resizebox{0.45\textwidth}{!}{
\includegraphics{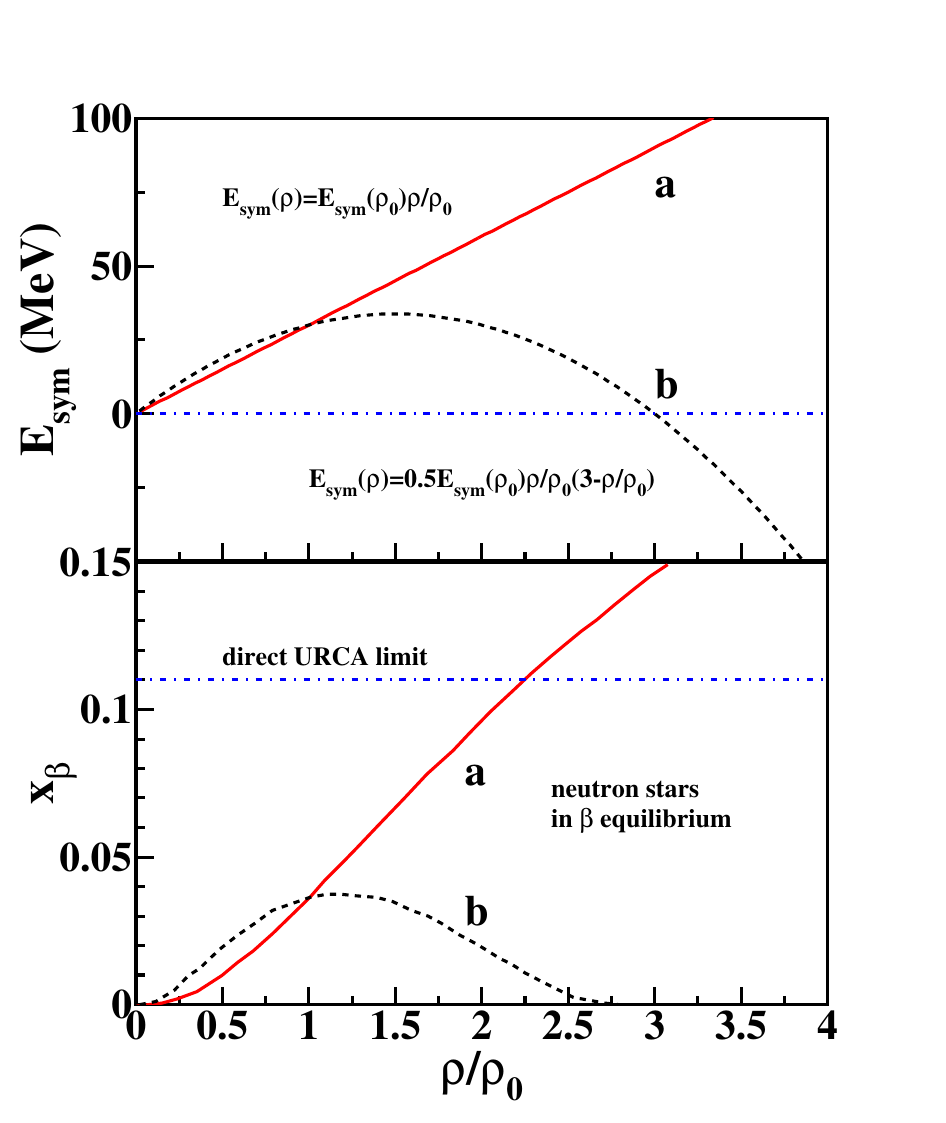}
}
\resizebox{0.45\textwidth}{!}{
\includegraphics{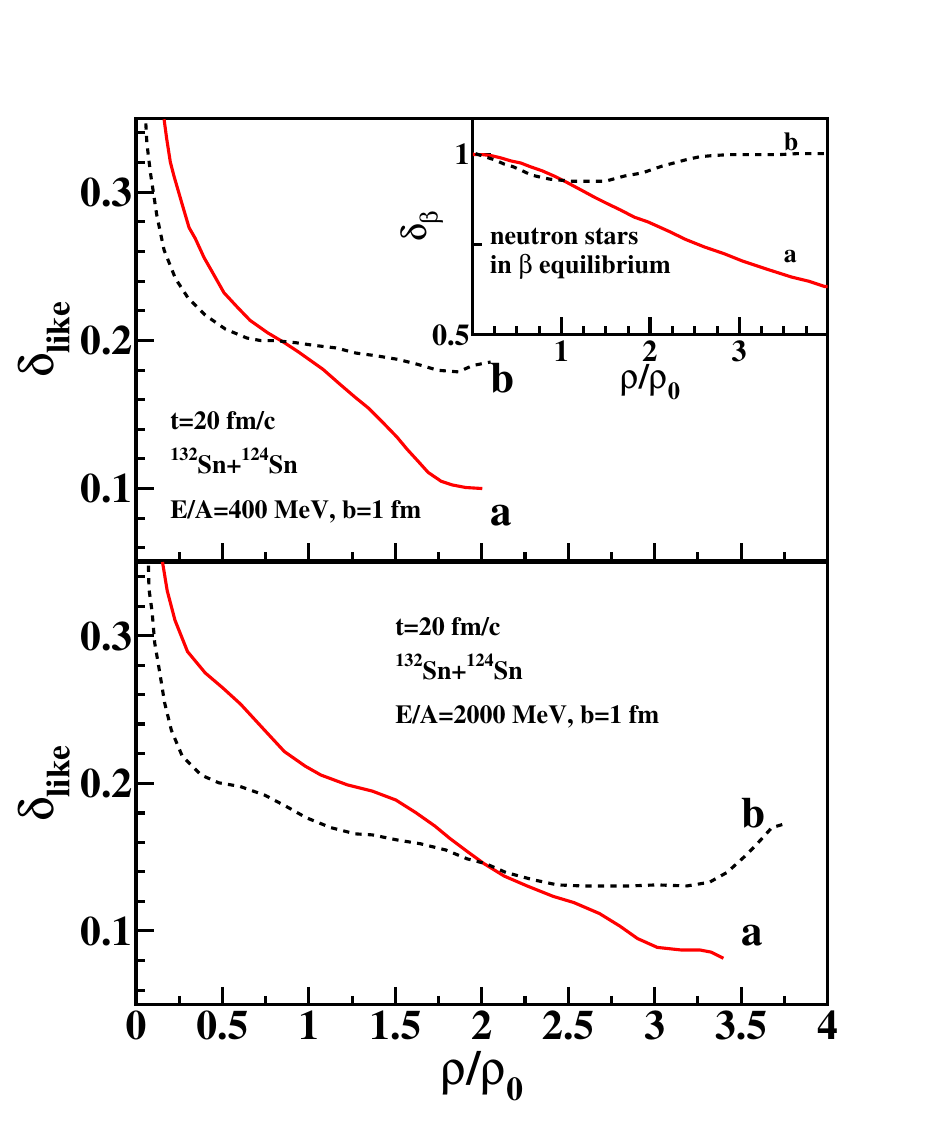}
}
\caption{Left: Upper window: Two representatives of the
nuclear symmetry energy as a function of density. Lower window: the
corresponding proton fractions in neutron stars at $\beta$
equilibrium. 
Right: Upper window: the isospin
asymmetry-density correlations at t=20 fm/c and $E_{\rm beam}=400$ MeV/nucleon in the central $^{132}$Sn+$^{124}$Sn reaction with the
nuclear symmetry energy $E^a_{\rm sym}$ and $E^b_{\rm sym}$,
respectively. Lower window: the same correlation as in the upper
window, but at 10 fm/c and $E_{\rm beam}=2$ GeV/nucleon. The corresponding
correlation in neutron stars is shown in the insert. Taken from Refs. \cite{lba02,li02npa}. }
\label{Xiao12}
\end{figure*}

Finally, to illustrate more clearly that the skewness $J_{\rm{sym}}$ of symmetry energy is not constrained much by the radii of canonical neutron stars regardless of their precision, shown in Fig. \ref{fut} are comparisons of the posterior PDFs of high-density EOS parameters with $R_{1.4}=11.9$ km, $\Delta R=0.1$ km and $J_{\rm{sym}}({\rm{max}})$ set at $800$ (green) and 1000 MeV (wine), respectively. As the $J_{\rm{sym}}({\rm{max}})$ is moved {\it a priori} from 800 to 1000 MeV, the peak of the posterior PDF($J_{\rm{sym}})$ moves outward to $J_{\rm{sym}}({\rm{max}})$=1000 MeV accordingly. It means that the $R$ data does not restrict this parameter. This finding is consistent with that found earlier in Ref. \cite{Xie20}. 

We notice that the PDF($L$) is essentially not affected by the uncertain $J_{\rm{sym}}({\rm{max}})$ because of the very week correlation between $L$ and $J_{\rm{sym}}$, while that of $K_{\rm{sym}}$ has an appreciable shit because of its much strong anti-correlation with $J_{\rm{sym}}$. The 2D PDFs (pairwise correlations) of high-density EOS parameters with $J_{\rm{sym}}=1000$ are shown in the bottom row. The correlations indicate clearly that the two peaks of approximately equal heights in PDF($K_{\rm{sym}})$ are due to the roughly equally strong $J_{\rm{sym}}$-$K_{\rm{sym}}$ and $K_{\rm{sym}}$-$L$ correlations, respectively. It is interesting to see that as the $J_{\rm{sym}}({\rm{max}})$ changes from 800 to 1000 MeV, the left peak in PDF($K_{\rm{sym}})$ shifts accordingly to a smaller value while the right peak due to the $K_{\rm{sym}}$-$L$ correlation stays at roughly the same location but now with a lower height.

\begin{figure*}[ht]
\centering
\resizebox{0.8\textwidth}{!}{
  \includegraphics{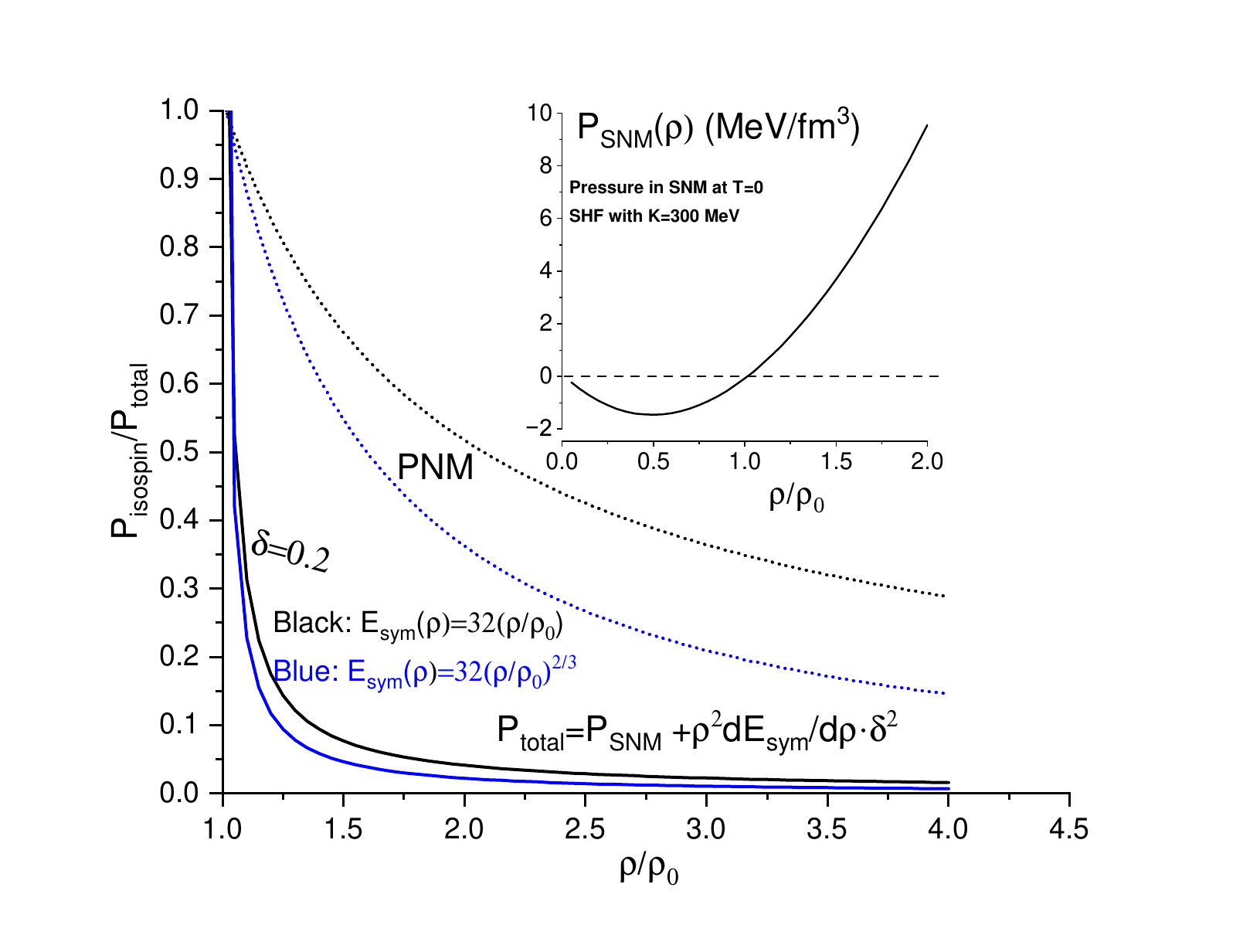}
  }
  \vspace{-0.5cm}
    \caption{Main panel: Fraction of isospin-dependent part over the total pressure in ANM at a fixed $\delta=0.2$ and in PNM ($\delta=1$) as functions of reduced density $\rho/\rho_0$ within SHF with the indicated parameterizations of symmetry energy and $K=300$ MeV. Inset: pressure in SNM within SHF with $K=300$ MeV.}\label{pf2}
    \end{figure*}

\section{Multi-messengers of high-density nuclear symmetry energy from heavy-ion collisions}
Heavy-ion collisions at intermediate-relativistic energies are the only means to create dense neutron-rich nucleonic matter in terrestrial laboratories, see, e.g., Refs. \cite{HW,Sor23,Lov22}. In this section, we first address generally the question on (1) what kind of high-density neutron-rich matter can be formed, (2) what kinds of observables in heavy-ion collisions may be useful for probing the high-density behavior of nuclear symmetry energy, (3) why it is so challenging to probe high-density symmetry energy with heavy-ion collisions, and (4) how hard it can be compared to using high-precision X-rays from isolated neutron stars and/or gravitational waves from their mergers.  
We then give four examples of heavy-ion probes of high-density symmetry energy: (1) the neutron-proton differential transverse flow, (2) the neutron-to-proton and neutron-to-charged particle elliptic flow ratios, (3) the ratios of charged pions, and (4) heavy strange particle production. Other sensitive observables exist and should be explored. A complete list and detailed theoretical discussions about the main probes of symmetry energy in heavy-ion collisions can be found in several earlier reviews, see, e.g., Refs. \cite{LCK08,LiBA14}. While more detailed discussions about the experimental aspects of extracting high-density symmetry energy with heavy-ion reactions can be found in Ref. \cite{Wolfgang}.  
\begin{enumerate}
    \item 
{\bf What kinds of isospin asymmetries and densities can actually be reached in typical heavy-ion reactions?} To answer this question quantitatively, we recall two examples in Fig.~\ref{Xiao12} where two characteristically different symmetry energy functionals are used for illustrations \cite{lba02,li02npa}. One has a linearly increasing symmetry energy while the other one features a super-soft behavior above about $1.5\rho_0$.
The corresponding proton fraction $x_{\beta}$ and isospin asymmetry $\delta_{\beta}$ in NSs at beta equilibrium are shown in the lower left window and the inset of the upper right window, respectively. The threshold for fast cooling of protoneutron stars through the direct URCA process at $x_{\beta}\approx 0.11$ is also indicated. Since some nucleons will be excited to baryon resonances once the beam energy is above the pion production threshold, we are dealing with a hadronic matter instead of a pure nucleonic matter. It was thus suggested to measure the effective isospin asymmetry $\delta_{\rm{like}}$ with \cite{lba02,li02npa}
\begin{equation}
\delta_{\rm{like}}\equiv \frac{(\rho_n)_{\rm{like}}-(\rho_p)_{\rm{like}}}
{(\rho_n)_{\rm{like}}+(\rho_p)_{\rm{like}}},
\end{equation}
where the neutron-like and proton-like densities for the $N+\pi+\Delta$ matter are defined as 
\begin{eqnarray}
(\rho_n)_{\rm{like}}&=&\rho_n+\frac{2}{3}\rho_{\Delta^0}+\frac{1}{3}\rho_{\Delta^+}
+\rho_{\Delta^-},\\\
(\rho_p)_{\rm{like}}&=&\rho_p+\frac{2}{3}\rho_{\Delta^+}+\frac{1}{3}\rho_{\Delta^0}
+\rho_{\Delta^{++}}.
\end{eqnarray}
At higher beam energies, heavier baryon resonances may appear, and their contributions can be taken into account similarly according to their respective couplings with neutrons and protons. 
Of course, the $\delta_{\rm{like}}$ reduces naturally to the isospin asymmetry $\delta$ of nucleonic matter as the beam energy becomes smaller than the pion production threshold. 

Shown in the right two windows are the $\delta_{\rm{like}}$ as functions of density in the two central $^{132}$Sn+$^{124}$Sn reactions with a beam energy of 400 MeV/nucleon (upper) and 2000 MeV/nucleon (lower) at the instant of 20 fm/c. The red and black curves are results obtained by using the relatively stiffer and softer symmetry energies shown in the upper-left window, respectively. 

While the higher densities being reached with higher impact energy provide the necessary condition to investigate the effects of high-density symmetry energy on observables of heavy-ion reactions, the results shown in Fig.~\ref{Xiao12} indicate that the effects of using the two different symmetry energy functions on $\delta_{\rm{like}}$ at a beam energy of 2000 MeV/nucleon (lower-right) are not larger than those at 400 MeV/nucleon (upper-right). This is consistent with the results shown in Fig. \ref{pf2} where the fraction of isospin-dependent part over the total pressure in ANM at a fixed $\delta=0.2$ and in PNM ($\delta=1$) are shown as functions of reduced density $\rho/\rho_0$ within SHF with the indicated parameterizations of symmetry energy and $K=300$ MeV. Even for PNM, the symmetry energy effect is not increasing significantly at high densities. For example, as we shall show in Fig. \ref{Li12B}, symmetry energy effects on the neutron-proton differential transverse flow saturate in reactions at beam energies above about 500 MeV/nucleon. All these indicate the difficulties in probing the high-density behavior of symmetry energy. Indeed, identifying promising messengers of high-density symmetry energy from heavy-ion collisions has been very challenging. There is no guarantee that high-energy heavy-ion collisions can finally help pin down the high-density behavior of nuclear symmetry energy before reaching the hadron-quark transition.

It is seen that the isospin asymmetry-density correlations in heavy-ion collisions are very similar to those obtained in NSs at beta equilibrium. This is not surprising as in both cases the dense matter is more neutron-rich with the softer symmetry energy (red) as the beta equilibrium in NSs and chemical equilibrium in heavy-ion collisions are controlled by the same $E_{\rm{sym}}(\rho)\cdot \delta^2$ term in the EOS of nuclear matter. The latter favors a higher value of $\delta$ when/where the $E_{\rm{sym}}(\rho)$ is lower in both NSs and heavy-ion collisions. While the isospin asymmetry-density correlations in NSs and heavy-ion collisions are very similar, the value of $\delta_{\rm{like}}$ in the dense region of heavy-ion collisions is much less than that reached in the core of NSs. This makes the extraction of high-density symmetry energy from heavy-ion collisions very challenging, although one can create dense matter under controlled conditions in terrestrial labs. Of course, as we have discussed earlier, extracting the high-density symmetry energy from observations of NSs has its own challenges.\\

\item
{\bf What kind of isospin-sensitive observables in heavy-ion collisions are we looking for?} 
In heavy-ion collisions, the isovector part of the single-nucleon potential, namely, the isovector/symmetry potential underlying the $E_{\rm{sym}}(\rho)$ governs the reaction dynamics. Thus, many observables in heavy-ion collisions may carry useful information about the symmetry potential/energy \cite{LCK08}. However, since the isovector/symmetry potential is normally much weaker than the isoscalar potential, one must look for observables that are most sensitive to the isovector potential with little dependence on the isoscalar potential and other agents, such as the in-medium nucleon-nucleon scattering cross sections. For these reasons, the relative yields and differential collective motions of neutrons and protons, as well as those of oppositely charged pions or light mirror nuclei have been proposed and tested as sensitive probes of the density dependence of nuclear symmetry energy.\\

Generally speaking, the symmetry energy effects, even in reaction systems with the highest neutron/proton ratio available, are less than 30\%. While it is far less challenging than detecting and analyzing gravitational waves from merging neutron stars in space, extracting the high-density behavior of nuclear symmetry energy from heavy-ion collisions has been very difficult so far, although there are already some very interesting results, as we shall illustrate with examples. \\

\item
{\bf Why is it so challenging to probe high-density symmetry energy with heavy-ion collisions?} 
First of all, the fraction of the isospin-dependent part over the total pressure during heavy-ion reactions is very small. More quantitatively, shown in Fig. \ref{pf2} is an example of this fraction in ANM at a fixed $\delta=0.2$ (typical isospin asymmetry for heavy-ion reactions) and in PNM as functions of density using SHF-based parameterizations for SNM EOS with $K=300$ MeV and the indicated symmetry energy functionals (one representing Siemens's scaling and the other one is a typical RMF model prediction). The inset illustrates the pressure in SNM with $K=300$ MeV. In this example, we use the simplest SHF-based parameterization for SNM EOS still widely used in transport models \cite{LiXie25,BD87}. At $\rho_0$ the fraction is one as $P_{\rm{SNM}}$ vanishes there. As the density increases, the fraction quickly decreases as the $P_{\rm{SNM}}$ grows much faster than the symmetry energy contribution. Nevertheless, at densities less than about $1.5\rho_0$ the fraction is above about 10\%. At higher densities, it quickly decreases to less than about 2\%, leading to very small effects on observables. To distinguish the two symmetry energy functions from the latter would require using special methods to cancel out the effects of $P_{\rm{SNM}}$. Very different from this situation, for NS matter at $\beta$-equilibrium because of the lepton contribution that is entirely determined by the density dependence of symmetry energy, see the third term in Eq.(\ref{pre}), the total pressure around $2\rho_0$ can be dominated by the isospin-dependent contribution as shown in Fig. 35 of Ref. \cite{LiBA19}. \\

Most of the information about the EOS and symmetry energy in particular from heavy-ion collisions is extracted from comparing experimental data with transport model simulations. The latter uses single-nucleon potentials, sometimes with momentum dependence as a basic input. One outstanding difficulty has been the model dependence in the analyses of heavy-ion reaction data using transport models. As we mentioned earlier, because of our poor knowledge about the isospin-dependence of strong interactions and the associated correlations, it is already very hard to predict the static properties of both finite nuclear many-body systems and infinite nuclear matter. Dynamical problems in nuclear reactions are even more difficult to handle. For example, in probing the high-density behavior of nuclear symmetry energy with the ratio of charged pions requires the knowledge of not only the isovector interactions of both nucleons and their resonances, such as $\Delta(1232)$ and $N^{*}(1440)$, but also in-medium properties of pions, such as their mean-field potentials and in-medium production thresholds. They are necessary inputs in transport model simulations of heavy-ion reactions. \\

In practice, various assumptions are often made. It is thus not surprising that there are still sometimes strong model dependencies in predicting isospin-sensitive observables. Realizing these challenges, the international community of transport model developers and users has in recent years been conducting systematic comparisons, see. e.g., Refs. \cite{TRS1,TRS2,TRS3,TRS4,TRS5,TRS6}. Indeed, some progress has been made. However, for some observables, such as pions, as we shall discuss in more detail, there are still some significant model dependencies \cite{HW}. Nevertheless, often within a given model, effects of the symmetry energy can still be revealed by keeping all other model ingredients the same. \\
\item
{\bf How hard is it to extract high-density symmetry energy from heavy-ion collisions compared to using X-rays from isolated neutron stars or gravitational waves from their mergers?} Generally speaking, within the current uncertainty range of model parameters, symmetry energy effects on most observables in heavy-ion collisions are less than 30\%, while the systematic errors from model to model may be compatible or even larger. To put these uncertainties in proper perspective, we notice that among the astrophysical observables, the radii of neutron stars are one of the most sensitive observables to the symmetry energy around $2\rho_0$. It is known that the radii of the same NS extracted from different analyses of X-ray data have some systematic errors (e.g., results from two independent analyses of the same data taken by the same collaboration may be somewhat different) and occasionally some observational data get updated more than once.\\

Similarly, the same tidal deformability of NSs involved in GW170817 reported by the LIGO/VIRGO Collaborations has been analyzed by using at least 20 different models within 2 years after its first release, leading to a radius for canonical NSs spreading between about 9 and 14 km. The latter also amounts to an approximately 30\% systematic error besides the individual statistical errors of 10 to 30\% \cite{LiBA19}. Thus, revealing the underlying high-density symmetry energy from both heavy-ion reaction data and astrophysical observations of neutron stars is very challenging. The following discussions should thus be understood within the context of having comparable model dependences as well as systematic and statistical errors in both astrophysics and nuclear physics.
\end{enumerate}
\subsection{\bf Example-1: the neutron-proton differential transverse flow} 
In a coordinate system where the beam is in the $z$ direction, and the reaction plane is in the $x-o-z$ plane, the first ($v_1$) and second ($v_2$) coefficients of the Fourier decomposition of the particle azimuthal angle $\phi$ distribution 
\begin{equation}
\frac{2\pi}{N}\frac{dN}{d\phi} = 1 + 2\sum_{n=1}^{\infty}v_n\cos{[n(\phi)]}
\end{equation}
are normally used to measure the strength of the directed (transverse) and elliptical flow \cite{pawel85,oll,art}, respectively.
Their values as functions of rapidity $y$ and transverse momentum $p_t$ (kinematic differential flows) can be evaluated from
\begin{eqnarray}
v_1(y,p_t)&=&\left<cos(\phi)\right>(y,p_t)=\frac{1}{n}\sum_{i=1}^{n}\frac{p_{ix}}{p_{it}},\\
v_2(y,p_{t})&=&\left<cos(2\phi)\right>(y,p_t)=\frac{1}{n}\sum_{i=1}^{n}\frac{p_{ix}^2-p_{iy}^2}{p_{it}^2}, 
\end{eqnarray}
where $p_{ix}$ and $p_{iy}$ are the x- and y-component of the $i^{{\rm th}}$ particle momentum, respectively. Either the transverse momentum $p_t$ or the rapidity $y$ can be marginalized to investigate the various integrated flows. 
While the integrated or kinematic differential flows for various particles are mostly useful for probing the EOS and transport properties of SNM, differences and/or ratios of various flow components between neutrons and protons or between light mirror nuclei (isospin differential flows) are more useful for probing the density dependence of nuclear symmetry energy. This is mainly because the effects of the isoscalar potentials are largely canceled out, while those due to the isovector potentials (that are positive for neutrons but negative for protons) are combined constructively in the isospin differential flows.

\begin{figure}[!hpbt]
\centering
\resizebox{0.46\textwidth}{!}{
\includegraphics[angle=-90]{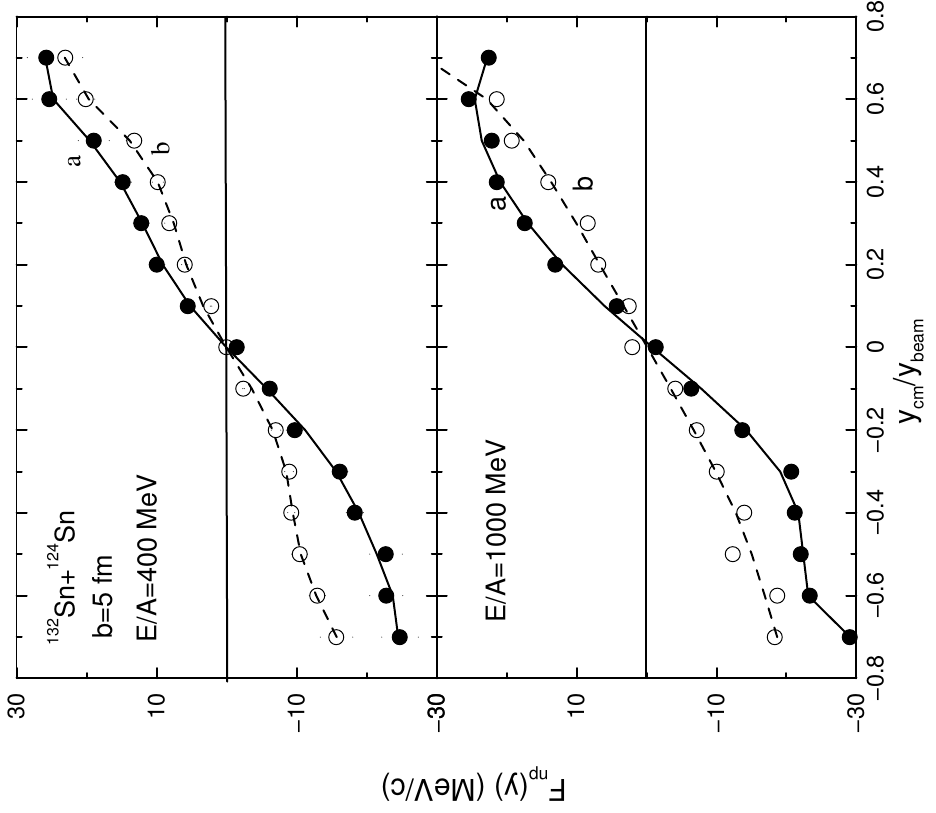}
}
\vspace{0.4 cm}
\caption{The neutron-proton differential transverse collective flow (i.e., $v_1$) versus reduced rapidity in the center of mass system } in the mid-central $^{132}Sn+^{124}Sn$ reactions at $E_{beam}/A=400$ MeV (upper window) and 1000 MeV (lower window) with the nuclear symmetry energy 
$E^a_{sym}$ and $E^b_{sym}$ shown in Fig. \ref{Xiao12}, respectively.
Taken from Refs. \cite{li00}. 
\label{Li12A}
\end{figure}

\begin{figure}[!hpbt]
\centering
\resizebox{0.46\textwidth}{!}{
\includegraphics[angle=-90]{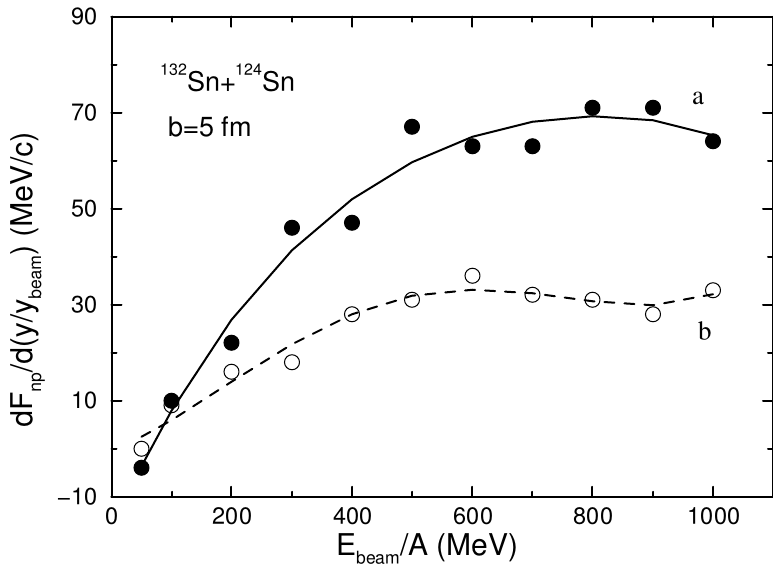}
}
\vspace{0.4 cm}
\caption{Excitation function of the slope parameter of the neutron-proton 
differential flow for the mid-central $^{132}Sn+^{124}Sn$ reaction.
Taken from Refs. \cite{li00}. }
\label{Li12B}
\end{figure}

The neutron-proton differential collective flow is defined as \cite{li00} 
\begin{equation}
F_{np}(y)\equiv\frac{1}{N(y)}\sum_{i=1}^{N(y)}p_{x_i}\tau_i,
\end{equation}
where $N(y)$ is the total number of free nucleons at the rapidity $y$, 
$p_{x_i}$ is the transverse momentum of particle $i$ in the reaction 
plane, and $\tau_i$ is $+1$ and $-1$ for neutrons and protons, respectively.
As it was demonstrated first in ref. \cite{li00}, the $F_{np}(y)$ combines constructively the in-plane transverse 
momenta generated by the isovector potentials while reducing 
significantly influences of the isoscalar potentials of both neutrons 
and protons. 

Shown in Fig.~\ref{Li12A} is the neutron-proton differential collective flow
in mid-central $^{132}Sn+^{124}Sn$ reactions at 400 MeV/nucleon 
and 1000 MeV/nucleon, respectively, using the two symmetry energy functionals shown in the upper left window of Fig.~\ref{Xiao12} within IBUU \cite{li00}.
Effects of the different high-density behaviors of the symmetry energy are clearly revealed. 

One normally measures the strength of the collective transverse flow 
with the slope of the average transverse momentum at mid-rapidity. Similarly, the slope of the neutron-proton differential flow at mid-rapidity was introduced to measure the strength of the neutron-proton differential transverse flow. As shown in Fig.~\ref{Li12B}, 
high-density nuclear symmetry energy has a strong influence on the excitation function of $dF_{np}/d(y/y_{beam})$, especially above about 400 MeV/nucleon.

\subsection{\bf Example-2: the ratios/differences of neutron-to-proton and neutron-to-charged particle elliptic flows}

Similar to the neutron-proton differential transverse flow, the differences or ratios of collective flows of neutrons and protons or light mirror nuclei have been investigated both theoretically and experimentally. For example, Cozma studied the effects of many ingredients in transport models and uncertainties of centrality cuts systematically in the data analyses on both the differential (transverse momentum dependent) and integrated elliptical flows of neutrons, protons, and hydrogens as well as their ratios and differences \cite{Cozma18}. 

\begin{figure}[!hpbt]
\centering
\resizebox{0.39\textwidth}{!}{
\includegraphics{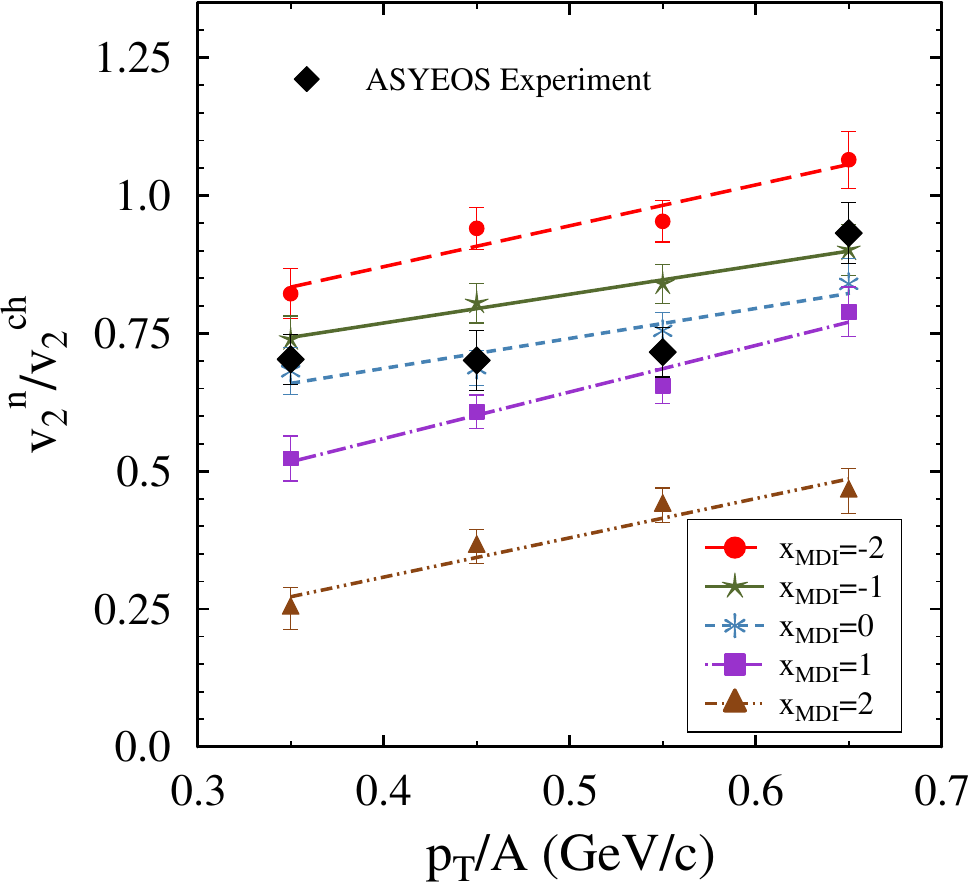}
}
\vspace{0.4 cm}
\caption{Transverse momentum dependent neutron-to-charged particles elliptic flow ratios in comparison with the ASY-EOS data~\cite{Rus16}.
Taken from ref.\cite{Cozma18}.}
\label{CozmaFigures1}
\end{figure}

Shown in  Fig.~\ref{CozmaFigures1} are the transverse momentum dependent neutron-to-charged particles elliptic flow ratios calculated within a newly updated version of the T\"{u}bingen QMD model with a modified MDI interaction in comparison with the ASY-EOS data~\cite{Rus16}. The $x_{\rm{MDI}}$ is a parameter characterizing the density dependence of nuclear symmetry energy used in the modified MDI interaction.  Qualitatively, as the $x_{\rm{MDI}}$ increases from -2 to +2, the slope parameter L decreases. 
The ASY-EOS data scatter between the calculated results with $x_{\rm{MDI}}$ between 1 and -1. 

\begin{figure}[!hpbt]
\centering
\resizebox{0.4\textwidth}{!}{
\includegraphics{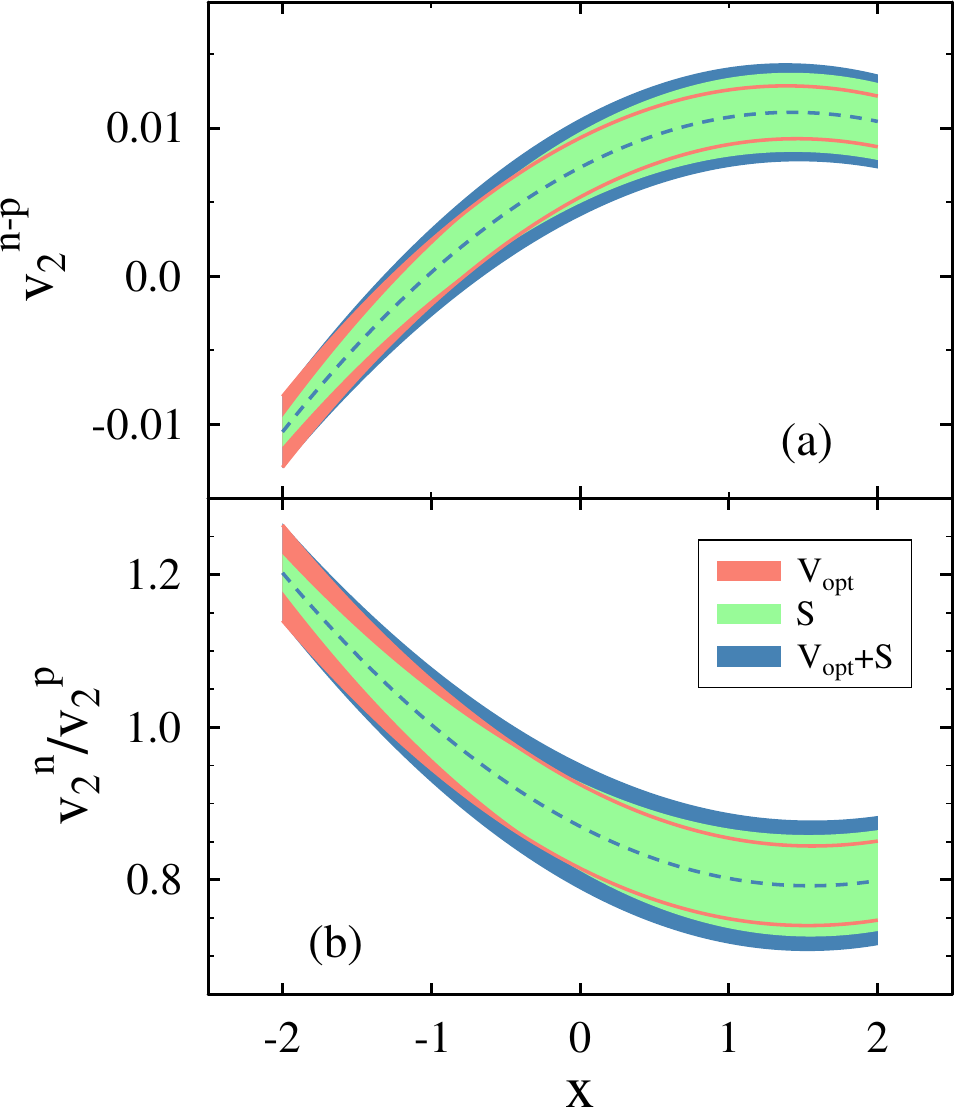}
}
\vspace{0.4 cm}
\caption{Variations of the ratio and difference of the integrated elliptical flows of neutrons and protons due to different choices for the optical potential ($V_{opt}$), parametrization of symmetry-energy ($S$) as well as the combined, quadratically added, uncertainty. Taken from ref.\cite{Cozma}.}
\label{CozmaFigures2}
\end{figure}

Shown in Fig.~\ref{CozmaFigures2} are the
variations of the ratio and difference of the integrated elliptical flows of neutrons and protons as a function of the $x_{\rm{MDI}}$ parameter with different choices for the optical potential ($V_{opt}$), parametrization of symmetry-energy ($S$) as well as a combined, quadratically added, uncertainty. While the spreads caused by the latter are significant, the strong dependence on the $x_{\rm{MDI}}$ parameter (e.g, the density dependence of nuclear symmetry energy) is overwhelming. Detailed analyses of the combined 
FOPI-LAND and ASY-EOS data within this model lead to 
$
L=85\pm\phantom{3}22(\mathrm{exp})\pm\phantom{2}20(\mathrm{th})\pm\phantom{1}12(\mathrm{sys})\,\,\mathrm{MeV} $
and $
K_{sym}=96\pm315(\mathrm{exp})\pm170(\mathrm{th})\pm166(\mathrm{sys})\,\,\mathrm{MeV}$ as mentioned earlier.
Within the specified uncertainties, these two parameters are correlated as one generally expects. 

\begin{figure*}[!hpbt]
\centering
\resizebox{0.6\textwidth}{!}{
\includegraphics{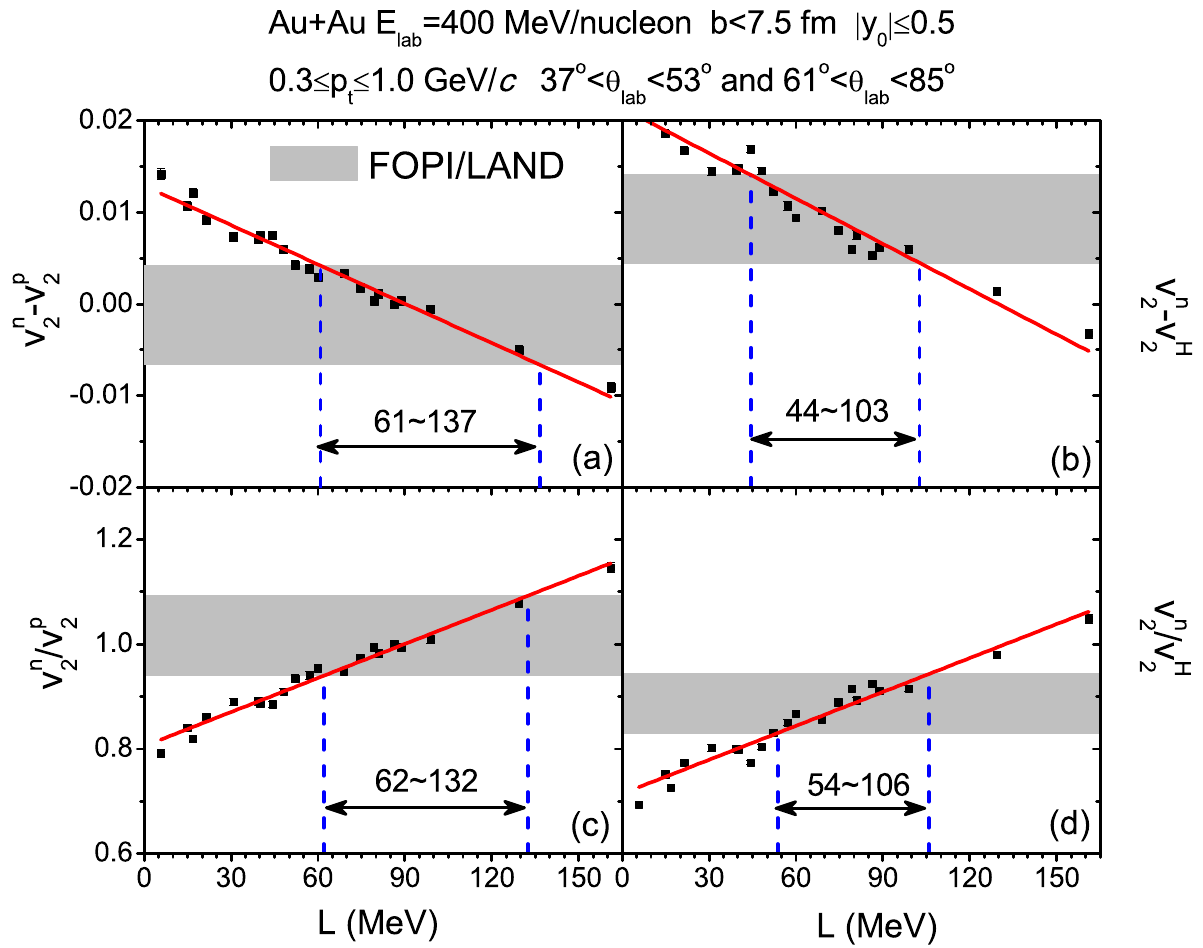}
}
\caption{ The elliptic flow differences $v_2^{n}-v_2^{p}$ between neutrons and protons (a) and
$v_2^{n}-v_2^{H}$ between neutrons and hydrogens (b) as well as theirs ratios $v_2^{n}/v_2^{p}$ (c) and $v_2^{n}/v_2^{H}$ (d)
produced in moderately central $^{197}$Au+$^{197}$Au collisions at
$E_{\rm lab}=400$~MeV/nucleon versus the slope parameter $L$ of nuclear symmetry energy. The gray shaded regions indicate the $p_t$ (transverse momentum) integrated experimental data \cite{Rus11} while the full squares denote UrQMD calculations with different Skyrme forces. Taken from Ref.~\cite{YWang}.}
\label{WangFigure}
\end{figure*}
Strong effects of the symmetry energy on the ratios or differences of neutron-to-proton and neutron-to-charged particle elliptic flows similar to the ones shown above have also been reported by other studies using various transport models and interactions. The results are qualitatively consistent, while there are quantitative differences. As an example, shown in Fig.~\ref{WangFigure} are the  elliptic flow differences $v_2^{n}-v_2^{p}$ between neutrons and protons (a) and $v_2^{n}-v_2^{H}$ between neutrons and hydrogens (b) as well as theirs ratios $v_2^{n}/v_2^{p}$ (c) and $v_2^{n}/v_2^{H}$ (d) produced in moderately central ($b<7.5$~fm) $^{197}$Au+$^{197}$Au collisions at
$E_{\rm lab}=400$~MeV/nucleon versus the slope parameter $L$ of nuclear symmetry energy. The gray shaded regions indicate the $p_t$ (transverse momentum) integrated experimental data \cite{Rus11} while the full squares denote UrQMD calculations with different Skyrme forces \cite{YWang}. Obviously, the studied ratios and differences of the elliptical flows of neutrons, protons, and hydrogens are sensitive to the variation of $L$.  A comparison with the data allowed an extraction of $L = 89 \pm 23$~MeV. 
Clearly, the results from the T\"{u}bingen QMD and UrQMD are generally consistent \cite{Cozma,YWang} while the former used much more elaborated density and momentum dependent isoscalar and isovector potentials. 

\begin{figure*}[htp] 
\centering
\resizebox{0.6\textwidth}{!}{
\includegraphics[angle=-90]{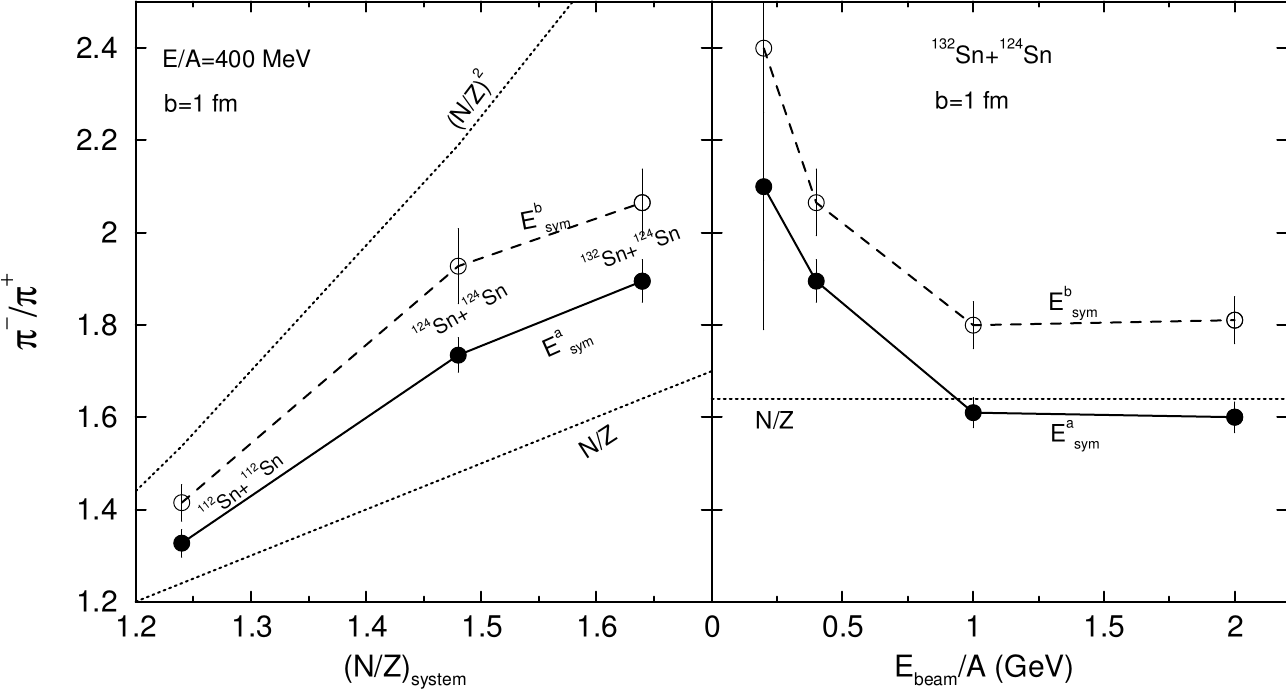}
}
\caption{The $(\pi^-/\pi^+)$ ratio as a function of the isospin asymmetry (left window) and beam energy (right window) of the reaction systems. Taken from ref. \cite{li03}.} 
\label{BALI-NZ}
\end{figure*}
\subsection{\bf Example-3: the ratio of charged pions }
It has long been known that the measurements of pion asymmetry (measured with the multiplicity ratio $\pi^-/\pi^+$ or difference $\pi^--\pi^+$) in various nuclear reactions 
are useful for extracting important information about the structure of radioactive nuclei, the sizes of neutron skins of heavy stable nuclei \cite{tel87,lom88,lihb91,antoni,Wei14,HLi19}
and the symmetry energy of neutron-rich matter \cite{lba02,uma,li03,gai04,Li-Yong,Yong06}. In particular, the $\pi^-/\pi^+$ ratio was proposed as a sensitive probe of the high-density symmetry energy \cite{lba02}. Partially stimulated by the strong physics motivations and the interesting data from the FOPI \cite{FOPI07} and SpiRIT \cite{SpiRIT:2021gtq} Collaborations, significant efforts have been devoted to understanding in detail effects of symmetry energy on the charged pion ratio over the last decade, see, e.g., Refs.\cite{XiaoPRL,MZhang,XiaoEPJA,Fen10,Xie13,Fer05,Son15,Coz16,Xu10,Xu13,Hon14,Guo15,Fen15,Yong17,Tsang-pion,AMD16,AMD18,AMD19,Cozma21,XLi,Zhen17,Zhen18}. However, strong model dependences and controversies about the necessary high-density symmetry energies to reproduce the same data using different transport models exist, see Refs. \cite{HW,TRS4,TRS5,TRS6}, for the latest review. 
While there are always the possibilities of having some mistakes/bugs in the transport codes used, most of the known differences are really due to our poor knowledge about the very complicated pion-nucleon-$\Delta$ dynamics in dense neutron-rich hadronic matter. In fact, static properties of such systems, such as those encountered in the core of neutron stars at $\beta$ equilibrium, are already very difficult to handle, and there are many examples of conflicting conclusions in the literature. Dynamical properties of such systems encountered in nuclear reactions are even more difficult to model. 

While the transport model developers and users continue the detailed code comparison project, and new data on pion production from dedicated experiments are being analyzed, we recall below a few key physics points about the pion-nucleon-$\Delta$ dynamics. We also comment on several issues regarding the ratio of charged pions as a probe of high-density nuclear symmetry energy.

\subsubsection{\bf Why is the $\pi^-/\pi^+$ ratio useful for probing nuclear symmetry energy?} 
The answer to this question depends on several factors. First of all, most pions are produced through the
decays of $\Delta(1232)$ and/or $N^{*}(1440)$ resonances in heavy-ion collisions at beam energies below about 2 GeV/nucleon. While $\pi^-$ mesons are mostly from the decays of $\Delta^{-}$ resonances formed in neutron-neutron collisions, $\pi^+$ mesons are mostly from proton-proton collisions. The $\pi^-/\pi^+$ ratio is expected to be sensitive to the n/p ratio in the participant region of the reaction. More quantitatively, assuming pions are all produced through $\Delta(1232)$ resonances in the first chance nucleon-nucleon scatterings and neglecting the influence of subsequent
pion rescatterings and reabsorptions, the primordial $\pi^-/\pi^+$ ratio was proposed to scale 
with the N/Z ratio of the participant region according to \cite{stock}
\begin{equation}
\pi^-/\pi^+=(5N^2+NZ)/(5Z^2+NZ)\approx (N/Z)^2.
\end{equation}
Often, the neutron/proton ratio $(N/Z)_{\rm{system}}$ of the colliding nuclei was used in some analyses. As discussed in detail in Ref. \cite{li03}, in more realistic situations based on transport model simulations, the actual $\pi^-/\pi^+$ ratio is between $(N/Z)_{\rm{system}}$ and $(N/Z)_{\rm{system}}^2$ depending on the density dependence of the symmetry energy. The latter determines the isospin fractionation in different density regions during the reaction as illustrated in Fig.~\ref{Xiao12}. As shown in Fig.~\ref{BALI-NZ}, the softer symmetry energy ($E^b_{\rm{sym}}$ shown in Fig.~\ref{Xiao12}) leads to a higher $\pi^-/\pi^+$ ratio. 

\begin{figure}[!hpbt]
\resizebox{0.51\textwidth}{!}{
\includegraphics[angle=-90]{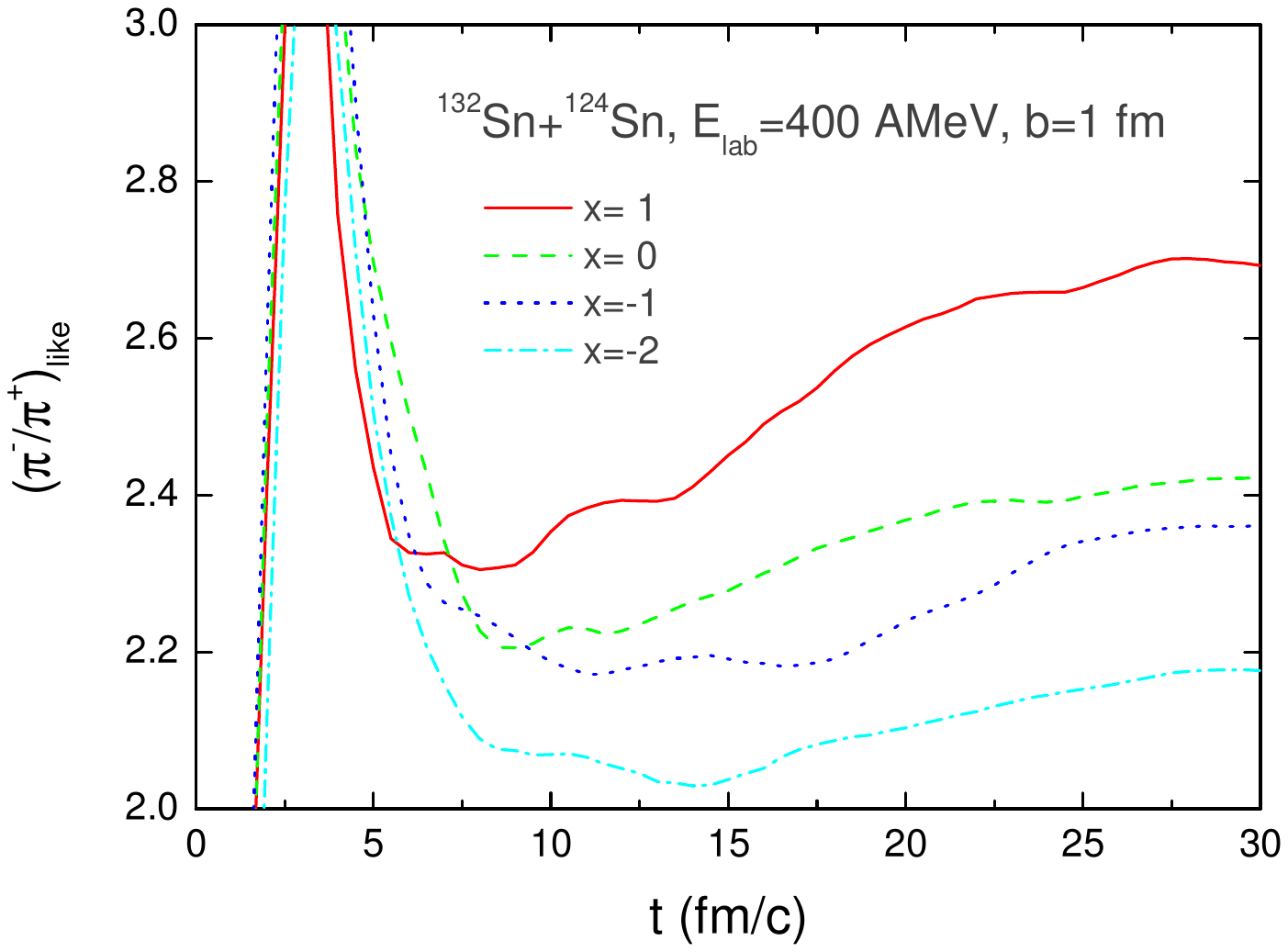}
}
\resizebox{0.53\textwidth}{!}{
\includegraphics[angle=-90]{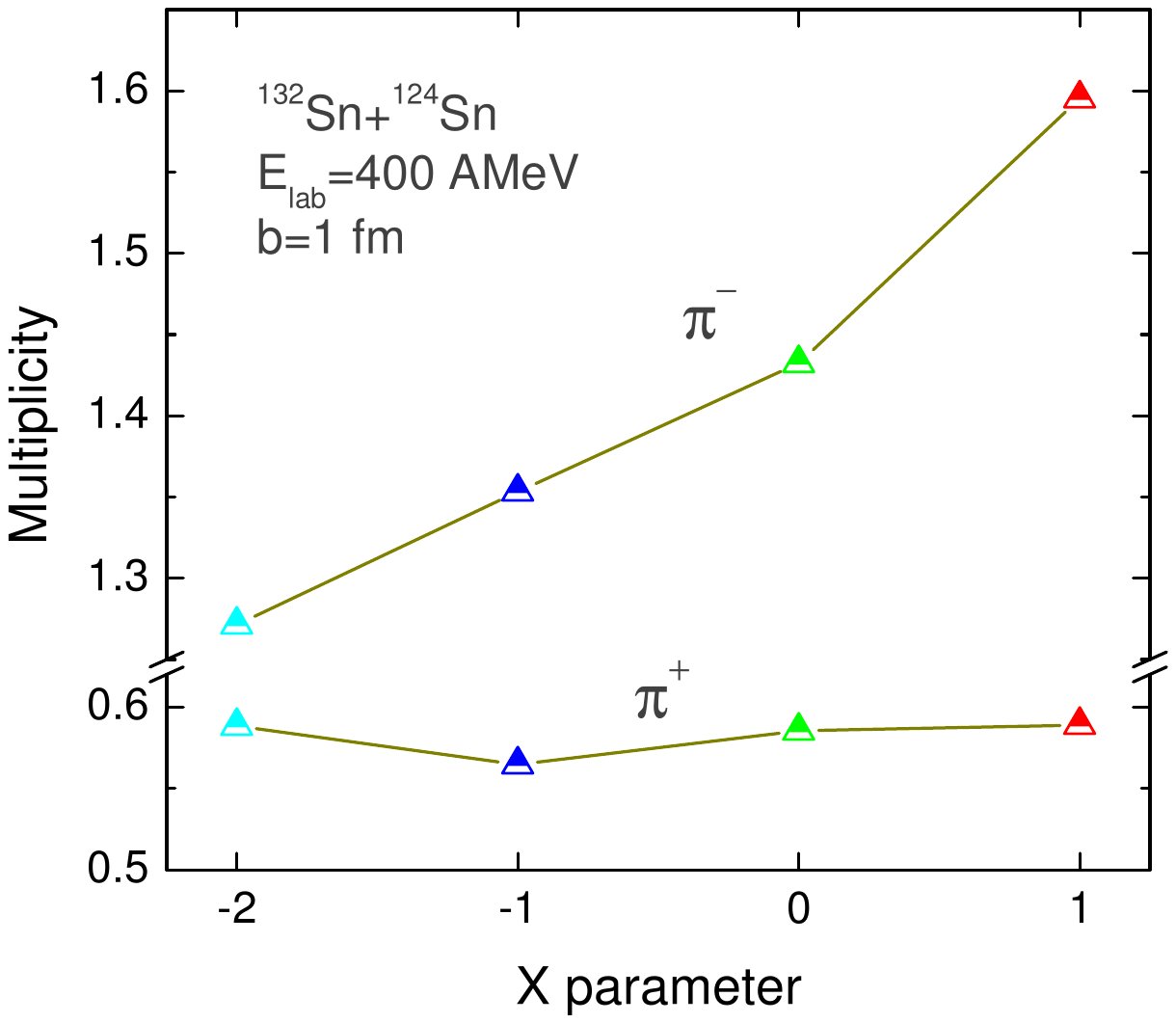}
}
\caption{(Upper: Evolution of the $(\pi^-/\pi^+)_{like}$ ratio in the reaction of $^{132}Sn+^{124}Sn$
at a beam energy of 400 MeV/nucleon and an impact parameter of 1
fm. Lower: The average multiplicity of $\pi^+$ and $\pi^-$ as a function of the $x$
parameter for the reaction of $^{132}Sn+^{124}Sn$ at a beam energy
of 400 MeV/nucleon and an impact parameter of 1 fm. Taken from ref. \cite{Li-Yong}.}
\label{Li-pion}
\end{figure}

Effects of various forms of the symmetry energy have been studied using different transport models. 
While there are quantitative differences, most transport models predict consistently that softer symmetry energies at suprasaturation densities lead to higher $\pi^-/\pi^+$ ratios, especially in more neutron-rich reactions. Considering the dynamics of resonance production and decays, one normally examines the $(\pi^-/\pi^+)_{\rm{like}}$ ratio 
\begin{equation}
(\pi^-/\pi^+)_{\rm{like}}\equiv \frac{\pi^-+\Delta^-+\frac{1}{3}\Delta^0}
{\pi^++\Delta^{++}+\frac{1}{3}\Delta^+}
\end{equation}
as a function of time \cite{li03}. This ratio naturally becomes the final $\pi^-/\pi^+$ ratio at the freeze-out after all resonances have decayed. As an example, shown in the upper window of Fig.\ref{Li-pion} are the evolutions of $(\pi^-/\pi^+)_{\rm{like}}$ from the IBUU04 \cite{LiBA04} simulations of $^{132}Sn+^{124}Sn$ reactions at a beam energy
of 400 MeV/nucleon and an impact parameter of 1 fm using the MDI interactions \cite{Das03}. The corresponding symmetry energy functionals with different $x$ parameters are shown in the left window of Fig.~\ref{Xiao}. It is seen that the $(\pi^-/\pi^+)_{\rm{like}}$ ratios freeze-out at different values depending on the $x$ parameter after approximately 30 fm/c. Moreover, as shown in the lower window, the multiplicity of $\pi^-$ is more sensitive to the variation of the symmetry energy than $\pi^+$ mesons, as one expects for the neutron-rich reaction system \cite{li03}. Thus, generally speaking, from the reaction dynamics point of view, with different density dependences of nuclear symmetry energy, the neutron/proton ratios of the participant regions are different. These differences are reflected in the final $\pi^-/\pi^+$ ratio experimentally measurable.  

\begin{figure*}
  \includegraphics[width=\textwidth]{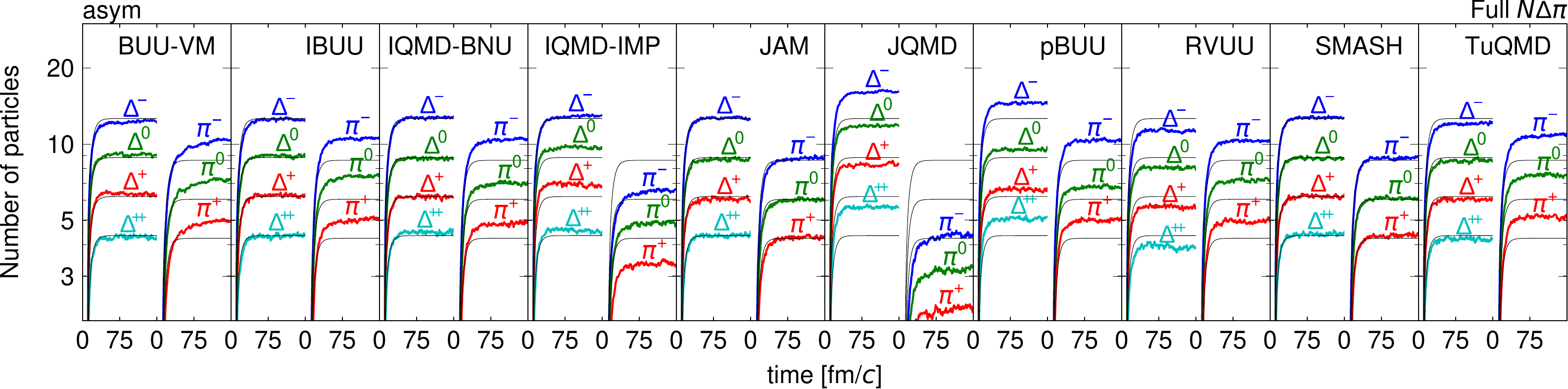}
  \caption{\label{trans-c} Time evolution of the numbers of $\Delta$ and $\pi$ in an asymmetric ($\delta=0.2$) full-$N\Delta\pi$ system in a box with periodic boundary condition using the cascade mode of the indicated 10 transport models without Pauli blocking for Fermions and Bose enhancement for pions.  Results from the rate equation are represented by thin lines. Taken from ref. \cite{TRS3}.}
\end{figure*}
From the statistical point of view, one also expects the $\pi^-/\pi^+$ ratio to be sensitive to the density dependences of nuclear symmetry energy. 
For example, within a statistical model for pion production \cite{nature,bona}, Bertsch et al. showed that the $\pi^-/\pi^+$ ratio is
proportional to ${\rm exp}\left[(\mu_n-\mu_p)/T\right]$,
where T is the temperature, $\mu_n$ and $\mu_p$ are the chemical potentials of neutrons and protons,
respectively. The latter can be written as \cite{thermal}
\begin{eqnarray}\label{sta}
\mu_n-\mu_p&=&V^n_{\rm{asy}}-V^p_{\rm{asy}}-V_{\rm{Coulomb}}\\ \nonumber
&+&T\left[{\rm ln}\frac{\rho_n}{\rho_p}+\sum_m\frac{m+1}{m}B_m(\frac{\lambda_T^3}{2})^m(\rho^m_n-\rho^m_p)\right],
\end{eqnarray}
where $V_{\rm{Coulomb}}$ is the Coulomb potential for protons, $\lambda_T$ is the thermal wavelength of
a nucleon and $B'_m$s are the inversion coefficients of the Fermi distribution function \cite{thermal}.
The difference in neutron and proton symmetry
potentials $V^n_{\rm{asy}}-V^p_{\rm{asy}}\approx 2U_{\rm{sym,1}}\delta$ is directly related to the 
strength of the nucleon symmetry potential $U_{\rm{sym,1}}$. Moreover, the kinetic part of the difference $\mu_n-\mu_p$ relates directly to the
isospin asymmetry $\rho_n/\rho_p$ or $\rho_n-\rho_p$. Thus, within statistical models for pion production in heavy-ion collisions, the $\pi^-/\pi^+$ ratio is also expected to be a good probe of nuclear symmetry energy. 

\subsubsection{\bf Why is it so difficult to extract the high-density nuclear symmetry energy using $\pi^-/\pi^+$ ratios in heavy-ion collisions?} The main physics reasons discussed above for using the $\pi^-/\pi^+$ ratios in heavy-ion collisions to probe the high-density symmetry energy are convincing to at least some experts in the community. However, there are still great difficulties in modeling the physics of $\pi+N+\Delta$ dynamics in dense neutron-rich matter. In addition, there are many technical challenges in extracting relatively small effects of nuclear symmetry energy through the nucleon isovector potential from observables that are mostly dominated by the nucleon isoscalar potential. In fact, many physics inputs are necessary to model the dynamics of $\pi$-nucleon-baryon resonances in neutron-rich matter. These include the in-medium pion production threshold, pion in-medium dispersion relation or mean-field potential, pion-baryon scattering cross sections, baryon resonance productions, scatterings, propagation in the strong mean-fields, absorptions and decays, Pauli blocking for Fermions and Bose enhancement for pions, etc. 

Not surprisingly, even in the simplified case of simulating $\pi$-nucleon-baryon resonances in a static box with periodic boundary conditions in the cascade mode without any mean-field and Pauli blocking is still very model dependent \cite{TRS3}. As an example, shown in Fig.~\ref{trans-c} are the time evolution of the numbers of $\Delta$ and $\pi$ in an asymmetric ($\delta=0.2$) full-$N\Delta\pi$ system initialized with nucleons at saturation density and at 60 MeV temperature in a box with periodic boundary condition using the cascade mode of the indicated 10 transport models without Pauli blocking for Fermions and Bose enhancement for pions.  The simulations are compared to two reference cases of a chemically equilibrated ideal gas mixture and of the rate equation \cite{TRS3}. It was found that the differences of the predicted multiplicities of pions and $\Delta$ resonances depend significantly on the different sizes of the time step and ways in ordering the sequence of reactions, such as collisions and decays, that take place in the same time step. Nevertheless, the uncertainty in the transport-code predictions of the final $\pi^-/\pi^+$ ratio, after letting the existing $\Delta$ resonances decay, was found to be within a few percent for the system initialized at neutron/proton=1.5 as some of the differences using different transport models get canceled out. 

The comparisons shown in Ref. \cite {TRS3} were done purposely without using any mean-field and Pauli blocking to identify step-by-step sources of model dependences. This effort will be continued to investigate ingredients more directly related to the density and momentum dependence of the symmetry (isovector) potential of nucleons and $\Delta$ resonances, as well as the
associated Pauli blocking in dense neutron-rich matter.  Before doing a thorough comparison using different transport models, existing results obtained using individual transport models are educational and illustrative of the importance of several model ingredients. We emphasize again that some of these ingredients are important not only for heavy-ion collisions but also for understanding the properties of neutron stars. 

The reason that their treatments in different transport models may be different is mainly due to our poor physics knowledge. For example, effects of different techniques of treating the Pauli blocking on the $\pi^-/\pi^+$ ratio were recently studied in Ref. \cite{AMD19}. They used a hybrid model for pion production in heavy-ion collisions by combining the antisymmetrized molecular dynamics (AMD) and a hadronic cascade model (Jet AA Microscopic transport model, JAM). Depending on whether clustering effects are considered and how the nucleon phase space distribution functions are evaluated, the Pauli blocking in the $NN\leftrightarrow N\Delta$ and $\Delta\rightarrow \pi N$ channels is found to have different effects on the $\pi^-/\pi^+$ ratio. As they discussed in detail in ref. \cite{AMD19}, because of large fluctuations in evaluating the phase space distribution in nuclear reactions using different techniques, the Pauli blocking factor is normally very model-dependent. 

\begin{figure*} [!hbt]
\centering 
\resizebox{0.3\textwidth}{!}{
\includegraphics[scale=0.6]{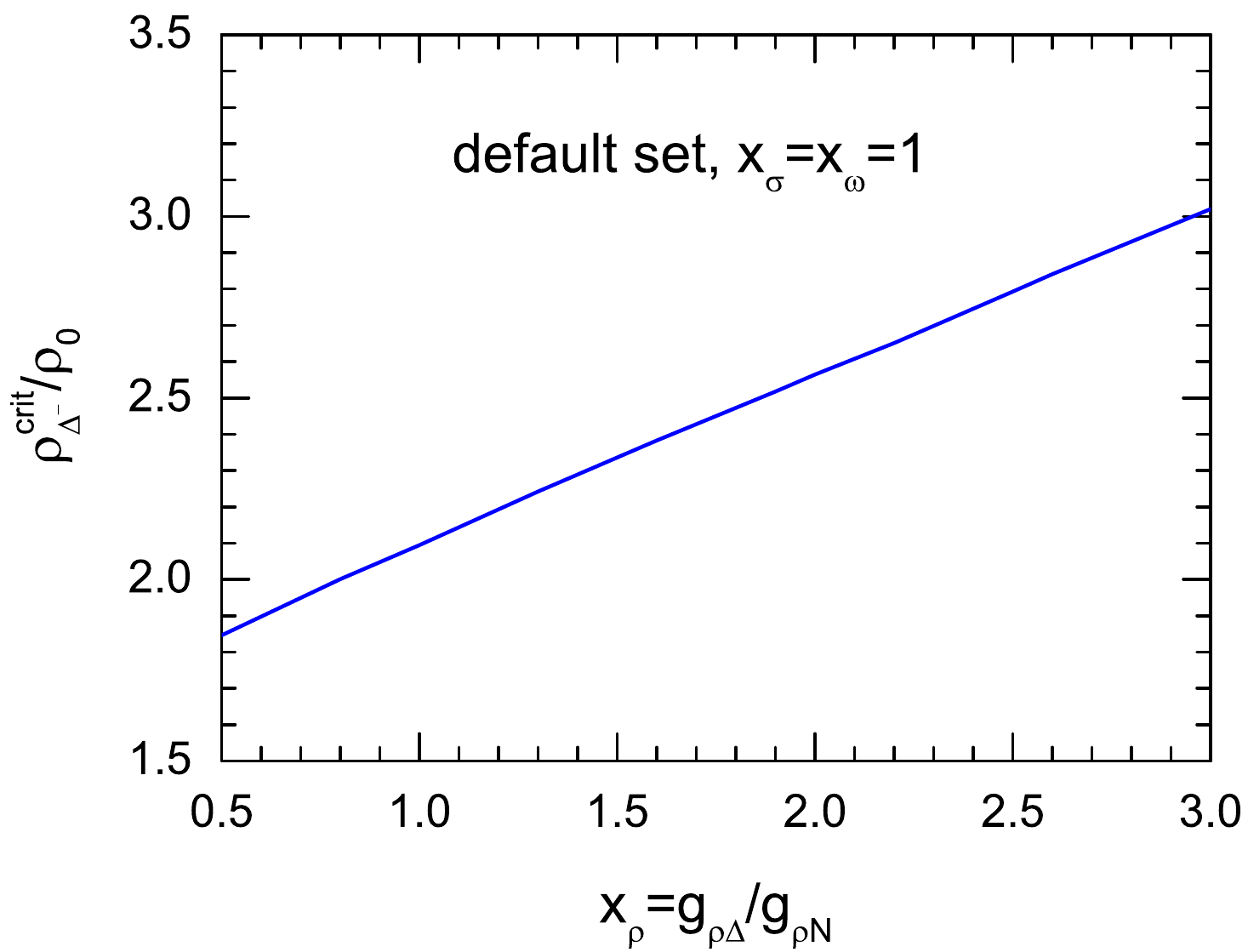} 
}
\resizebox{0.35\textwidth}{!}{
\includegraphics[scale=0.6]{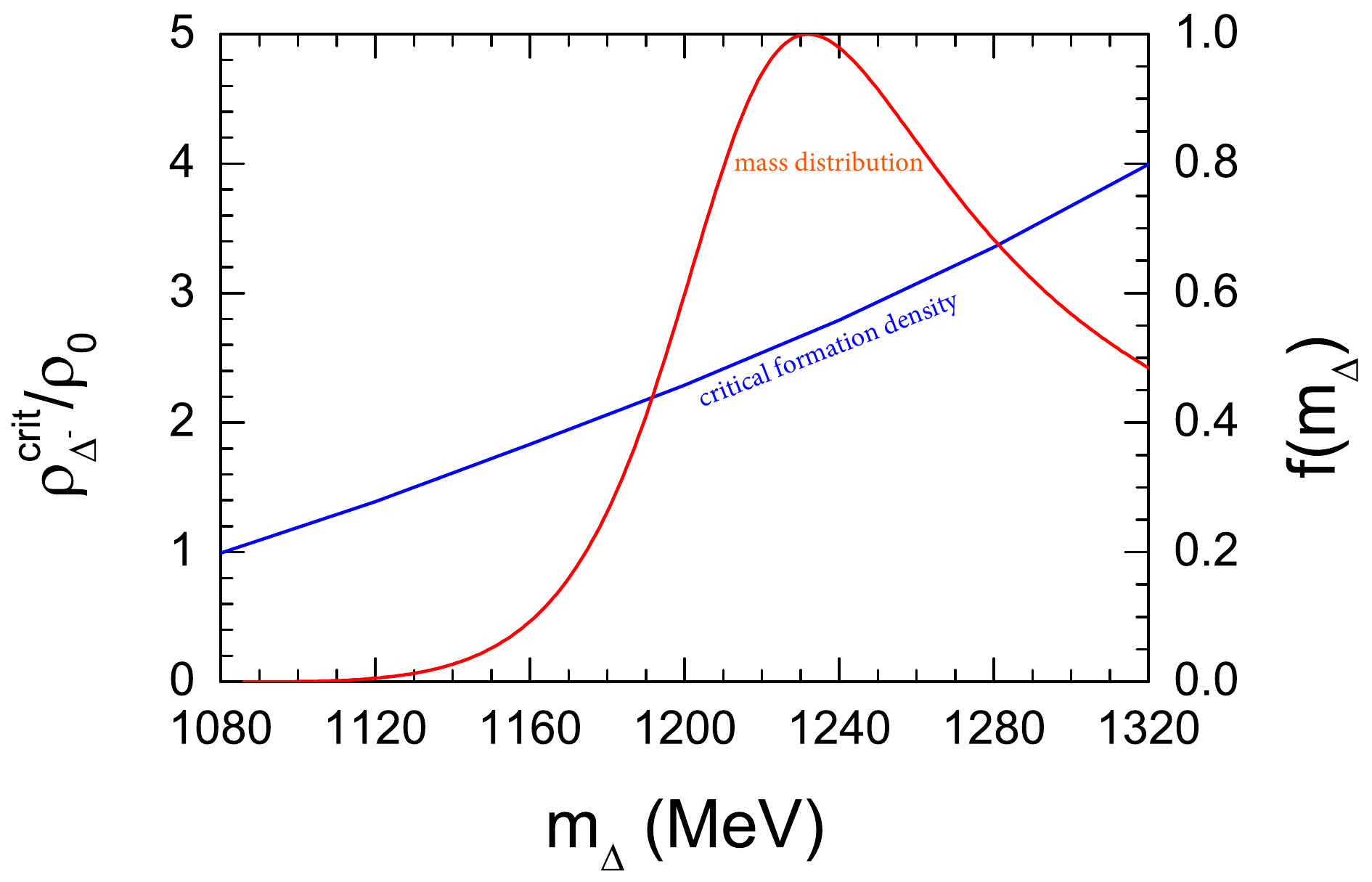} 
}
\resizebox{0.3\textwidth}{!}{
\includegraphics[scale=0.6]{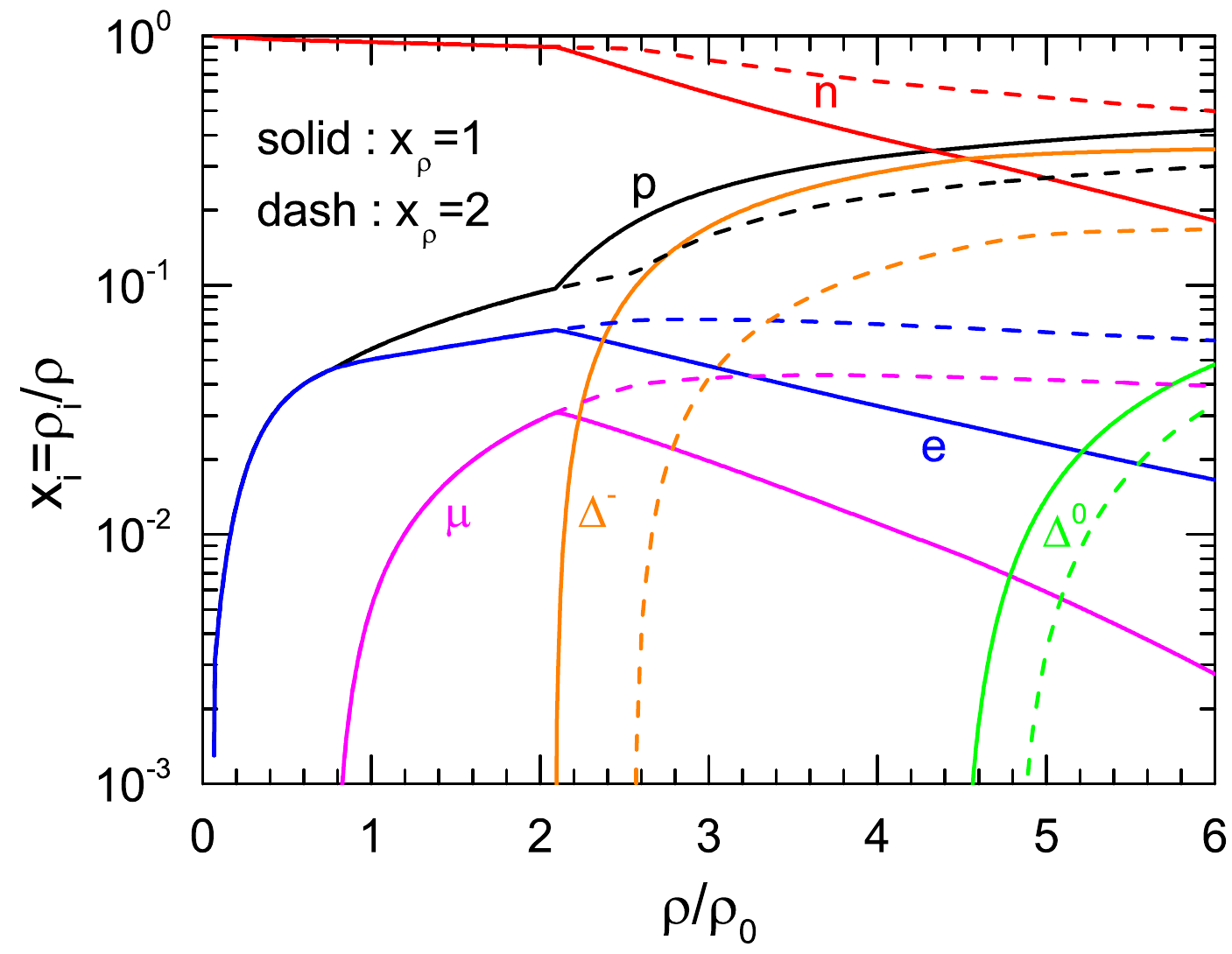} 
}
\caption{Left: the critical density for $\Delta(1232)$ formation in neutron stars as a function of the ratio of isovector coupling constants $x_{\rho}\equiv g_{\rho\Delta}/g_{\rho\textrm{N}}$ for $\Delta$ with a peak mass of 1232 MeV. Middle: the critical $\Delta$ formation density (blue) and the distribution (red) of $\Delta$ mass.} Right: the composition of neutron stars considering $\Delta$ formation within a RMF model. Taken from ref.\cite{Cai-Delta}. \label{Cai-Delta}
\end{figure*}

\subsubsection{\bf What do we know about the isovector potential of $\Delta(1232)$ resonance?}
Another important ingredient critical for predicting the $\pi^-/\pi^+$ ratio is the relative strength of $\Delta$ isovector potential with respect to that of nucleons.  Moreover, this ingredient has strong astrophysical ramifications and has been a longstanding issue in nuclear many-body theories. In fact, the spectroscopy as well as interactions of $\Delta(1232)$ resonances in neutron stars and during heavy-ion collisions have been studied extensively for a long time, see, e.g., 
Refs. \cite{Brown75,Oset81,Tak85,Oset87,Con90,Bal90,Jon92,Baldo94,pion-book,Metag,Mosel1,Mosel2}. Nevertheless, there are interesting new issues unique in neutron-rich matter that can be probed with nuclear reactions induced by rare isotopes \cite{Lenske,GSI-FRS}. Chief among the most uncertain and important physics responsible for many of the unresolved issues is the isovector potential of $\Delta$ resonances due to the $\tau_3(\Delta)\cdot \tau_3(N)$ term in the $\Delta-N$ interaction \cite{Lenske}. While there are indications from analyses of electron-nucleus, photoabsorption and pion-nucleus scatterings that the $\Delta$ isoscalar potential $V_{\Delta}$ is up to about 30 MeV more attractive than the 
nucleon isoscalar potential at $\rho_0$, essentially nothing is known about its isovector potential in isospin-asymmetric nuclear matter \cite{Mig90,VJ,Bog82,Liz97,Oli00,Kos97,Xia03,ChY07}.
Moreover, without knowing the density dependence of these potentials reduces their usefulness in simulating heavy-ion collisions or neutron stars.

The role of $\Delta$ resonances in neutron stars has long been regarded as an important issue, see, e.g., Section 8.11 of the book in Ref. \cite{Stu}. Indeed, a 10\% decrease of the isoscalar potential of \D resonances with respect to that of nucleons was found to affect the mass-radius relation of neutron stars \cite{Sch10}. Using a RMF model giving a symmetry energy of $E_{\mathrm{sym}}(\rho_0)=36.8$\,MeV and $L(\rho_0)\gtrsim 90$\,MeV at $\rho_0$ and assuming a universal baryon-meson coupling scheme in which the $\Delta$-meson couplings are set equal to the nucleon-meson couplings
($g_{\sigma\Delta}/g_{\sigma\textrm{N}}=g_{\omega\Delta}/g_{\omega\textrm{N}}=g_{\rho\Delta}/g_{\rho\textrm{N}}=1$),
the critical density \rc above which the first $\Delta^{-}(1232)$
appears was found to be above $9\rho_0$ about (25-40) years ago\,\cite{Gle85,Gle91,Gle00}. However, using softer symmetry energies with $E_{\mathrm{sym}}(\rho_0)$ and $L(\rho_0)$ values more consistent with their experimental constraints just became available in recent years, the \rc has been found consistently to be as low as $\rho_0$ based on several independent studies \cite{Dra14,Cai-Delta,Lav10,Drago14,AngLi,Kim,India,Armen}. Moreover, as shown in the left window of Fig.~\ref{Cai-Delta}, the \rc increase almost linearly with the unknown ratio of isovector coupling constants $x_{\rho}\equiv g_{\rho\Delta}/g_{\rho\textrm{N}}$ for $\Delta$ with a peak mass of 1232 MeV. As shown in the middle window, the low-mass $\Delta$ can appear even at $\rho_0$. This will affect the composition of neutron stars as shown in the right window of Fig.~\ref{Cai-Delta}.
It is also well known that the appearance of $\Delta$ in neutron stars may lead to the so-called $\Delta$-puzzle, namely, abundant populations of \D resonances soften the EOS too much to support massive neutron stars around 2.0M$_{\odot}$ observed. 

To our best knowledge, there is presently no information about how to experimentally determine the $x_{\rho}$ itself.
Since \D resonances and nucleons have an isospin 3/2 and 1/2, respectively, the total isospin is 1 or 2 for the N$\Delta$ pair, while it
is 1 or 0 for the NN pair. Because of the isospin conservation, the \D production can only happen in the total isospin 1 NN channel.
Therefore, the abundances and properties of \D resonances are sensitive to the isospin asymmetry of the system, as neutron-neutron pairs always have an isospin 1 while neutron-proton pairs can have an isospin 1 or 0. Because of the $\tau_3(\Delta)\cdot \tau_3(N)$ term in the $\Delta-N$ interactions, the isovector potentials of $\Delta^-$ and $\Delta^{++}$ are opposite in sign, i.e., $V_{\rm{asy}}(\Delta^-)=-V_{\rm{asy}}(\Delta^{++})$. The $\pi^-/\pi^+$ ratio in heavy-ion collisions through the reaction channels $(\pi^-+n\leftrightarrow\Delta^-)$ and $(\pi^++p\leftrightarrow\Delta^{++})$ is thus expected to be affected by the \D isovector potentials. Therefore, because of their ramifications in both nuclear physics and astrophysics, both the isoscalar and isovector parts of the $\Delta$ potential deserve further investigations. Naturally, the $\pi^-/\pi^+$ ratio is a potentially useful probe.

\begin{figure}
\begin{center}
\resizebox{0.45\textwidth}{!}{
\includegraphics{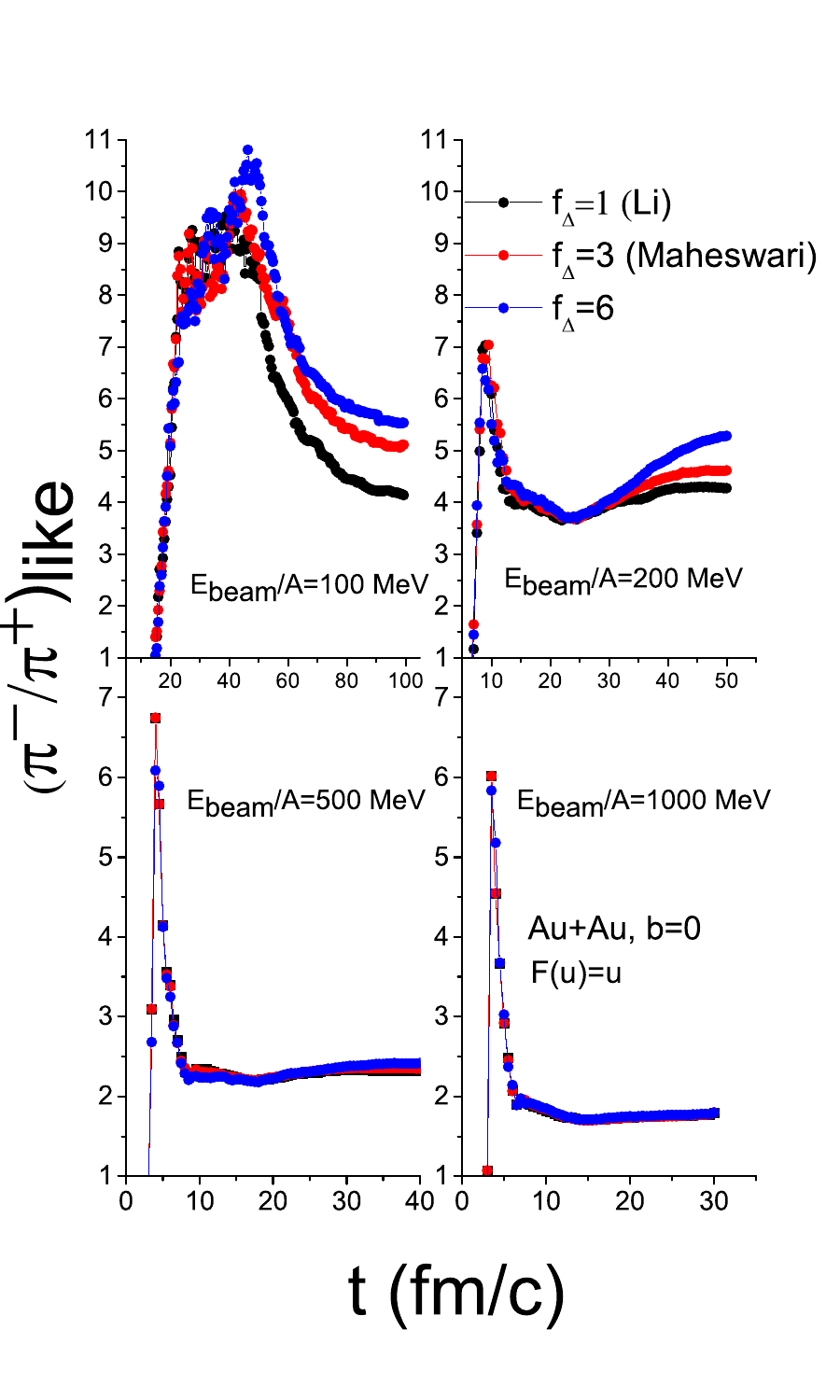} 
}
\vspace{-0.3cm}
\caption{\label{rpievo} The $(\pi^-/\pi^+)_{\rm{like}}$ ratio as a function of time in head-on Au+Au reactions at a beam energy of 100, 200, 500 and 1000 MeV/nucleon
with a symmetry potential proportional to the reduced density $u=\rho/\rho_0$ but 3 different $f_{\Delta}$ values. Taken from Ref.\cite{Li-D}.}\label{FDelta}
\end{center}
\end{figure}

\begin{figure*} [!hbt]
\centering 
\resizebox{1.\textwidth}{!}{
\includegraphics[scale=2.8]{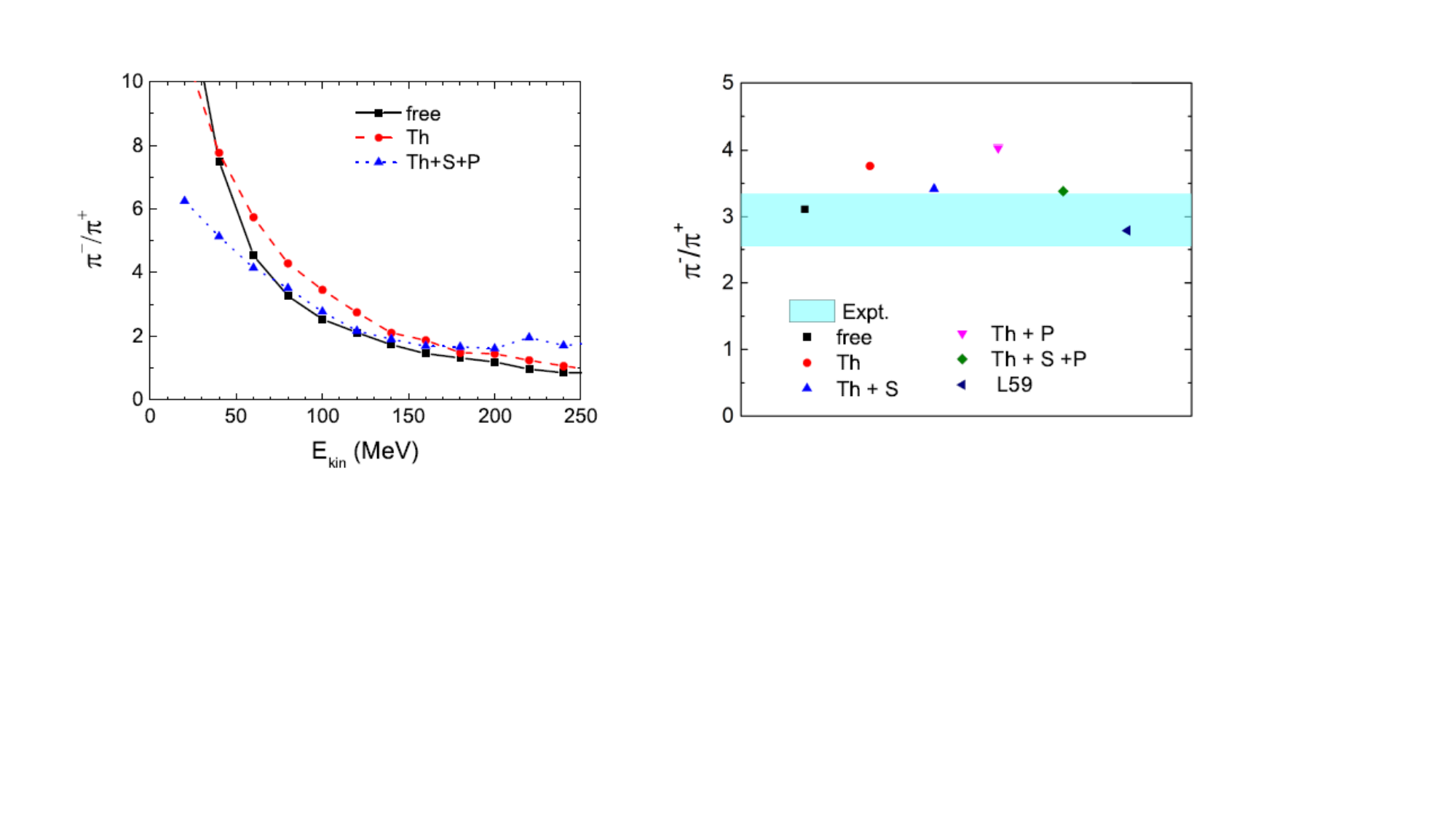} 
}
\vspace{-4.3cm}
\caption{Left: The $\pi^{-}/\pi^{+}$ ratio in Au + Au collisions at $E/A=400~\mathrm{MeV}$ and impact parameter of $1.4~\mathrm{fm}$ as a function of kinetic energy in the center-of-mass frame for the three cases of without any medium effects (free), with the in-medium threshold effect (Th), and with both the threshold and pion potential effects (Th+S+P). 
Right: The $\pi^{-}/\pi^{+}$ ratio in Au+Au collisions at impact parameter of 1.4 fm and energy of $E/A=400~\mathrm{MeV}$ from the NL$\rho$ model in different cases. The experimental data from the FOPI collaboration are shown as the cyan band. Taken from refs.\cite{Zhen17,Zhen18}
} \label{Fig-Zhang}
\end{figure*}

\subsubsection{\bf At what beam energy is the $\Delta$ potential important for modeling heavy-ion collisions?}
In simulating heavy-ion collisions, one normally assumes that the isovector potential for \D resonances is a factor $f_{\Delta}$ times that of nucleons without knowing how to fix the $f_{\Delta}$ itself \cite{lba02,Coz16,Mah98,Catania,Li-D}. As an example, assuming the $\Delta$ is a molecule consisting of a nucleon and a pion that does not have a potential, the isovector potential of the $\Delta$ resonance is an average of that for neutrons and protons with weights given by the square of the Clebsch-Gordon coefficients in the $\Delta\leftrightarrow \pi N$ processes conserving the total isospin \cite{lba02}
\begin{eqnarray}\label{dpot1}
V_{\rm{asy}}(\Delta^-)&=&V_{\rm{asy}}(n),\nonumber\\
V_{\rm{asy}}(\Delta^0)&=&\frac{2}{3}V_{\rm{asy}}(n)+\frac{1}{3}v_{\rm{asy}}(p)=\frac{1}{3}V_{\rm{asy}}(n),\nonumber\\
V_{\rm{asy}}(\Delta^+)&=&\frac{1}{3}V_{\rm{asy}}(n)+\frac{2}{3}v_{\rm{asy}}(p)=-\frac{1}{3}V_{\rm{asy}}(n),\nonumber\\
V_{\rm{asy}}(\Delta^{++})&=&V_{\rm{asy}}(p)=-V_{\rm{asy}}(n).
\end{eqnarray}
Assuming that baryon isovector potentials are all due to the $\rho$ meson exchange, neglecting possible contributions of the $\delta$-meson, within a nonlinear RMF model, it has been shown in Ref. \cite{Cai-Delta} through Eqs. (18-19) therein that the above prescription is equivalent to setting $x_{\rho}=1/3$. 

The above relations were used as a basis, and a ``$\Delta$-probing factor $f_{\Delta}$" was multiplied to them to explore effects of the $\Delta$ isovector potential on pion observables in heavy-ion collisions in Ref. \cite{Li-D}. As shown in Fig.~\ref{FDelta}, the $\Delta$ isovector potential affects the $(\pi^-/\pi^+)_{\rm{like}}$ ratio only at beam energies below the pion production threshold of about 300 MeV/nucleon (in free space without considering the Fermi motion of nucleons). This was found to be due to the fact that only low-mass \D resonances live long enough to feel the mean-field effects \cite{Li-D}. The $\Delta$ mean lifetime $\tau_{\Delta}=\hbar /\Gamma(m_{\Delta})$ is determined by its in-medium width \cite{kit}. For \D resonances near the peak 1232 MeV of its mass distribution, their mean lifetime is only about 1.7 fm/c. This is too short for these $\Delta(1232)$ resonances to feel any mean-field effect before they decay into nucleons and pions. On the other hand, low-mass \D resonances near their threshold mass at $m_{\pi}+m_N$ live much longer. They are thus more strongly affected by their mean-field potentials through the $-\nabla U\cdot dt$ term in each timestep of length $dt$ during the reaction. Therefore, the $\Delta$ potential is only important for modeling heavy-ion collisions at beam energies around and below the pion production threshold.
We notice that the study of Ref. \cite{Li-D} did not consider the time dilation effects of $\tau_{\Delta}$. The latter may make faster-moving \D resonances formed at higher energies live longer and feel more mean-field effects.  \\
\begin{figure*}[ht]
\centering 
\resizebox{0.45\textwidth}{!}{
\includegraphics[width=0.8\columnwidth,scale=0.45]{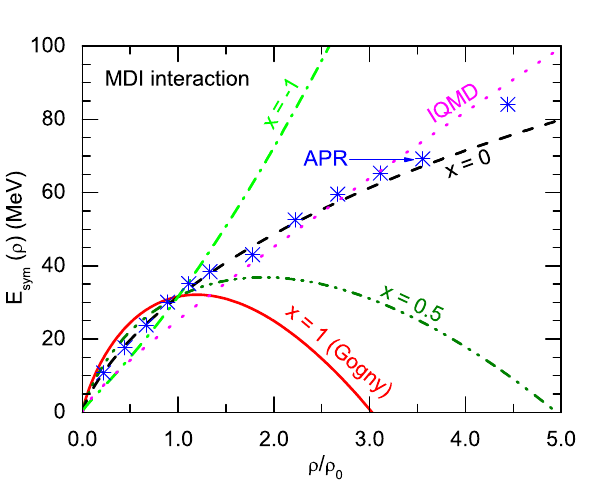}
}
\resizebox{0.42\textwidth}{!}{
\includegraphics[width=0.8\columnwidth,scale=0.45]{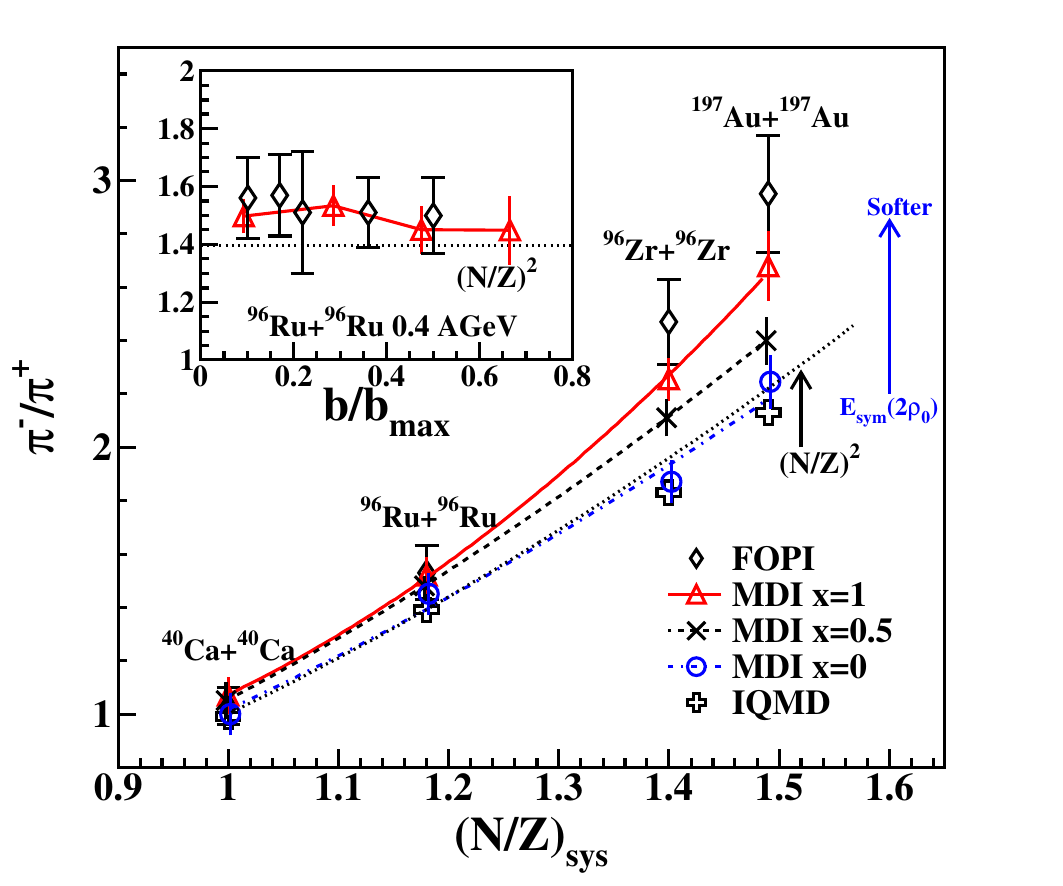}
}
\caption{Left: Density dependence of nuclear
symmetry energy \esym predicted by APR (stars),
used in IQMD (dotted line) and that using the MDI interaction with x=1, 0, 0.5 and -1, respectively.
Right: The $\pi^-/\pi^+$ ratio as a
function of the neutron/proton ratio of the reaction system at 0.4
AGeV with the reduced impact parameter of $b/b_{max}\leq 0.15$. The
inset is the impact parameter dependence of the $\pi^-/\pi^+$ ratio
for the $^{96}$Ru+$^{96}$Ru reaction at 0.4 AGeV. Taken from ref.\cite{XiaoPRL}.}\label{Xiao}
\end{figure*}
\subsubsection{\bf What do we know about the pion potential and its effects on the $\pi^-/\pi^+$ ratio?} 
Over the last decade, some of the physics ingredients not considered in IBUU04 were found to significantly affect the final 
$\pi^{-}/\pi^{+}$ ratio. For example, Refs. \cite{Zhen17,Zhen18} studied effects of the pion in-medium effects on the charged pion ratio in Au+Au collision at $E/A=400~\mathrm{MeV}$ within the RVUU model with the nucleon mean-field potentials based on the relativistic NL$\rho$ model. 
Shown in the left window of Fig.~\ref{Fig-Zhang} is the differential pion ratio as a function of pion kinetic energy for the three cases: (1) without any medium effects (free), (2) with the in-medium threshold effect (Th), and (3) with both the threshold and pion potential effects (Th+S+P).  It is seen that the low-energy part of the pion ratio is strongly affected by the pion in-medium effects considered. 

The right window shows the in-medium effects on the total ratio of charged pions. Quantitatively, results from six different cases are compared with the FOPI data: (i) without the threshold and pion in-medium effects (free), namely, nucleons, $\Delta$ resonances and pions are treated as free particles in all reactions; (ii) with only the threshold effect (Th); (iii) with the threshold effect and the pion $s$-wave potential (S); (iv) with the threshold effect and the pion $p$-wave potential (Th+P); (v) with the threshold effect and both the pion $s$-wave and $p$-wave potentials (Th+S+P); (vi) same as case (v) but with the coupling constant $f_{\rho}=0.95~\mathrm{fm}^2$ of the isovector-vector $\rho$ meson to nucleon in the NL$\rho$ model reduced to $f_{\rho}=0.43~\mathrm{fm}^2$. They found that the threshold effect substantially increases the $\pi^-/\pi^+$ ratio by about $20\%$.  For the effects of pion potentials, it is seen that the pion $s$-wave potential reduces and the $p$-wave potential enhances the $\pi^-/\pi^+$ ratio. Consequently, including both potentials leads to a significant decrease ($\sim 10\%$) of the $\pi^-/\pi^+$ ratio. They explained that the effect of $s$-wave potential can be easily understood since at densities below about $2\rho_0$, the effective masses of $\pi^-$ and $\pi^0$ increase while the $\pi^+$ effective mass slightly decreases.  The enhancement of the $\pi^-/\pi^+$ ratio after the inclusion of the pion $p$-wave potential is due to the softer dispersion relation of $\pi^{-}$ in the pion branch. The pion in-medium effects on the $\pi^{-}/\pi^{+}$ ratio obtained in their study are qualitatively consistent with the results found in Refs.~\cite{Xu10,Xu13} based on a thermal model. 

It is also seen that their prediction on the $\pi^-/\pi^+$ ratio using the RVUU model (originally giving $L=84$ MeV by default) with both the threshold effect and the pion in-medium effect is slightly larger than the upper bound of the experimental data. Nevertheless, reducing the symmetry energy slope parameter from 84 MeV to $L=59~\mathrm{MeV}$, the $\pi^-/\pi^+$ ratio is reduced and becomes consistent with the experimental results. 

We notice that there are ongoing efforts to further investigate pion and $\Delta$ potentials as well as their connections with nucleon potentials using observables of heavy-ion collisions, see, e.g., Refs. \cite{Nat25,Dan25}. Effects of pion potential are normally intertwined with other issues in simulating heavy-ion collisions. However, not all studies have necessarily incorporated the pion potential consistently in all processes. For instance, the pion potential may also affect how the $\Delta$ potential is related to the nucleon potential. Thus, there is presently no community consensus on the form and effects of pion potential on the $\pi^-/\pi^+$ ratio in heavy-ion collisions.

Moreover, electromagnetic potentials also affect the $\pi^-/\pi^+$ ratio, especially at low kinetic energies during heavy-ion collisions, see, e.g., Refs.\cite{nature,Wolf,Lib79,Ben,Gyu81,Sch82} for examples of earlier studies. They need to be carefully incorporated in transport models \cite{Pawel,Li95,OuLi,Wei17,Stone1,Stone2} to extract reliable information about the nuclear EOS, especially the relatively weak nucleon isovector potential and/or pion potential. Effects of the latter may be compatible with those due to the electromagnetic field, depending on the isospin asymmetry and density reached, as well as the geometry (impact parameter, orientation, and deformations, etc) during the reaction. 

\subsubsection{\bf Where are we now on using the $\pi^-/\pi^+$ ratio as a probe of high density symmetry energy?}
Given the discussions above, it is easy to understand that no community-wide consensus has been reached about using the $\pi^{-}/\pi^{+}$ ratio from heavy-ion reactions to probe the high-density nuclear symmetry energy. Nevertheless, some interesting information about model ingredients affecting the $\pi^{-}/\pi^{+}$ ratio has been revealed in several studies. 

Of course, conclusions regarding the high-density behavior of nuclear symmetry energy based solely on comparing individual transport model calculations with the available data have been model-dependent and sometimes controversial. For example, as shown in Fig.~\ref{Xiao}, a rather stimulating finding in an earlier analysis using the IBUU04 code is that the FOPI pion production data favors a super-soft symmetry energy corresponding to the original prediction of the GHF with $x=1$ (red cure shown in the left window) \cite{XiaoPRL}.  As shown on the right, using the super-soft symmetry energy with $x=1$, the predicted $\pi^{-}/\pi^{+}$ ratio is close to but still below the FOPI Au+Au data at a beam energy of 400 MeV/nucleon. While it is largely agreed that a softer symmetry energy at suprasaturation densities leads to a higher $\pi^{-}/\pi^{+}$ ratio, the quantitative results from the IBUU04 calculations can be reproduced by some other transport models but challenged by others, see, e.g., Refs. \cite{Fen10,Xie13}. As shown in the left window of Fig.~\ref{Xiao}, the $E_{\rm{sym}}(2\rho_0)$ with $x=1$ is only about 25 MeV, which is approximately half of what Siemens' $\rho^{2/3}$ scaling and several microscopic theories have predicted, as shown in Fig. \ref{Esym-survey}. 

Moreover, more recent analyses of the SpiRIT data on pion production have extracted a value of $L= 79.9\pm 37.6$ MeV with $E_{\rm{sym}}(\rho_0)= 35.3\pm 2.8$ MeV using Cozma's QMD code \cite{SpiRIT:2021gtq}. However, since the results of calculations with seven transport codes scatter broadly around the SpiRIT data on pion production \cite{TRS5}, no model-independent conclusion has been drawn about the high-density behavior of symmetry energy from analyzing this data either.
Altogether, it is true that we presently can't draw any model-independent conclusion regarding the high-density symmetry energy based on comparing transport model calculations with the available pion data. 

In principle, model-independent conclusions can only be drawn when all the uncertainties regarding both nuclear in-medium effects and transport model simulations of heavy-ion collisions at intermediate energies are much more thoroughly understood. In practice, the history of heavy-ion physics indicates that this is necessarily a long process. Then, is it really hopeless to learn anything useful anytime soon about the high-density symmetry energy from pions or generally any observable in heavy-ion collisions? 

Indeed, there are numerous unquantified uncertainties, making it challenging to analyze heavy-ion reaction data without relying on complex dynamical models. Nevertheless, strong encouragement can be drawn from the remarkable success of detecting and analyzing gravitational waves from the occasional NS mergers occurring in distant regions of space, as demonstrated by the astrophysics community. In the shared pursuit of inferring the EOS of supradense nuclear matter, analyses of both heavy-ion collision data and observations from NS mergers entail significant uncertainties, assumptions, and model dependence. For instance, it is well known that the lower limits of the tidal deformabilities or radii of the two NSs involved in GW170817 are highly dependent on the model, assumptions, and prior knowledge, as discussed in detail in Ref. \cite{Tide-NS}. To the best of our knowledge, the challenges associated with interpreting observables from heavy-ion collisions are no greater than those encountered when analyzing NS mergers. Therefore, we should remain optimistic!

\begin{figure}[thb]
\centering
  \resizebox{0.45\textwidth}{!}{
  \includegraphics{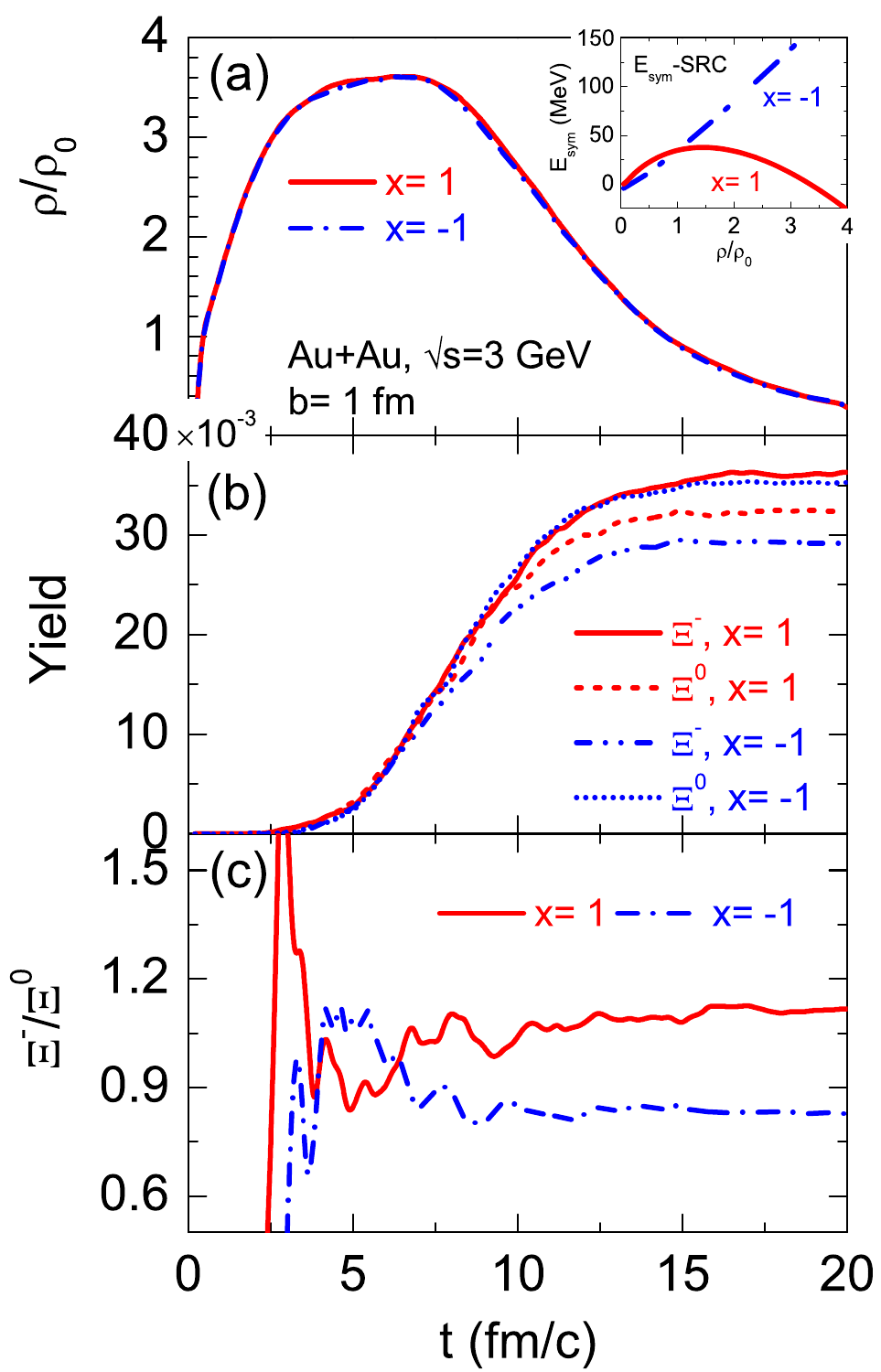}
  }
\caption{Evolutions of central compression densities (a), yields of $\Xi^{-}$ and $\Xi^{0}$ (b) and the ratios of $\Xi^{-}/\Xi^{0}$ (c) in central Au+Au reactions at $\sqrt{s_{NN}}$ = 3 GeV with a soft ( $x = 1$) and a stiff ($x = -1$) symmetry energy shown in the inset of panel (a). Taken from Ref. \cite{Yong22}.} \label{evo}
\end{figure}
\subsection{\bf Example-4: Strange particle ratios $K_{s}^{0}/K^{+}$, $\Sigma^{-}/\Sigma^{+}$ and $\Xi^{-}/\Xi^{0}$ from relativistic heavy-ion collisions}
The isospin asymmetry of the nucleonic component of dense matter formed in heavy-ion reactions varies with beam energy, and once the quark deconfinement happens or hyperons start forming, the meaning and definition of $E_{\mathrm{sym}}(\rho)$ may need to be generalized (see discussions in Section \Ref{G-Esym}). Thus, heavy-ion reactions at energies that create hadronic matter as dense as possible but not high enough to form the quark-gluon plasma (QGP) yet are particularly useful for probing the high-density $E_{\mathrm{sym}}(\rho)$. Of course, one wishes to make the dense matter as neutron-rich as possible using probably high-energy radioactive beams. 

Very interestingly, the STAR Collaboration found that the number of constituent quarks (NCQ) scaling in collective flow disappears and all collective flow data can be well described by hadronic transport models using nuclear mean-field potentials in $\sqrt{s_{NN}}$ = 3 GeV Au+Au reactions \cite{STAR1,STAR2}. More recently, they found that, as the collision energy increases, a gradual evolution to NCQ scaling (which is a known signature of QGP formation) is observed. Their data on the beam-energy dependence of elliptic flow for all hadrons studied provides evidence for the onset of dominant partonic interactions by $\sqrt{s_{NN}}$=4.5 GeV \cite{STAR25}. Therefore, central heavy-ion collisions below $\sqrt{s_{NN}}$ = 3 GeV are safely expected to form dense hadronic matter at baryon density less than about $(3.6-4.0)\rho_0$ depending on the hadronic EOS used, as shown in the upper window of Fig. \ref{evo} \cite{Yong22}.

There are strong interests in probing the high-density $E_{\mathrm{sym}}(\rho)$ using strange particles as an integral part of the science missions of several intermediate-relativistic heavy-ion reaction facilities under construction, see, e.g, Refs. \cite{Wolfgang,Xiao14,Hong14,GSI,Senger}. Since strange particles are rarely absorbed by the surrounding medium on their way out, they have long been used in probing dense matter EOS, see, e.g., Refs. \cite{raf82,jor1985,fuchs2001,kaon2006,ditoro,fuchs2006,feng11,liq05,qm2018,adam20,casyong2021}.
In particular, kaons have been thoroughly studied in the literature \cite{jor1985,fuchs2001,kaon2006,ditoro,fuchs2006,feng11}. Moreover, the singly strange $\Lambda$ and $\Sigma$ hyperons have been studied experimentally, see, e.g., Ref. \cite{chung2001}. Their connections to the nuclear EOS were explored theoretically \cite{liq05,fengnpa2013,fengcpl2021}. It was also proposed that 
the $K^0/K^+$ inclusive yield ratio is a more sensitive probe of high-density $E_{\mathrm{sym}}(\rho)$ than the $\pi^{-}/\pi^+$ ratio \cite{ditoro}. Indeed, the FOPI Collaboration investigated the double $K^+/K^0$ ratio in two isobar reactions (
Ru + Ru and Zr + Zr) at a beam energy of 1.528 A GeV \cite{FOPI-kaon}. Unfortunately, in comparison with thermal and transport model calculations, no clear conclusion was made about the high-density $E_{\mathrm{sym}}(\rho)$ from these studies. For more detailed discussions of early work on the $K_{s}^{0}/K^{+}$ and $\Sigma^{-}/\Sigma^{+}$ ratios, we refer the reader to an earlier review in Section 7.15 of Ref. \cite{LCK08}.

The doubly strange $\Xi$ production has also been continuously studied over the last two decades, see, e.g., Refs. \cite{exp2003,exp2009,exp2015,urqmd2014,urqmd2016,buu2018,ko2002npa,ko2004npa,ko2004plb,chen2004plb,li2012prc}. It is produced mainly from collisions of two singly strange particles. Its fraction in the central participant region was found to be more than twice that of $K^{+}$ or $\Lambda+\Sigma^{0}$ \cite{casyong2021}. It may thus be more useful than singly strange hadrons in probing the EOS of dense matter \cite{casyong2021}. However, it has been recognized that its dependence on the SNM EOS is less known. Moreover, its elementary production cross sections are largely uncertain. To reduce these difficulties and see the relative effects of nuclear symmetry energy, a comparative study of the $\Xi^{-}/\Xi^{0}$ ratio in comparison with the isospin multiplet ratios of other particles, namely, the $n/p$, $\pi^{-}/\pi^{+}$, $K_{s}^{0}/K^{+}$, and $\Sigma^{-}/\Sigma^{+}$ ratios was carried out recently in Ref. \cite{Yong22} for $\sqrt{s_{NN}}$ = 3 GeV Au+Au reactions using the ART (A Relativistic Transport) model \cite{ART1,art2001}. The latter has been extensively used and continuously improved in several aspects \cite{ko2002npa,ko2004npa,ko2004plb,chen2004plb,li2012prc,deu2009} by the community. It is used as the hadronic afterburner in the publicly available AMPT (A Multiphase Transport) package \cite{AMPT2005} for simulating relativistic heavy-ion collisions from RHIC-BES to LHC energies, see Ref. \cite{Lin21-review} for a recent review. 

To investigate the effects of high-density $E_{\mathrm{sym}}(\rho)$ on the $\Xi^{-}/\Xi^{0}$ ratio, the momentum-dependent isoscalar and isovector single-nucleon mean-field potentials \cite{Das03} were adopted in Ref. \cite{Yong22}. The mean-field potentials for strange baryons $\Lambda$, $\Sigma$, $\Xi$ were obtained by adopting the quark counting rule asserting that these strange baryons interact with other baryons only through their non-strange (2/3, 2/3, 1/3) constituents \cite{chung2001,mos74}. Moreover, the known decay branching ratios of different isospin multiplets of these strange baryons were used to determine the relationships between their mean-field potentials and those for neutrons and protons, respectively. More specifically, one has \cite{Yong22}
\begin{eqnarray}
U_{\Lambda} &=& 2/3(1/3U_{n}+2/3U_{p}), \nonumber\\
U_{\Sigma^{-}} &=& 2/3U_{n}, \nonumber\\
U_{\Sigma^{0}} &=& 2/3(1/3U_{n}+2/3U_{p}), \nonumber\\
U_{\Sigma^{+}} &=& 2/3(1/2U_{n}+1/2U_{p}), \nonumber\\
U_{\Xi^{-}} &=& 1/3(1/3U_{n}+2/3U_{p}), \nonumber\\
U_{\Xi^{0}} &=& 1/3(1/3U_{n}+2/3U_{p}).
\end{eqnarray}
In this study, the form of kaon potential was taken from Ref.~\cite{ligq97}. However, for pions, no nuclear mean-field except the Coulomb potentials were used. Besides the strangeness exchange reactions $\bar{K}$+$Y$ $\leftrightarrow$ $\pi$+$\Xi$ (Y = $\Lambda$ or $\Sigma$) that are already in the AMPT package, the isospin-averaged cross sections were used for simulating the $Y$+$Y$ $\leftrightarrow$ $N$+$\Xi$ reactions and $\Xi$ productions via the $Y$+$N$ $\rightarrow$ $N$+$\Xi$+$K$ processes \cite{urqmd2014,urqmd2016,buu2018,li2012prc}.

Shown in Fig.~\ref{evo} are the $\Xi^{-}$ and $\Xi^{0}$ yields (panel b) and their ratio (panel c) as functions of the reaction time. While both yields saturate at about t = (15-20) fm/c, their ratio is seen to saturate much earlier, revealing key information about the early high-density phase of the reaction. Moreover, the yields depend obviously on the symmetry energy parameter $x$. As it was discussed in Ref. \cite{Yong22}, the $\Xi^{-,0}$ are mostly produced through the $\Lambda\Sigma^{-,+}\rightarrow\Xi^{-,0}$ reaction channels. Since pions involved in the $\pi^{-,+}N(n,p)\rightarrow\Sigma^{-,+}$ reactions are mostly from nn (pp)$\rightarrow\pi^{-,+}$ scatterings, practically there is an effective reaction path of nn (pp)$\rightarrow\Xi^{-,0}$. As discussed earlier and shown in Fig.~\ref{Xiao12}, because of the isospin fractionation phenomenon,
a soft $E_{\mathrm{sym}}(\rho)$ above $\rho_0$ make the compressed region more neutron-rich, first resulting in more (less) $\pi^- (\pi^+)$ particles and finally a higher $\Xi^{-}/\Xi^{0}$ ratio. Indeed, as shown in the panel(c), using a soft $E_{\mathrm{sym}}(\rho)$ with $x=1$ the $\Xi^{-}/\Xi^{0}$ ratio is about $30\%$ higher compared to the stiff case ($x=-1$). 

\begin{figure}[thb]
\centering
\includegraphics[width=0.45\textwidth]{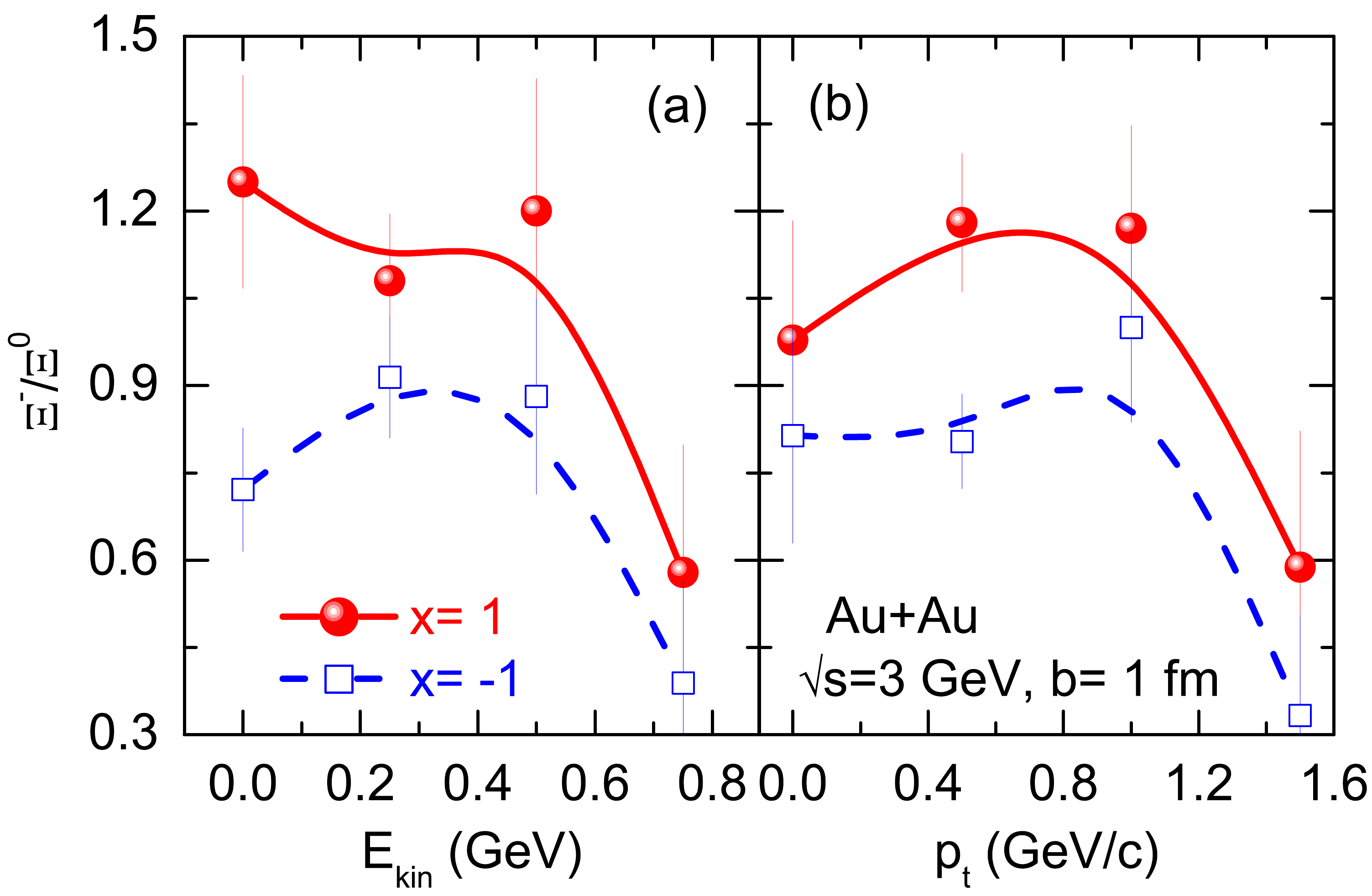}
\caption{Kinetic energy (a) and transverse momentum (b) distributions of the doubly strange baryon $\Xi^{-}/\Xi^{0}$ ratio in the central Au+Au reactions at $\sqrt{s_{NN}}$ = 3 GeV with the stiff ($x=-1$) and soft ($x=1$) symmetry energies, respectively. The symbols with error bars are results of simulations, while the curves are drawn to guide the eye.} Taken from Ref. \cite{Yong22}. \label{rcas}
\end{figure}

\begin{figure}
   \resizebox{0.45\textwidth}{!}{
  \includegraphics{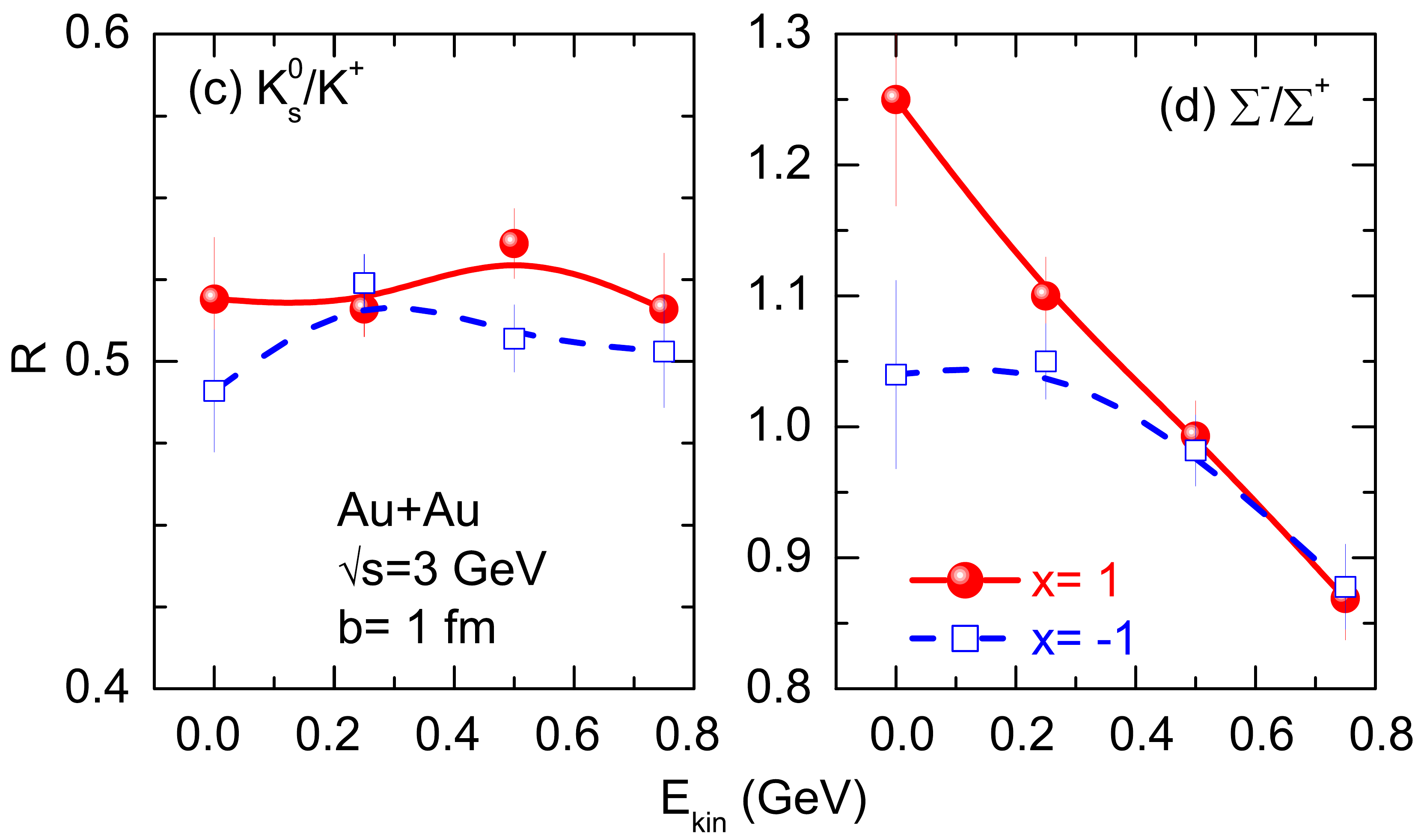}
    }
 \caption{The Kinetic energy distributions of $K_{s}^{0}/K^{+}$, and $\Sigma^{-}/\Sigma^{+}$ ratios in the central Au+Au reactions with the stiff ($x=-1$) and soft ($x=1$) symmetry energies at $\sqrt{s_{NN}}$ = 3 GeV. The symbols with error bars are results of simulations, while the curves are drawn to guide the eye.} Taken from Ref. \cite{Yong22}.\label{nonstrange}    
\end{figure}

The $E_{\mathrm{sym}}(\rho)$ effect is more clearly seen in the differential $\Xi^{-}/\Xi^{0}$ ratio as a function of kinetic energy or transverse momentum in Fig. \ref{rcas}. For a comparison, the transverse momentum dependences of $K_{s}^{0}/K^{+}$ and $\Sigma^{-}/\Sigma^{+}$ ratios are shown in Fig.~\ref{nonstrange}. While the $E_{\mathrm{sym}}(\rho)$ affect the $K_{s}^{0}/K^{+}$ ratio by only about 6\%, its effect on the $\Sigma^{-}/\Sigma^{+}$ ratio is as much as 20\% at low kinetic energies. However, the latter is still much less than the effect on the differential $\Xi^{-}/\Xi^{0}$ ratio over a broad range of kinetic energy and/or transverse momentum. 

As discussed in more detail in Ref. \cite{Yong22}, the elementary reaction nn (pp) $\rightarrow\Sigma^{-,+}$ dominates the $\Sigma$ production via the intermediate step $\pi+N (\rm {n~ or~ p})\rightarrow\Sigma$. The $E_{\mathrm{sym}}(\rho)$ information carried by the primordial pion ratio $\pi^{-}/\pi^{+}$ is then passed to the $\Sigma^{-}/\Sigma^{+}$ ratio. These $\Sigma$ hyperons mostly have low kinetic energies as a result of the reaction kinematics, while the more energetic ones are produced through reactions involving two baryons that carry less isospin information than the primordial pions. As we discussed earlier in this review, the primordial $\pi^{-}/\pi^{+}$ ratio is expected to be proportional to $(n/p)^2_{\rm{like}}$ of the participant region. The later is higher with the softer ($x=1$) $E_{\mathrm{sym}}(\rho)$ at suprasaturation densities. After the energetic pions have been converted to strange mesons and hyperons, the remaining ones will have a reduced $\pi^{-}/\pi^{+}$ ratio. While the opposite is expected for the newly produced particles through the scatterings involving pions. Indeed, the $K_{s}^{0}/K^{+}$, and $\Sigma^{-}/\Sigma^{+}$ ratios are higher with the softer ($x=1$) $E_{\mathrm{sym}}(\rho)$. This isospin information is subsequently passed to the $\Xi^{-}/\Xi^{0}$ ratio mostly through the intermediate $\Lambda\Sigma^{-,+}\rightarrow\Xi^{-,0}$ reaction channels \cite{Yong22}.

Based on the findings summarized above, one expects the $\Sigma^{-}/\Sigma^{+}$ and $\Xi^{-}/\Xi^{0}$ ratios in high-energy heavy-ion collisions to be useful for probing the high-density behavior of $E_{\mathrm{sym}}(\rho)$. We emphasize here that for these observables to work effectively as a robust probe of high-density $E_{\mathrm{sym}}(\rho)$, understanding well the $\pi+N+\Delta$ dynamics in heavy-ion reactions at intermediate beam energies discussed in the previous subsection is a prerequisite. This is because the isospin effects on the ratios of heavy strange particles are mostly through the $\pi^-/\pi^+$ ratio originally from the elementary nn/pp scatterings in dense matter. Thus, understanding the $\pi^-/\pi^+$ ratio at relatively low beam energies is still the most critical task at this time. 

\begin{table}[ht]
\centering
\caption{Benchmark values of symmetry-energy parameters from Siemens’ scaling compared with empirical constraints.}
\begin{tabular}{@{}lcc@{}}
\hline\hline
Parameter & Siemens’ $\rho^{2/3}$ scaling & Empirical constraints \\ 
\hline
$E_{\mathrm{sym}}(\rho_0)$ & $\sim 31$ MeV & $31 \pm 3$ MeV \\
$L(\rho_0)$ & $2E_{\mathrm{sym}}(\rho_0) \approx 60$ MeV & $60 \pm 20$ MeV \\
$K_{\mathrm{sym}}(\rho_0)$ & $-2E_{\mathrm{sym}}(\rho_0) \approx -60$ MeV & $-100 \pm 100$ MeV  \\
$E_{\mathrm{sym}}(2\rho_0)$ & $1.58 E_{\mathrm{sym}}(\rho_0) \approx 50$ MeV & $45$–$55$ MeV \\
$J_{\mathrm{sym}}(\rho_0)$ & $\sim 8E_{\mathrm{sym}}(\rho_0) \approx 250$ MeV & Poorly constrained \\ 
\hline\hline
\end{tabular}
\label{tab:benchmarks}
\vspace{-0.5cm}
\end{table}
\section{Summary and outlook}
In summary, the EOS of dense neutron-rich matter remains one of the most challenging problems in nuclear physics and astrophysics. The nuclear symmetry energy, $E_{\mathrm{sym}}(\rho)$, which quantifies the energy cost of converting protons into neutrons, is particularly uncertain above $\rho_0$. Its density dependence has profound implications: it affects neutron-star radii, tidal deformability, oscillation modes, cooling rates, and gravitational-wave signals from binary mergers, as well as many aspects of nuclear structure and heavy-ion collisions. 

While nuclear experiments, astrophysical observations, and microscopic nuclear many-body theories constrain $E_{\mathrm{sym}}(\rho)$ up to about $2\rho_0$, predictions diverge broadly at higher densities. Ongoing and planned heavy-ion reaction programs (e.g., GANIL, RIKEN, FRIB+FRIB400, SIS+FAIR, NICA, CSR+HIAF, RAON) and precision multimessenger astrophysics (e.g., NICER, LIGO/VIRGO, future X-ray missions, and more advanced gravitational wave detectors) are expected to significantly improve constraints on the high-density behavior of $E_{\mathrm{sym}}(\rho)$. 

Siemens' $\rho^{2/3}$ scaling for $E_{\mathrm{sym}}(\rho)$ up to about $2\rho_0$ provides a benchmark for theory and experiment.
As summarized in Table \ref{tab:benchmarks}, there is strong empirical evidence supporting Siemens' $\rho^{2/3}$ scaling. In particular, surveys of analyses of terrestrial experiments including nuclear masses, neutron-skins of heavy nuclei, isospin diffusion, electrical dipole polarizability, differences in charge radii of mirror nuclei, etc, combined with measurements of neutron-star radii using X-rays and tidal deformations through gravitational waves, consistently yield
$
E_{\mathrm{sym}}(\rho_0) \approx 31 \pm 3 \, \mathrm{MeV}$ and $L(\rho_0) \approx 60 \pm 20 \, \mathrm{MeV}.
$
These values satisfy $L \simeq 2 E_{\mathrm{sym}}(\rho_0)$, consistent with $\rho^{2/3}$ scaling. Experimentally extracted values and theoretical predictions for $K_{\mathrm{sym}}$ and $E_{\mathrm{sym}}(2\rho_0)$ also agree broadly with the scaling. However, Siemens’ $\rho^{2/3}$ scaling is expected to break down at higher densities. Microscopic causes include: (1) Strong density dependence of the tensor force in the neutron-proton isosinglet channel, (2) Contributions from three-body forces, (3) Isospin-dependent nucleon short-range correlations altering kinetic symmetry energy, and (4) Relativistic effects and non-quadratic momentum dependence of the isoscalar potential. Interestingly, transport-model analyses of relative flows of nucleons and light clusters in heavy-ion reactions at 400 MeV/nucleon (e.g., ASY-EOS, FOPI-LAND) show hints of stiffening symmetry energy starting around $2\rho_0$, consistent with an expected deviation from the $\rho^{2/3}$ scaling at suprasaturation densities.

There are significant challenges to better constrain the high-density behavior of nuclear symmetry energy. In particular, we emphasize the following:
\begin{itemize}
  \item Large uncertainties in curvature ($K_{\mathrm{sym}}$) and skewness ($J_{\mathrm{sym}}$) parameters, critical for high-density extrapolations of symmetry energy.
  \item Poorly constrained spin–isospin dependent tensor forces and three-body forces as well as the resulting nucleon short-range correlations in dense neutron-rich matter, which strongly influence $E_{\mathrm{sym}}(\rho)$ at suprasaturation densities.

\item Medium dependence of strong interactions as well as possible appearance of new phases and particles (e.g., hyperons, $\Delta$ resonances, dark matter).

  \item Discrepancies among transport models in extracting $E_{\mathrm{sym}}(\rho)$ especially at high densities from heavy-ion collisions.
  \item Limited data and lack of consensus on neutron–proton effective mass splitting in neutron-rich matter, which feeds back into transport model calculations and EOS constraints from nuclear reactions.
  \item Astrophysical data limited in variety and precision: current radius/tidal-deformability measurements mainly probe $E_{\mathrm{sym}}$ around ($2$–$3)\rho_0$, not deeper neutron-star cores. 
  \item Intrinsic composition degeneracy of TOV equations enables many different EOS models to reproduce equally well the same mass-radius sequence as long as the same EOS is constructed, regardless of what ingredients are considered.
\end{itemize}
Looking forward, combining multimessengers from both astrophysical observations and terrestrial nuclear experiments is the most promising path to finally pin down the high-density symmetry energy and thus the EOS of supradense neutron-rich matter. High-energy heavy-ion programs at various facilities are poised to make major contributions to this important scientific endeavor.

\section*{Disclaimer}
Given the limitations of the present author’s knowledge, this review is necessarily incomplete and may reflect certain biases in its coverage. Nevertheless, it is hoped that the discussion presented here will provide a useful contribution to this rapidly evolving field, in which many important and intriguing questions remain to be explored.

\section*{Acknowledgement}
We would like to thank Bao-Jun Cai, Lie-Wen Chen, Pawel Danielewicz, Massimo Ditoro, Subal Das Gupta, Farrukh J. Fattoyev, Wen-Jun Guo, Xavier Grundler, Xiao-Tao He, Brayden Jeanotte, Wei-Zhou Jiang, Che-Ming Ko, Plamen Krastev, Ang Li, Xiao-Hua Li, James Lattimer, William G. Lynch, Macon Magno, Yu-Gang Ma, Joe Natowitz, William G. Newton, Li Ou, Jake Richter, Andrew Steiner, Wolfgang Trautmann, Betty Tsang, Isaac Vidana, De-Hua Wen, T.R. Whitehead, Hermann Wolter, Wen-Jie Xie, Zhigang Xiao, Chang Xu, Jun Xu, Nu Xu, Gao-Chan Yong and Nai-Bo Zhang for helpful discussions on some of the issues reviewed in this paper.

\section*{Statements and Declarations}

\begin{itemize}

\item {\bf Funding:} This work was supported in part by the U.S. Department of Energy, Office of Science, under Award Number DE-SC0013702, the CUSTIPEN (China-U.S. Theory Institute for Physics with Exotic Nuclei) under the US Department of Energy Grant No. DE-SC0009971.

\item {\bf Data availability statement:} The author declares that the data supporting the findings of this study are available within the paper.

\item {\bf No financial interest:} The author has no relevant financial or non-financial interests to disclose. 

\item {\bf No competeting interest:} The sole author has no competing interests to declare that are relevant to the content of this article. The sole author of this particle certifies that he has no affiliations with or involvement in any organization or entity with any financial interest or non-financial interest in the subject matter or materials discussed in this manuscript.

\item {\bf Author contribution:} This review article is solely authored by Bao-An Li.
\end{itemize}


\newcommand{\apjl}{Astrophys. J. Lett.\ }
\newcommand{\apj}{Astrophys. J. \ }
\newcommand{\prl}{Phys. Rev. Lett.\ }
\newcommand{\prc}{Phys. Rev. C\ }
\newcommand{\prd}{Phys. Rev. D\ }
\newcommand{\mnras}{Mon. Not. R. Astron. Soc.\ }
\newcommand{\aap}{Astron. Astrophys.\ }
\newcommand{\nphysa}{Nucl. Phys. A\ }
\newcommand{\physrep}{Phys. Rep.\ }
\newcommand{\nat}{Nature\ }


\end{document}